\documentclass[aps,amsmath,twocolumn,floatfix,superscriptaddress,pra,footinbib,
10pt]{revtex4-2}
\usepackage{physics}
\usepackage{graphicx} 
\usepackage{dcolumn} 
\usepackage{bm} 
\usepackage{amsfonts} 
\usepackage{braket} 
\usepackage{ulem} 
\usepackage[usenames,dvipsnames]{color}
\usepackage[%
  colorlinks=true,
  urlcolor=blue,
  linkcolor=blue,
  citecolor=blue
]{hyperref}
\usepackage{siunitx}
\usepackage{mhchem}
\usepackage{comment}
\usepackage{mathrsfs}

\newcommand{\Asf}{\mathsf{A}}
\newcommand{\ee}{\mathrm{e}}
\newcommand{\ii}{\mathrm{i}}
\newcommand{\mD}{\hat{\mathcal{D}}}
\newcommand{\mL}{\hat{\mathcal{L}}}
\newcommand{\hsb}{\hat{H}_{sb}}
\newcommand{\rhoness}{\hat{\rho}_\mathrm{ness}}
\newcommand{\rhofg}{\hat{\rho}_\mathrm{FG}}
\newcommand{\lamb}{\hat{\Lambda}^\mathrm{LS}}

\newcommand{\tred}[1]{\textcolor{red}{#1}}

\begin{document}

\title{Floquet theory and applications in open quantum and classical systems}%

\author{Masahiro Sato}%
\email{sato.phys@chiba-u.jp}
\affiliation{Department of Physics, Chiba University, Chiba 263-8522, Japan}

\author{Tatsuhiko N. Ikeda}
\email{tatsuhiko\_ikeda@zen.ac.jp}
\affiliation{Faculty of Social Informatics, ZEN University, Zushi, Kanagawa, 249-0007, Japan}
\affiliation{RIKEN Center for Quantum Computing, Wako, Saitama 351-0198, Japan}

\date{\today}%

\begin{abstract}
This article reviews theoretical methods for analyzing Floquet engineering (FE) phenomena in open (dissipative) quantum or classical systems, with an emphasis on our recent results. 
In many theoretical studies for FE in quantum systems, researchers have used the Floquet theory for closed (isolated) quantum systems, that is based on the Schr\"odinger equation. 
However, if we consider the FE in materials driven by an oscillating field like a laser, a weak but finite interaction between a target system and an environment (bath) is inevitable. 
In this article, we describe these periodically driven dissipative systems by means of the quantum master (GKSL) equation. In particular, we show that a nonequilibrium steady state appears after a long driving due to the balance between the energy injection by the 
driving field and the release to the bath.  
In addition to quantum systems, if we try to simply apply Floquet theory to periodically driven classical systems, it failed because the equation of motion (EOM) is generally nonlinear, and the Floquet theorem can be applied only to linear differential equations. Instead, by considering the distribution function of the classical variables (i.e., Fokker-Planck equation), one can arrive at the effective EOM for the driven systems. We illustrate the essence of the Floquet theory for classical systems. On top of fundamentals of the Floquet theory, we review representative examples of FEs (Floquet topological insulators, inverse Faraday effects in metals and magnets, Kapitza pendulum, etc.) and dissipation-assisted FEs.  
\end{abstract}
\maketitle

\tableofcontents

\section{Introduction}
Floquet engineering (FE) refers to the control of physical properties of target systems by applying a time-periodic external field~\cite{Goldman2014,Bukov2015,Holthaus2015,Eckardt2017,Oka2019,Sato2021book,Tsuji2024}. Thanks to the development of laser science and technology, one can use electromagnetic waves in a broad frequency range from terahertz (THz) to X-ray. 
In principle, it is possible to perform FE in solids by applying an intense laser to them~\cite{Oka2019,Sato2021book,Tsuji2024}.   
In this paper, 
we use the term ``Floquet theory'' as a general term for theories that analyze time-periodic systems based on the Floquet theorem~\cite{Shirley1965,Sambe1973}.
Over the past few decades, Floquet theory and engineering have penetrated the field of condensed-matter and statistical physics~\cite{Oka2019,Bukov2015}. This is mainly attributed the two factors: (i) (as we mentioned above) the development of laser techniques enables us to generate time-periodic systems by applying intense laser to materials, and (ii) several notable theoretical studies have shown that intriguing quantum states (Floquet topological states~\cite{Oka2009,Kitagawa2010,Wang2013,Jotzu2014,McIver2020}, magnetic states~\cite{Kirilyuk2010,Oka2019,Sato2021book}, ultra-cold atomic states~\cite{Choi2017,Zhang2017}, etc.) can be created/controlled by applying laser, in principle. 
In most of the theoretical studies, researchers have used the Floquet theory for closed (isolated) quantum systems, namely, they have analyzed Schr\"odinger equation with a time-periodic Hamiltonian~\cite{Oka2009,Bukov2015}. 
These studies have succeeded in predicting the above results for the FE states. 

\begin{figure}[t]
\begin{center}
\includegraphics[width=9cm]{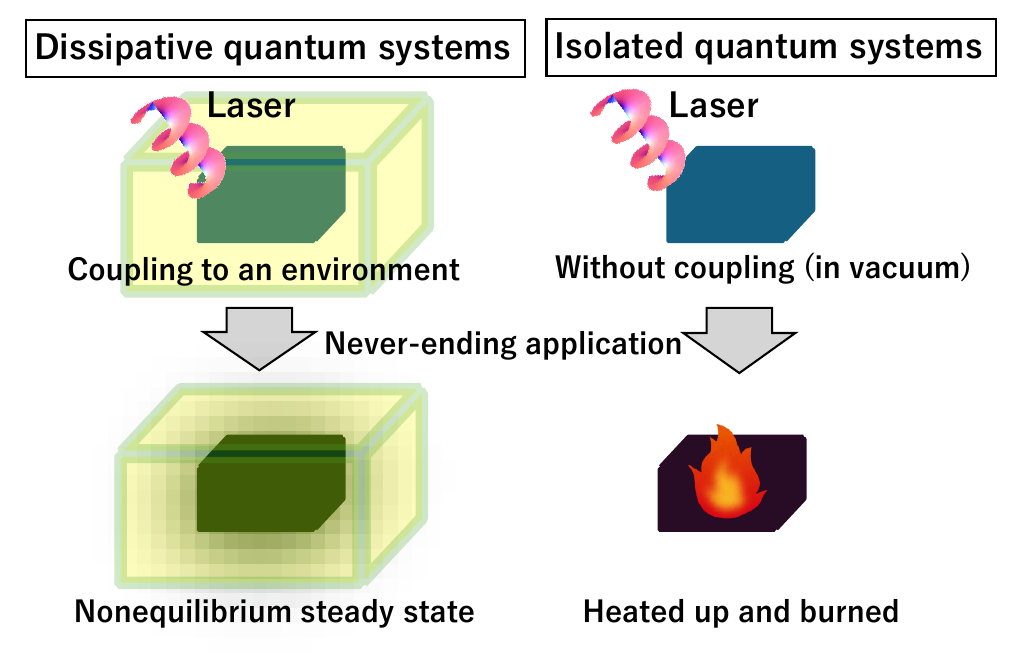}
\caption{Schematic images of isolated (closed) and dissipative (open) systems driven by laser. In the closed system, a long enough application of laser makes the system burned, while a nonequilibrium steady state (NESS) is expected to appear in the open (dissipative) system.}
\label{fig:LongLaserApplication}
\end{center}
\end{figure}

However, if we consider the FE in materials irradiated by an intense laser, a weak but finite coupling between a target system and an environment (bath) is inevitable, and it usually affects time evolution of physical quantities in the system because a part of energy generally flows out of the system into the bath, and some conservation laws are broken by the system-bath coupling. Therefore, it is important to develop the Floquet theory for open (dissipative) systems coupled to an environment~\cite{Dehghani2014,Aoki2014,Seetharam2015,Ikeda2020,Ikeda2021}. For electron systems in solids, typical system-bath interactions are electron-electron, electron-phonon, electron spin-phonon, electron-impurity, electron spin-nclear spin interactions, and so on. 
Recently, we have developed some theoretical methods for describing time-periodic dissipative systems. 
Figure~\ref{fig:LongLaserApplication} shows typical behavior of isolated (closed)~\cite{Abanin2015,Mori2016,Kuwahara2016,Abanin2017,Peng2021} and dissipative (open) periodically driven systems after long enough application of a periodic field like laser. 
In isolated systems (as we will discuss soon), they will finally heat up after a long drive, and the FE fails in a long time range~\cite{Lazarides2014,DAlessio2014,Kim2014}. 
On the other hand, if we consider dissipative systems coupled to an environment, a nonequilibrium steady state (NESS) is expected to be realized after a long driving. This is because of the balance between the energy injection by the external field and the release to the environment.

In this paper, we review some Floquet theories for dissipative quantum or classical systems, showing representative examples of FEs. In particular, we will stress the usefulness of the Floquet theories from the practical point of view. 

There are several theoretical approaches to attack dissipative systems subject to a periodic field. For quantum systems, this article focuses on the method based on the quantum master (GKSL) equation~\cite{Lindblad1976,Gorini1976,BreuerBook,Alicki2007book}, 
a representative equation of motion for the density matrix describing a generic class of Markovian dissipative systems. 
Using the GKSL equation, we mainly analyze NESS in driven dissipative systems~\cite{Ikeda2020,Ikeda2021}. 
There are several advantages of using the GKSL equation. For example, the GKSL equation is generally useful for calculating the real-time evolution of arbitrary physical observables. Moreover, if the equation is accurately solved, one can obtain nonequilibrium physical quantities in both weak (perturbative) and strong (non-perturbative) driving regimes. 
We note that the nonequilibrium Green's function method~\cite{Aoki2014,Tsuji2009,Haug2008,Stefanucci2013}
is another powerful method to investigate the dissipative systems. The Green's function method is generally effective to compute the correlation functions in many-body systems and to give the physical interpretation of nonequilibrium phenomena, especially, in the perturbative regime. 

For periodically driven classical systems, 
we can also develop the Floquet theory by combining the approaches of the Langevin~\cite{Kubo1991} and Fokker-Planck~\cite{Risken1996} equations. 
We will illustrate how to apply the Floquet theorem to nonlinear classical systems~\cite{Higashikawa2018}.

The remaining part of this paper is organized as follows. 
We first review the fundamental contents of FE in closed systems in Sec.~\ref{sec:Closed}, focusing on the high-frequency expansion method. 
As concrete examples of many-body FEs, we discuss the Floquet topological insulators, the inverse Faraday effect in metal, and FEs in magnets. 
The techniques and concepts in Sec.~\ref{sec:Closed} are useful even in considering open systems. Section~\ref{sec:Open} is devoted to the Floquet theory in open quantum systems, applying the GKSL equation. 
In particular, we argue how NESS is theoretically described in the GKSL equation formalism. We also discuss dissipation-assisted FE, by considering two simple models. 
In Sec.~\ref{sec:Open_classical}, we consider the Floquet theory in open classical systems. The equations of motion in classical systems are often nonlinear with respect to mechanical variables; therefore, the Floquet theorem cannot be applied directly. However, we overcome this difficulty through the distribution function of the variables, the Fokker-Planck equation. We apply this method to the Kapitza pendulum and the semi-classical equation of motion for spin systems (Landau-Lifshitz-Gilbert equation) to demonstrate its efficiency. 
Finally, we briefly summarize the full contents in Sec.~\ref{sec:Summary}.

\section{Floquet theory for closed systems\label{sec:Closed}}
Before discussing open (dissipative) systems, we review the Floquet theory for closed quantum systems. The result for closed systems also serves as the basis for open systems. 

\subsection{Floquet theorem}
\begin{figure}[t]
\begin{center}
\includegraphics
[width=9cm]{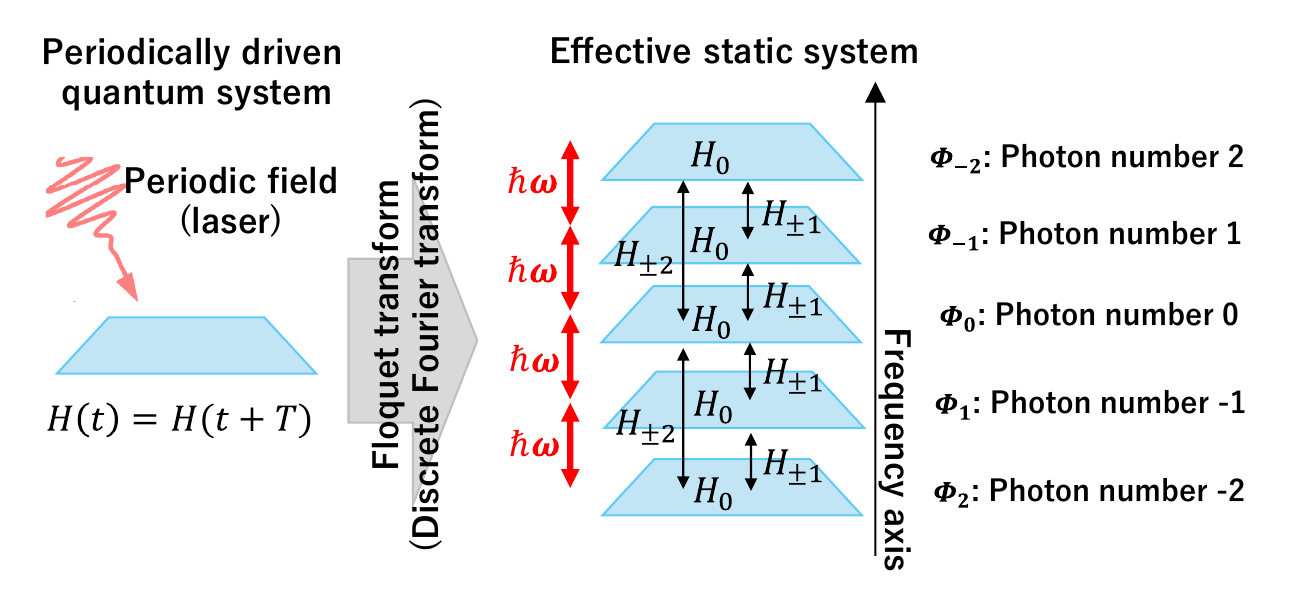}
\caption{Image of a time-periodic system (e.g., a laser-driven system) and the effective eigenvalue problem of Eq.~\eqref{eq:GeneralEigen} mapped from the original Schr\"odinger equation of Eq.~\eqref{eq:Sch}. }
\label{fig:FloquetMap}
\end{center}
\end{figure}

We start from the Schr\"odinger equation for the time-periodic system:
\begin{align}
i\hbar \frac{\partial}{\partial t} \ket{\psi(t)}=\hat H(t) \ket{\psi(t)},
\label{eq:Sch}
\end{align}
where the Hamiltonian satisfies $\hat H(t)=\hat H(t+T)$ and $T$ is the period. 
If we consider a laser-driven system, $T=2\pi/\omega$ with $\omega$ being the (angular) frequency of the laser. 
In the following, we will often set $\hbar=1$. 
For this time-dependent system, the Floquet theorem dictates that each linearly-independent solution $\ket{\psi(t)}$ is written as~\cite{Floquet1883,Shirley1965,Sambe1973}
\begin{align}
\ket{\psi(t)}=\exp(-i\epsilon t) \ket{\phi(t)},
\label{eq:Floquet}
\end{align}
where the real number $\epsilon$ is called the quasienergy and $\ket{\phi(t)}$ is a periodic state vector, $\ket{\phi(t)}=\ket{\phi(t+T)}$, called the Floquet state. 
The Floquet theorem is also referred to as the time version of the Bloch theorem.
The periodic natures of $\hat H(t)$ and $\ket{\phi(t)}$ allow us to define their Fourier transforms as follows: 
\begin{align}
\label{eq:Fourier}
\hat H(t)=\sum_{m\in \mathbb{Z}} e^{-im\omega t} \hat H_m, \hspace{0.5cm}
\ket{ \phi(t)}=\sum_{m\in \mathbb{Z}} e^{-im\omega t}\ket{\phi_m}.
\end{align}
The inverse transformation is given by 
\begin{align}
\label{eq:Fourier2}
\hat H_m &= T^{-1} \int_0^T dt \,\, e^{im\omega t} \hat H(t),\nonumber\\
\ket{\phi_m} &= T^{-1} \int_0^T dt \,\,e^{im\omega t} \ket{\phi(t)}. 
\end{align}
Substituting these expressions into the Schr\"odinger equation, 
we arrive at the following eigenvalue problem for quasi energy $\epsilon$: 
\begin{align}
\label{eq:GeneralEigen}
\sum_{n\in \mathbb{Z}}(\hat H_{m+n}-m\omega \delta_{m,n})\ket{\phi_n} &= \epsilon \ket{\phi_m}. 
\end{align}
One can more deeply understand the physical meaning of this equation 
if we express it in the following matrix form: 
\begin{widetext}
\begin{equation}
\label{eq:Matrix}
\left(\begin{matrix}
\ddots &  \ddots                 & \ddots               & \ddots  &                 &                               &             \\
\cdots & \hat H_0-2\hbar \omega & \hat H_1                    & \hat H_2    & \hat H_3                    & \cdots                &              \\
\cdots & \hat H_{-1}                    & \hat H_0-\hbar\omega & \hat H_1    & \hat H_2                    & \hat H_3                    & \cdots    \\
\cdots & \hat H_{-2}                    & \hat H_{-1}                 & \hat H_0    & \hat H_1                    & \hat H_2                    & \cdots    \\
\cdots & \hat H_{-3}                    & \hat H_{-2}                 & \hat H_{-1} & \hat H_0+\hbar\omega & \hat H_1                    & \cdots    \\
         & \cdots                   & \hat H_{-3}                 & \hat H_{-2} & \hat H_{-1}                 & \hat H_0+2\hbar\omega &  \cdots   \\
        &                             &                          & \ddots & \ddots                 & \ddots                 & \ddots \\
\end{matrix}\right)
\left(\begin{matrix}
\vdots \\
\ket{\phi_2} \\
\ket{\phi_1} \\
\ket{\phi_0} \\
\ket{\phi_{-1}} \\
\ket{\phi_{-2}} \\
\vdots \\
\end{matrix}\right)
=\epsilon
\left(\begin{matrix}
\vdots \\
\ket{\phi_2} \\
\ket{\phi_1} \\
\ket{\phi_0} \\
\ket{\phi_{-1}} \\
\ket{\phi_{-2}} \\
\vdots \\
\end{matrix}\right).
\end{equation}
\end{widetext}
We have put $\hbar$ back on the left-hand side of Eq.~\eqref{eq:Matrix}. 
In the matrix on the left side, the diagonal part is given by 
$\hat H_0+n\hbar \omega$ ($n$: an integer). Therefore, the corresponding state vector $\ket{\phi_{-n}}$ may be viewed as the $n$-photon dressed state if we suppose that the time periodicity of the Hamiltonian comes from an external electromagnetic wave with frequency $\omega$. The intuitive picture for the eigenvalue problem of Eqs.~\eqref{eq:GeneralEigen} and \eqref{eq:Matrix} is given by Fig.~\ref{fig:FloquetMap}. 

We should note that we have not used any approximation during the mapping from the Schr\"odinger equation to the eigenvalue problem of Eq.~\eqref{eq:Matrix}. Namely, Eqs.~\eqref{eq:GeneralEigen} and \eqref{eq:Matrix} always hold if we consider any sort of time-periodic quantum system. An important point is that the time-dependent problem of Eq.~\eqref{eq:Sch} is exactly mapped to an eigenvalue problem, which has no explicit time dependence. Therefore, one can use various established techniques in equilibrium statistical physics to solve the "nonequilibrium" system of Eq.~\eqref{eq:Matrix}. 
On the other hand, we should also note that the resultant eigenvalue problem has a new index of photon number $n$. The width of the Hilbert subspace with a fixed photon number is the same as that of the original time-dependent problem in Eq.~\eqref{eq:Sch}. 
Namely, we have to consider a new static system with the photon number axis 
instead of computing the explicit time evolution of the wave function.

\subsection{High-frequency expansion}
\label{sec:High-frequency}
In order to obtain the effective Hamiltonian on the subspace with a fixed photon number, 
here we adopt one approximation. We first assume that the one-photon energy $\hbar|\omega|$ is much larger than the energy scale of the system ($\hbar|\omega|$ is much larger than the eigenenergies of $\hat H_{\pm n}$) and then we apply the degenerate perturbation theory focusing on a subspace with a fixed photon number. The unperturbed part is the diagonal photon energy $m\hbar\omega$, while all other parts $\hat H_{\pm n}$ are perturbative terms.  
The resultant effective Hamiltonian for the subspace with each photon number (Floquet Hamiltonian) is given by the power expansion of $1/\omega$~\cite{Eckardt2015,Mikami2016}: 
\begin{align}
    \hat H_{\rm eff}=\sum_{j=0,1,2,\cdots} \hat H^{(j)},
    \label{eq:Perturbation}
\end{align}
where each term $\hat H^{(j)}$ is proportional to $(1/\omega)^{j}$. The lower order terms are computed as
\begin{widetext}
\begin{align}
\label{eq:0th_1/omega}
\hat H^{(0)} =&\hat H_0,\\ 
\hat H^{(1)} =& 
-\sum_{m=1}^\infty 
\frac{[\hat H_{+m},\,\, \hat H_{-m}]}{m\hbar\omega},\label{eq:1st_1/omega}\\
\hat H^{(2)} =& \sum_{\substack{m=-\infty \\ (m\neq 0)}}^\infty\frac{[[\hat H_{-m},\hat H_0],\hat H_m]}{2m^2(\hbar\omega)^2}
+\sum_{\substack{m=-\infty \\(m\neq 0)}}^\infty \sum_{\substack{n=-\infty \\(n\neq 0,m)}}^\infty 
\frac{[[\hat H_{-m},\hat H_{m-n}],\hat H_n]}{3mn(\hbar\omega)^2},
\label{eq:2nd_1/omega}\\
\hat H^{(3)} =& \sum_{\substack{m=-\infty \\ (m\neq 0)}}^\infty\frac{[[[\hat H_{-m},\hat H_0],\hat H_0],\hat H_m]}{2m^3(\hbar\omega)^3}
+\sum_{\substack{m=-\infty \\(m\neq 0)}}^\infty \sum_{\substack{n=-\infty \\(n\neq 0,m)}}^\infty 
\frac{[[[\hat H_{-m},\hat H_{0}],\hat H_{m-n}],\hat H_n]}{3m^2n(\hbar\omega)^3}\nonumber\\
&+\sum_{\substack{m=-\infty \\(m\neq 0)}}^\infty \sum_{\substack{n=-\infty \\(n\neq 0,m)}}^\infty 
\frac{[[[\hat H_{-m},\hat H_{m-n}],\hat H_{0}],\hat H_n]}{4mn^2(\hbar\omega)^3}
-\sum_{\substack{m=-\infty \\(m\neq 0)}}^\infty \sum_{\substack{n=-\infty \\(n\neq 0)}}^\infty 
\frac{[[[\hat H_{-m},\hat H_{m}],\hat H_{-n}],\hat H_n]}{12mn^2(\hbar\omega)^3}
\nonumber\\
&+\sum_{\substack{m=-\infty \\(m\neq 0)}}^\infty \sum_{\substack{n=-\infty \\(n\neq 0,m)}}^\infty 
\frac{[[\hat H_{-m},\hat H_{0}],[\hat H_{m-n},\hat H_{n}]]}{12m^2n(\hbar\omega)^3}
+\sum_{\substack{m=-\infty \\(m\neq 0)}}^\infty \sum_{\substack{n=-\infty \\(n\neq 0)}}^\infty \sum_{\substack{\ell=-\infty \\(\ell\neq 0,m,n)}}^\infty
\frac{[[[\hat H_{-m},\hat H_{m-\ell}],\hat H_{\ell-n}],\hat H_n]}{6\ell mn(\hbar\omega)^3}\nonumber\\
&+\sum_{\substack{m=-\infty \\(m\neq 0)}}^\infty \sum_{\substack{n=-\infty \\(n\neq 0,m)}}^\infty \sum_{\substack{\ell=-\infty \\(\ell\neq 0,m-n)}}^\infty
\frac{[[[\hat H_{-m},\hat H_{m-n-\ell}],\hat H_{\ell}],\hat H_n]}{24\ell mn(\hbar\omega)^3}
+\sum_{\substack{m=-\infty \\(m\neq 0)}}^\infty \sum_{\substack{n=-\infty \\(n\neq 0)}}^\infty \sum_{\substack{\ell=-\infty \\(\ell\neq 0,m,n)}}^\infty
\frac{[[\hat H_{-m},\hat H_{m-\ell}],[\hat H_{\ell-n},\hat H_{n}]]}{24\ell mn(\hbar\omega)^3}.
\label{eq:3rd_1/omega}
\end{align}
\end{widetext}
In addition to the above perturbation theory, there are a few methods to obtain the same high-frequency (Floquet-Magnus) expansion of the Floquet Hamiltonian~\cite{Casas2001,Rahav2003,Blanes2009,Mananga2011,Mikami2016}. 
A famous method is as follows. 
First, we assume that the time evolution operator $\hat U(t)$ is decomposed into the following form: 
\begin{align}
    \hat U(t_2-t_1)&\equiv T_{t}\exp\left(-i\int^{t_2}_{t_1} d\tau\hat H(\tau) \right)
    \nonumber\\
    &= e^{-i\hat G(t_2)}e^{-i\hat H_{\rm eff}(t_2-t_1)}e^{i\hat G(t_1)},
    \label{eq:3decomp}
\end{align}
where the symbol $T_t$ means the time-ordered product, $\hat H_{\rm eff}$ is the time-independent Floquet Hamiltonian, 
and the periodic operator $\hat G(t)=\hat G(t+T)$, called kick (micro-motion) operator, describes the fast oscillating dynamics. 
Second, substituting Eq.~\eqref{eq:3decomp} into the EOM of the time evolution operator (equivalent to Schr\"odinger equation), we obtain the differential equation for $\hat H_{\rm eff}$ and $\hat G(t)$. Finally, if we solve the differential equation in a recursive way~\cite{Casas2001,Rahav2003,Blanes2009,Mananga2011}, 
we obtain the $1/\omega$ expansion formulas for $\hat H_{\rm eff}$ and $\hat G(t)$. The result of $\hat H_{\rm eff}$ is the same as Eq.~\eqref{eq:Perturbation}. The micro-motion operator is expanded as $\hat G(t)=\sum_{j=1} \hat G^{(j)}(t)$ and its lower order terms are given by 
\begin{align}
\label{eq:1nd2rdkicked}
i\hat G^{(1)}(t) =&-\sum_{\substack{m=-\infty \\ (m\neq 0)}}^\infty \frac{\hat H_m}{m\hbar\omega}e^{-im\omega t}
\\
i\hat G^{(2)}(t)=&\sum_{\substack{m=-\infty \\ (m\neq 0)}}^\infty
\frac{[\hat H_m,\hat H_0]}{m^2(\hbar\omega)^2}e^{-im\omega t}
\nonumber\\
&+
\sum_{\substack{m=-\infty \\(m\neq 0)}}^\infty \sum_{\substack{n=-\infty \\(n\neq 0,m)}}^\infty
\frac{[\hat H_n,\hat H_{m-n}]}{2mn(\hbar\omega)^2}e^{-im\omega t}.
\end{align}

\subsection{Floquet Hamiltonian and thermalization}
\label{sec:FloquetETH}
Here, we make several remarks on how to interpret the Floquet Hamiltonian obtained by the high-frequency expansion, focusing mainly on many-body systems. For concreteness, we consider a stroboscopic dynamics from $t_1=t$ to $t_2=nT+t$ with $n=0,1,2,\dots$, for which the time evolution is given by $\hat{U}(nT)=e^{-i\hat{H}_\mathrm{eff}' nT}$ with $\hat{H}_\mathrm{eff}'=e^{-i\hat{G}(t)}\hat{H}_\mathrm{eff}e^{i\hat{G}(t)}=\hat{H}_\mathrm{eff}+O(\omega^{-1})$. This means that the unitary evolution is generated by the time-independent Hamiltonian $\hat{H}_\mathrm{eff}'$ and no transitions occur between its eigenstates. Therefore, equilibration or thermalization can occur effectively in the sense of isolated quantum systems~\cite{Polkovnikov2011,Eisert2015,Dalessio2016,Gogolin2016,Mori2018}.

It is important to note that the high-frequency expansion of $\hat H_{\rm eff}$ is not convergent in generic many-body systems because its convergence radius shrinks to zero in the thermodynamic limit~\cite{Blanes2009}. The effective Hamiltonian obtained from $\hat{U}(nT)$ is no longer a local Hamiltonian, and each many-body eigenstate behaves as a featureless random vector. This eigenstate-level thermalization to infinite temperature is called the Floquet eigenstate thermalization hypothesis (ETH)~\cite{Lazarides2014,DAlessio2014,Kim2014}, which is believed to hold in generic interacting systems aside from certain exceptional cases~\cite{Prosen1998,DAlessio2013, Else2016,Yao2017,Heyl2019, Medenjak2020, Haldar2021,Emonts2022,Ikeda2024}. The Floquet ETH implies Floquet thermalization, i.e., isolated many-body systems subject to continuous periodic drive will eventually heat up to the featureless infinite temperature state.

Nonetheless, the heating rate, or the energy increase per Floquet cycle, turned out to be exponentially small in the range $\omega\gg g$, where $g$ is the typical local energy of the system~\cite{Kuwahara2016,Abanin2017}. While the heating can be neglected, i.e., up to time $\sim e^{O(\omega/g)}$, the system's evolution is well approximated by a truncated effective Hamiltonian $\hat H_{\rm eff}^{(n_0)}=\sum_{j=0}^{n_0}\hat H^{(j)}$, which is local. If $\hat H_{\rm eff}^{(n_0)}$ is a generic nonitegrable Hamiltonian, the system equilibrates to the thermal state with respect to $\hat H_{\rm eff}^{(n_0)}$~\cite{Dalessio2016,Mori2018}, i.e., the Floquet-Gibbs (FG) state whose partition function is given by
\begin{align}
\label{eq:FG}
    \hat\rho_{\rm FG}(t)&=\frac{1}{Z_{\rm FG}}
    \sum_j e^{-\beta_{\rm FG}\epsilon_j}
    |\phi_j(t)\rangle\langle\phi_j(t)|\\
    Z_{\rm FG}&={\rm Tr}
    [e^{-\beta_{\rm FG}\hat H_{\rm eff}^{(n_0)}}],
\end{align}
where $\epsilon_j$ are the eigenvalues of $\hat H_{\rm eff}^{(n_0)}$.
This effective thermalization before reaching infinite temperature is called the Floquet prethermalization~\cite{Kuwahara2016,Abanin2017}, which is illustrated in Fig.~\ref{fig:LongLaserApplication}. If $\hat H_{\rm eff}^{(n_0)}$ includes a qualitatively novel term induced by the periodic external field, we can expect that interesting engineering takes place. Note that it is difficult to predict/control the value of the effective inverse temperature $\beta_{\rm FG}$.

\begin{figure}[t]
\begin{center}
\includegraphics[width=9cm]{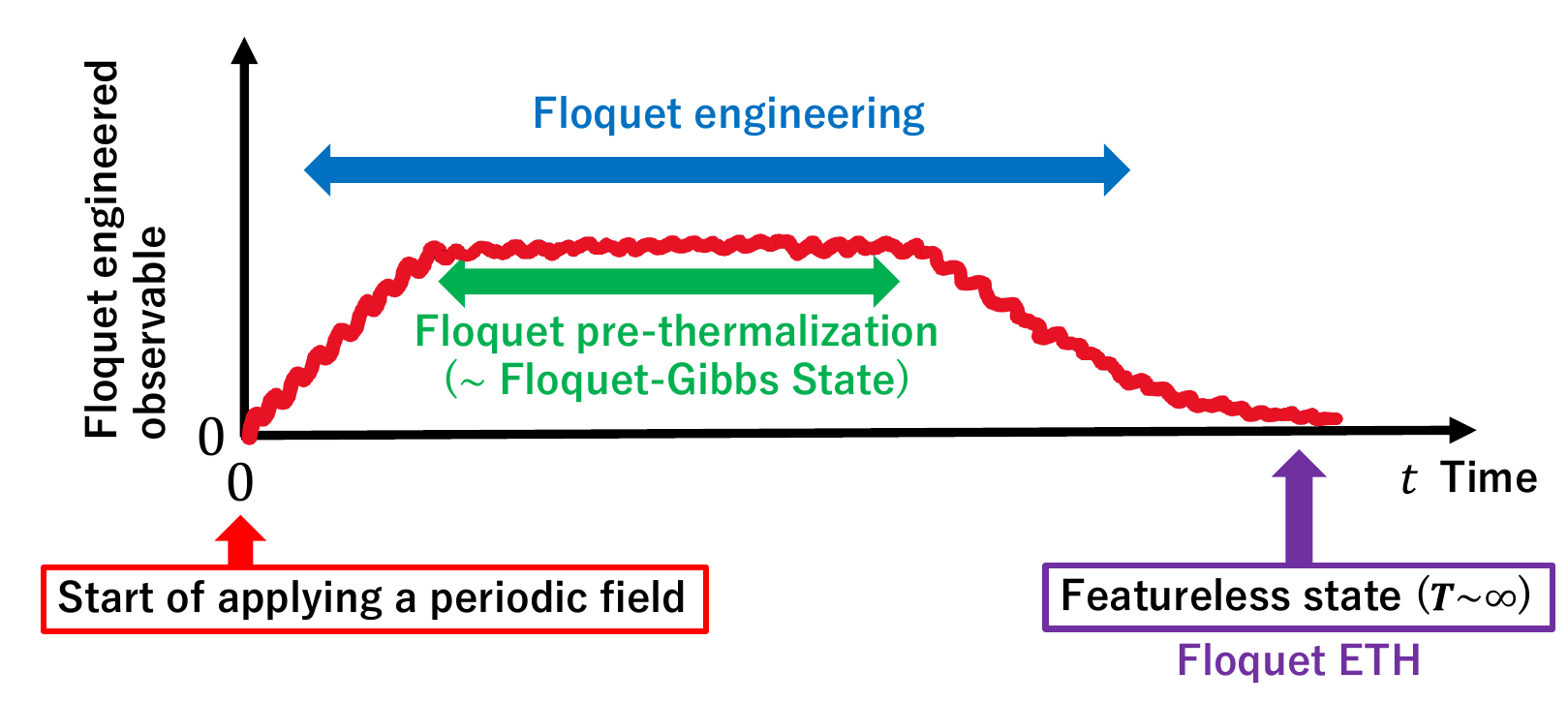}
\caption{Typical time evolution of a Floquet engineered quantity in closed quantum systems. We start the application of periodic external field at $t=0$. The physical quantity first grows up and approaches a constant value (Floquet pre-thermalization), whereas the engineered quantity finally disappears because the system approaches a featureless state after a long driving. FE is at least partially succeeded during the blue regime including the pre-thermalization (green) range 
as the engineered quantity is finite there.}
\label{fig:FE_Closed}
\end{center}
\end{figure}

We should be aware that the FE in closed systems is a transient phenomenon, as shown in Fig.~\ref{fig:FE_Closed}. As time goes on, Floquet heating, albeit as small as $e^{-O(\omega/g)}$ in one cycle, accumulates and the temperature gradually increases to reach infinity~\cite{Mallayya2019,Ikeda2021b}. Correspondingly, Floquet-engineered quantities become most significant before the Floquet heating matters, $nT\lesssim (\hbar/g)e^{O(\omega/g)}$, and will eventually vanish. We should note the following trade-off relation: The larger the engineering nontriviality (the difference between $\hat{H}_\mathrm{eff}$ and the simple time average $\hat{H}_0$) tends to be, the shorter the lifetime ${\sim}e^{O(\omega/g)}$ becomes.

\subsection{Examples of Floquet engineered states}
In this section, we briefly review a few examples of laser-driven many-body systems. 
Using the high-frequency expansion, we discuss what kind of Floquet engineering is expected in those models.

\subsubsection{Floquet topological insulator}
\label{sec:FloquetTI}
\begin{figure}[t]
\begin{center}
\includegraphics[width=9cm]{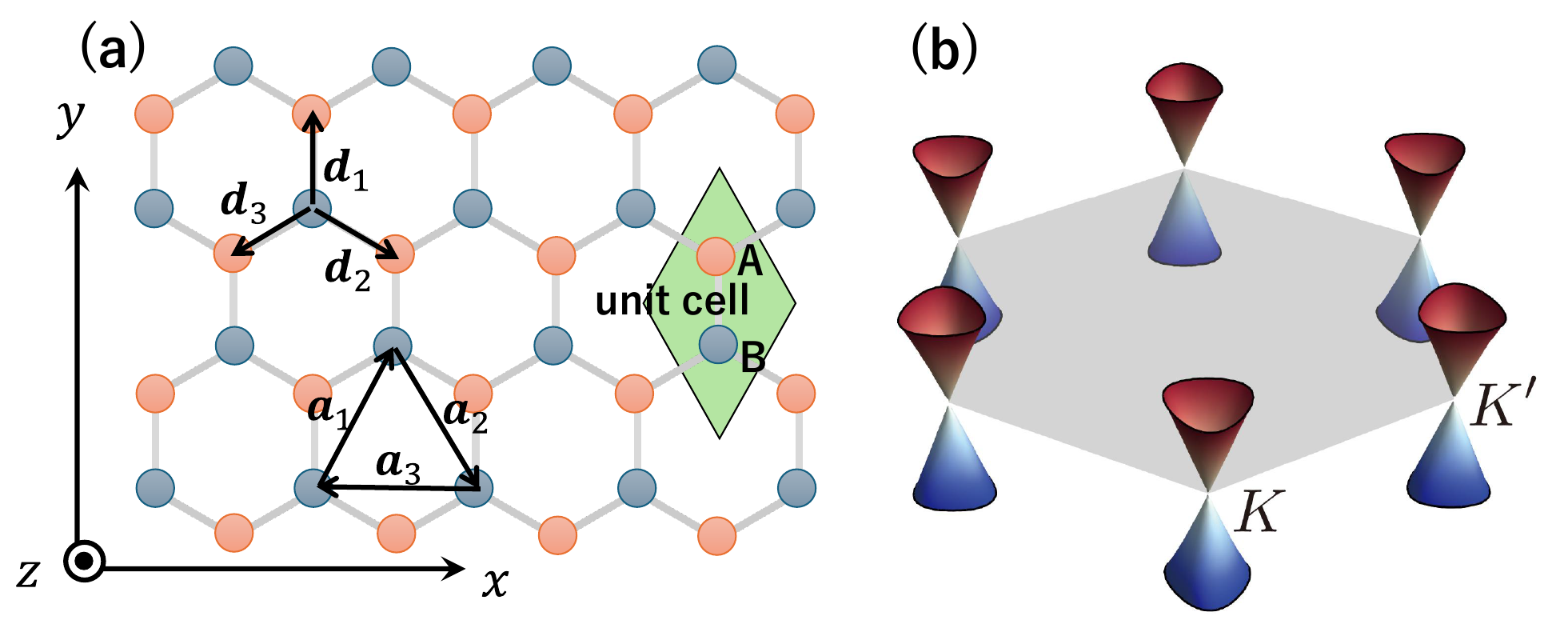}
\caption{(a) Honeycomb lattice structure of graphene with a staggered potential. Green diamond including A and B sublattice sites represents the unit cell. Vectors ${\bm d}_{1,2,3}$ are the unit vectors connecting two nearest-neighboring sites, while ${\bm a}_{1,2,3}$ are the vectors connecting two second-nearest-neighboring sites. 
(b) Energy dispersion of the model~\eqref{eq:Graphene} with $\mu_{\rm s}=0$ around the half-filled level. The gray honeycomb regime corresponds to the Brillouin zone.}
\label{fig:Graphene}
\end{center}
\end{figure}

Floquet topological insulator (TI) is one of the most famous Floquet engineered states~\cite{Oka2009,Kitagawa2010,Kitagawa2011}. 
The setup of the Floquet TI is simple: A circularly polarized laser is applied to a graphene-like system of Eq.~\eqref{fig:Graphene}, i.e., a two-dimensional (2D) Dirac electron system. 
Before the laser application, the Hamiltonian is defined as
\begin{align}
        \hat{H}=-t_1\sum_{\langle\bm{r},\bm{r}'\rangle}
        (a^\dagger_{\bm{r}}b_{\bm{r}'}+ b^\dagger_{\bm{r}'}a_{\bm{r}})
        \,+\mu_{\rm s}
        \left[\sum_{\bm{r}\in A}a^\dagger_{\bm{r}}a_{\bm{r}}-\sum_{\bm{r}\in B}\,b^\dagger_{\bm{r}}b_{\bm{r}}
        \, \right],
\label{eq:Graphene}
\end{align}
where $a^\dagger_{\bm{r}}$ ($b^\dagger_{\bm{r}}$) is the electron creation operator on the site $\bm r$ in A (B) sublattice, and we have ignored the spin degree of freedom. 
The first term is the kinetic energy of the nearest-neighboring hopping with amplitude $t_1$ and the second term is a staggered chemical potential with strength $\mu_{\rm s}$. For $\mu_{\rm s}=0$, we have two gapless Dirac-cone energy dispersions around the $K$ and $K'$ points as shown in Fig.~\ref{fig:Graphene} (b)
while a finite potential $\mu_{\rm s}$ induces a mass gap at the $K$ and $K'$ points.  

To take into account the effect of circularly polarized laser, we adopt the so-called Peierls's formalism. Namely, the hopping term is modified as 
\begin{align}
\label{eq:Peierls}
a^\dagger_{\bm{r}}b_{\bm{r}'} &\rightarrow e^{i\phi_{\bm r,\bm r'}(t)}a^\dagger_{\bm{r}}b_{\bm{r}'},
\end{align}
where the phase factor $\phi_{\bm r,\bm r'}(t)$ is given by 
\begin{align}
\label{eq:Phase}
\phi_{\bm r,\bm r'}(t)&=-\frac{e}{\hbar c}\int_{\bm r}^{\bm r'} d{\bm x} \cdot\bm A({\bm x},t),
\end{align}
with $\bm A({\bm x},t)$ being the vector potential of the laser. 
A circularly polarized laser with frequency $\omega$ 
is given by the vector potential $\bm{A}(t)=\,A\left(\cos\omega t,\sin\omega t,0 \right)$ whose electric field 
$\bm{E}(t)=E_0(-\sin\omega t,\cos\omega t,0)$ is in the $x$-$y$ plane with the strength $E_0=-\frac{\omega}{c}A$. We note that a left-handed laser is changed into the right-handed type by sign change $\omega\to -\omega$ and vice versa. 
Substituting $\bm{A}(t)$ in the Peierl's phase factor $\phi_{\bm r,\bm r'}(t)$, we obtain the laser-driven Hamiltonian satisfying $\hat H(t)=\hat H(t+T)$. The Fourier components of the time-dependent Hamiltonian are therefore estimated as 
\begin{align}
\hat{H}_0 &=-t_1   
        J_0(\tilde A)\sum_{\langle\bm{r},\bm{r}'\rangle}(a^\dagger_{\bm{r}}b_{\bm{r}'}+ b^\dagger_{\bm{r}'}a_{\bm{r}})\nonumber\\
        &+\mu_{\rm s}\left[\sum_{\bm{r}\in A} a^\dagger_{\bm{r}}a_{\bm{r}}
        -\sum_{\bm{r}\in B}b^\dagger_{\bm{r}}b_{\bm{r}}\right]
        \\
        \hat{H}_{n\neq 0}&=-t_1 i^{-n} J_{n}(\tilde A)\sum_{\bm{r},\,j}e^{in\beta_j}  \nonumber\\
        &\times\left[
        (-1)^n a^\dagger_{\bm{r}}b_{\bm{r}+\bm{d}_j}+
        b^\dagger_{\bm{r}+\bm{d}_j}a_{\bm{r}}\right],
\end{align}
where $\tilde A=-\frac{e}{\hbar c}A$ and $\beta_j$ is the angle between ${\bm d}_j$ and the $x$ axis. 
We have used the equalities $e^{ix\sin\phi}=\sum_{n=-\infty}^\infty J_n(x)e^{in\phi}$ and 
$e^{ix\cos\phi}=\sum_{n=-\infty}^\infty i^n J_n(x)e^{in\phi}$ with $J_n(x)$ being Bessel functions. Using these Fourier components, we can compute the second leading term of the high-frequency expansion, $\hat H^{(2)}=\sum^\infty_{n=1}\frac{[\hat{H}_{-n},\hat{H}_n]}{n\hbar\omega}$. 
As a result, up to $\hat H^{(2)}$, we have the Floquet Hamiltonian,
\begin{align}
\hat{H}_{\mathrm{eff}}=&
-t_1^{\mathrm{eff}}\sum_{\langle\bm{r},\bm{r}'\rangle}
(a^\dagger_{\bm{r}}b_{\bm{r}'}+ b^\dagger_{\bm{r}'}a_{\bm{r}})
\nonumber\\
&+it_2^{\mathrm{eff}}\sum_{\langle\langle\bm{r},\bm{r}'\rangle\rangle}\tau_{\bm{r}\bm{r}'}(a^\dagger_{\bm{r}}a_{\bm{r}'}+ b^\dagger_{\bm{r}}b_{\bm{r}'})\nonumber\\
&+\mu_{\rm s}
        \left[\sum_{\bm{r}\in A}a^\dagger_{\bm{r}}a_{\bm{r}}-\sum_{\bm{r}\in B}\,b^\dagger_{\bm{r}}b_{\bm{r}}
        \, \right],
\label{eq:FloquetTopo}
\end{align}
where the nearest-neighboring and second-nearest-neighboring hopping amplitudes are respectively given by
\begin{gather}
t_1^{\mathrm{eff}}=t_1J_0(\tilde A),\,\,\,\,\,\,\,\,\,\,\,\,
t_2^{\mathrm{eff}}=-2\sum^\infty_{n=1}
\frac{t_1^2J^2_n(\tilde A)}{n\hbar\omega} \sin\left( \frac{2\pi n}{3} \right).
\end{gather}
The sum of the second nearest-neighboring hopping $\sum_{\langle\langle\bm{r},\bm{r}'\rangle\rangle}$ is taken in all closed triangle paths defined by ${\bm a}_{1,2,3}$ in Fig.~\ref{eq:Graphene} (a). The parameter $\tau_{\bm{r}\bm{r}'}$ is defined as follows: It takes $+1$ ($-1$) when the triangle path rotates clockwise (anticlockwise). 
The effective Hamiltonian of Eq.~\eqref{eq:FloquetTopo} is equivalent to the famous Haldane honeycomb model~\cite{Haldane1988}. 

The amplitude of the nearest-neighboring hopping is modified from $t_1$ to $t_1J_0(\tilde A)$. 
The Bessel function $J_0(\tilde A)=J_0(-\tilde A)$ is monotonically decreasing as function of $|\tilde A|$ around $|\tilde A|=0$, and $J_0(|\tilde A|)=0$ at $|\tilde A|\sim 2.4$. 
Therefore, the electron band width generally decreases with laser application, and this phenomenon is called dynamical localization~\cite{Dunlap1986,Klappauf1998,Kayanuma2008}. In solid electron systems, it is difficult to reach the value $|\tilde A|\sim 2.4$, and a partial dynamical localization has been observed~\cite{Ishikawa2014}. On the other hand, clear dynamical localization has been realized in ultra-cold atomic systems~\cite{Moore1994}. 

The more interesting point is that a circularly polarized laser induces clockwise and anticlockwise second-nearest neighboring hoppings with imaginary amplitude $i t^{\rm eff}_2$. This hopping therefore breaks time-reversal symmetry. In fact, the hopping amplitude $i t^{\rm eff}_2$ is changed to $-i t^{\rm eff}_2$ by $\omega\to-\omega$. 

It is known that when the hopping $i t^{\rm eff}_2$ is sufficiently larger than the staggered potential $\mu_{\rm s}$, the ground state of $\hat{H}_{\mathrm{eff}}$ is in the Chern insulating phase. Therefore, we can expect a chiral gapless edge mode and Hall conductivity emerges if we apply strong enough circularly polarized laser to a 2D Dirac electron system like graphene. This is the basic scinario of the Floquet TI.
Recent experimental studies have reported that (i) circularly-polarized-laser induces a mass gap~\cite{Wang2013} and (ii) a finite (not quantized) Hall conductivity (i.e., anomalous Hall effect) emerges under application of a circularly polarized laser~\cite{McIver2020}. 

However, we should note the theoretical argument that when we measure the dc or ac Hall conductivities in circularly-polarized laser driven Dirac electron systems, there are some mechanisms of generating a finite Hall current in addition to that of Floquet TI~\cite{Sato2019a,Murotani2023,Murotani2025a,Murotani2025b}. 
For instance, phot-induced inverse spin Hall effect, field-induced circular photogalvanic effect~\cite{Murotani2023,Murotani2025a,Murotani2025b}, 
and photo-induced anomalous Hall effect can simultaneously occur. Thus, it is difficult to detect the characteristics of the Floquet TI by separating them from other effects in graphene, unlike clean ultracold atomic systems~\cite{Jotzu2014}. 

Finally, we note that if we consider the setup that THz or GHz circularly polarized wave is applied to the Kitaev honeycomb model with a multiferroic coupling~\cite{Sato2014}, 
the high-frequency expanded Hamiltonian up to $1/\omega$ order becomes equivalent to a Kitaev model with a three-spin interaction, whose ground state is known to a topological insulator of Majorana fermion with a gapless chiral edge mode. This can be regarded as a quantum spin liquid version of Floquet TI.

\subsubsection{Inverse Faraday effect}
\label{sec:IFE}
\begin{figure}[t]
\begin{center}
\includegraphics[width=9cm]{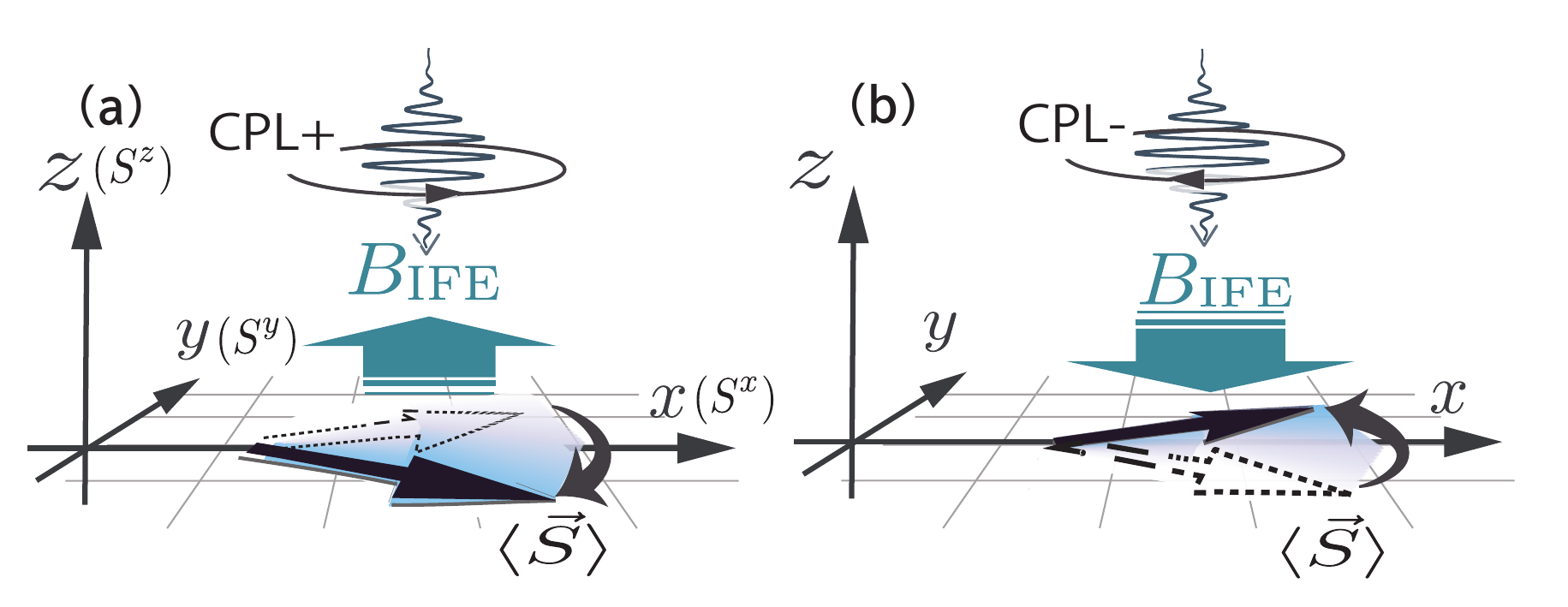}
\caption{Image of inverse Faraday effect. The photo-induced effective magnetic field $\bm B_{\rm IFE}$ changes its sign by changing the helicity of light from (a) left circularly polarization (CPL+) to (b) right circularly polarization (CPL-). If the system initially possesses a magnetization along the $x$ direction, a precession motion occurs due to $\bm B_{\rm IFE}$. 
Reproduced from Ref.~\cite{Tanaka2024} with permission.}
\label{fig:IFE}
\end{center}
\end{figure}

The inverse Faraday effect (IFE) is well known in the field of magneto-optics~\cite{Pershan1966,vanderZiel1965,Kimel2005,Hansteen2006,Satoh2010,Makino2012}. 
This refers to the phenomenon that when a circularly polarized light is applied to a solid electron system, an effective magnetic field or magnetization emerges parallel to the direction of light propagation. As we will explain soon, the IFE can be viewed as a typical FE in solids~\cite{Pershan1966,Berritta2016,Tanaka2020,Tanaka2024,Banerjee2022}, 
especially if we use low-frequency light such as THz or infrared laser.  

To explain the IFE, we start from a spin-orbit (SO) coupled electron system on a square lattice~\cite{Tanaka2020,Tanaka2024}. The Hamiltonian is represented as $\hat{H}_{\rm R} = \hat{H}_{\rm K}+\hat{H}_{\rm so}$, where 
\begin{align}
\label{eq:Rashba1}
  \hat{H}_{\rm K} =& -t_1\sum_{\bm{r}}\sum_{\sigma=\uparrow\downarrow\,}\sum_{j=x,y}
        c_{\bm{r}+\bm{e}_j,\sigma}^\dagger c_{\bm{r},\sigma} +{\rm h.c.}\\
  \hat{H}_{\mathrm{so}}=&-\frac{\alpha}{2}\sum_{\bm{r},\sigma,\sigma'}
\Big(i(\sigma_y)_{\sigma\sigma'}\,c_{\bm{r}+\bm{e}_x,\sigma}^\dagger c_{\bm{r},\sigma'}\nonumber\\
        &-i(\sigma_x)_{\sigma\sigma'}\,c_{\bm{r}+\bm{e}_y,\sigma}^\dagger \,c_{\bm{r},\sigma'}\Big)\,+{\rm h.c.}.
        \label{eq:Rashba2}
\end{align}
The operator $c_{\bm{r},\sigma}^\dagger$ is the electron creation operator with spin $\sigma$ on the site $\bm r$, ${\bm e}_{x(y)}$ denotes the vector connecting nearest-neighboring sites along the $x$ ($y$) direction, and $\sigma^{y,z}$ are Pauli matrices. 
The first term $\hat{H}_{\rm K}$ is a standard spin-independent hopping, while the second term $\hat{H}_{\rm so}$ is a Rashba type SO coupling with strength $\alpha$, that is, a spin-dependent hopping. 
Note that an SO coupling is necessary to generate an effective magnetic field by applying an electric field of light because electron spins cannot be directly coupled to an electric field, and an SO coupling connects the spin and light through the light-charge coupling. Here, for simplicity, we consider the photo-induced magnetization of electron spins. On the other hand, in real experiments, the IFE includes orbital magnetization~\cite{Berritta2016,Tazuke2025} as well.

Similarly to the case of Floquet TI, we introduce the effect of circularly polarized light by using Peierls's method. The vector potential is given by $\bm A(t)=A(-\sin\omega t, \cos\omega t, 0)$. In the wave vector $\bm k$ space, the Hamiltonian of the laser-driven Rashba electron system is given by 
\begin{align}
        \hat{H}_{\rm R}(t)&=\sum_{\bm{k}}
        \left(
                c^\dagger_{\bm{k},\uparrow}\,\,
                c^\dagger_{\bm{k},\downarrow}\,
        \right)
        \left(
                \begin{array}{cc}
                        \,\varepsilon_{\bm{k},A}(t) & \gamma_{\bm{k},A}(t) \\
                        \,\gamma^*_{\bm{k},A}(t)  &  \varepsilon_{\bm{k},A}(t)
                \end{array}
        \right)
        \left(
                \begin{array}{c}
                        c_{\bm{k},\uparrow}\\
                        c_{\bm{k},\downarrow}
                \end{array}
        \right)\nonumber\\
        &=\sum_{\bm k}\hat H_{\rm k}(t),
        \label{eq:Rashba_k}
\end{align}
where the electron operator in $\bm k$ space is defined as $c_{\bm{r},\sigma}=\frac{1}{\sqrt{N}}\sum_{\bm{k}}e^{i\bm{k}\cdot\bm{r}}c_{\bm{k},\sigma}$
with $N$ being the number of total sites and the matrix elements on the right-hand side are given by $\varepsilon_{\bm{k},A}(t) =-2t_1[\cos(k_x+\tilde A_x(t))+\cos(k_y+\tilde A_y(t))]$ and 
$\gamma_{\bm{k},A}(t) =\alpha[i \sin(k_x+\tilde A_x(t))+\sin(k_y+\tilde A_y(t))]$. 
Here, we have again defined
$\tilde {\bm A}(t)=\frac{e}{\hbar c}\bm A(t)$. 
Using this expression, one can compute the Fourier components $\hat H_n$ as 
\begin{align}
        \hat{H}_n &= J_n(\tilde A) \sum_{\bm{k}}
        \left(c^\dagger_{\bm{k},\uparrow},
                c^\dagger_{\bm{k},\downarrow}
        \right)
        \bm{h}^{(n)}_{\bm{k}}
        \left(
                \begin{array}{c}
                        c_{\bm{k},\uparrow}\\
                        c_{\bm{k},\downarrow}
                \end{array}
        \right),
\end{align}
where the matrix elements are given by $h^{(n)}_{\bm{k}\uparrow\uparrow}= h^{(n)}_{\bm{k}\downarrow\downarrow}=-2t_1 [i^n\,\cos\left(k_x-\frac{n\pi}{2} \right) + \cos\left(k_y+\frac{n\pi}{2} \right) ]$ and 
$h^{(n)}_{\bm{k}\uparrow\downarrow}=(h^{(-n)}_{\bm{k}\downarrow\uparrow})^*=\alpha \left[i^{n+1}\,\sin\left(k_x-\frac{n\pi}{2} \right) 
+ \sin\left( k_y+\frac{n\pi}{2} \right) \right]$, 
and $J_n(\tilde A)$ are the Bessel functions. 

From the leading term of the high-frequency expansion, namely, the time-averaged Hamiltonian $\hat H_0$, we again observe the reduction of the hopping amplitude. Therefore, a partial dynamical localization is expected to occur. A non-trivial FE is expected from the second leading term as  
\begin{align}
       \sum^\infty_{n=1}\frac{[\hat{H}_{-n},\hat{H}_n]}{n\hbar\omega} 
        &=\sum_{\bm k}B_{\mathrm{eff}}(\bm{k})\hat{S}^z_{\bm k},
\end{align}
where we have defined the spin operator in the $\bm k$ space as 
$\hat{S}^{x,y,z}_{\bm k}=\frac{1}{2}C^\dagger_{\bm k}\sigma^{x,y,z}C_{\bm k}$ with $C_{\bm k}^\dagger=(c^\dagger_{\bm{k},\uparrow}, c^\dagger_{\bm{k},\downarrow})$ and 
the effective magnetic field $B_{\mathrm{eff}}(\bm{k})$ is estimated as 
\begin{align}
        B_{\mathrm{eff}}(\bm{k})&=
        2(|h^{(+n)}_{\bm{k}\uparrow\downarrow}|^2 
        - |h^{(-n)}_{\bm{k}\uparrow\downarrow}|^2)
        \notag\\
        &=8\alpha^2\sum^\infty_{n=1}\frac{(-1)^n J^2_n(\tilde A)}{n\hbar\omega}
        \cos\frac{(n+1)\pi}{2}\nonumber\\
        &\times \sin\left( k_x-\frac{n\pi}{2} \right)\sin\left( k_y+\frac{n\pi}{2} \right).
        \label{eq:Beff}
\end{align}
One finds that the field $B_{\mathrm{eff}}(\bm{k})$ changes its sign via $\omega\to-\omega$. 
Therefore, we can expect an magnetic field and magnetization appear when we apply a circularly polarized laser to 2D SO-coupled electrons. This is the scenario of IFE as FE. 

We should note that the IFE in real experiments includes at least two phenomena: One is the above FE and another is the magnetic Raman scattering~\cite{Reid2010,Higuchi2011,Kanda2011,Fleury1968,Devereaux2007,Lemmens2003}, 
in which a part of the angular momentum (spin) of incident photons is transferred to electron spins in solid through an inelastic light-matter interaction. In fact, the effect of Raman scattering seems to be dominant in most experimental studies of IFEs. In particular, if we use a high-frequency pulse like ultraviolet or visible light, the time period of pulse is too short for FE to take place, whereas the magnetic Raman scattering can occur in such a short period. To observe the FE explained above, we should use lower-frequency laser pulses such as THz or infrared electromagnetic waves.

\subsubsection{Quantum spin systems and multiferroics}
\label{sec:spin_multiferro}
In this subsection, we consider magnetic insulators (quantum spin systems) irradiated by THz light or GHz wave~\cite{Hirori2011,Liu2017,Li2020,Kampfrath2011,Mashkovich2021,Huang2024,Zhang2024,Zhang2025}, 
whose photon energy is comparable to magnetic excitation energies. Usual magnets are coupled to only the ac magnetic field of light, but in the last decades, multiferroic materials~\cite{Kimura2003,Katsura2005,Mostovoy2006,Sergienko2006,Tokura2014} 
have been actively studied, in which electric polarization strongly interacts with electron spin degrees of freedom. Therefore, spins in multiferroics can be coupled to not only the magnetic field but also the electric field through the polarization-light coupling~\cite{Pimenov2006,Takahashi2012,Kubacka2013,Furukawa2010,Huvonen2009}.
The interaction between spins and electric polarization is called magnetoelectric (ME) coupling, cross-correlation, etc.  
Here, we focus on magnetic insulators with an ME coupling under the application of GHz or THz wave. 
Its generic Hamiltonian~\cite{Sato2016} 
is expressed as  
\begin{align}
\label{eq:Multiferro}
\hat H(t) &=  \hat H_{\rm mag}
-{\bm B}(t)\cdot\hat{\bm S}-{\bm E}(t)\cdot \hat{\bm P},
\end{align}
where $\hat H_{\rm mag}$ denotes the static magnetic interactions like Heisenberg exchange interactions, $\hat{\bm S}$ is the total spin, $\hat{\bm P}$ is the total electric polarization, and ${\bm B}(t)$ and ${\bm E}(t)$ are respectively the magnetic and electric fields of the applied electromagnetic (EM) wave. We have defined the "magnetic" field ${\bm B}(t)$ including parameters of the $g$ factor and the Bohr magneton $\mu_{\rm B}$. In multiferroic materials, the polarization $\hat{\bm P}$ is given by a function of spins. 

We suppose that the applied EM wave is circularly polarized in GHz or THz range and its EM fields are in the $x$-$y$ plane: $\bm E(t)=E_0(\cos(\omega t), \sin(\omega t),0)$ and $\bm B(t)=B_0(\sin(\omega t),-\cos(\omega t),0)$, where $\omega$ is the frequency of the wave. In this case, the Fourier components of the Hamiltonian are given by $\hat H_0= \hat H_{\rm mag}$ and $\hat H_{\pm 1}=-\frac{1}{2}(E_0\hat P^{\pm}\pm i B_0\hat S^{\pm})$, where we have defined $\hat S^{\pm}=\hat S^x\pm i \hat S^y$ and $\hat P^{\pm}=\hat P^x\pm i \hat P^y$. Using these, we can obtain the effective Floquet Hamiltonian~\cite{Sato2016} up to $1/\omega$ order as follows: 
\begin{align}
\label{eq:Eff_Multiferro}
\hat H_{\rm eff} =&  \hat H_{\rm mag}-\frac{1}{2\hbar\omega}\Big(B_0^2 \hat S^z-iE_0^2 [\hat P^x,\hat P^y]\nonumber\\
&-iE_0B_0( [\hat P^x,\hat S^x]+ [\hat P^y,\hat S^y])\Big).
\end{align}
The final term proportional to $E_0B_0$ appears only when the system is coupled to both the electric and magnetic fields. 
The second term $-\frac{1}{2\hbar\omega}B_0^2 \hat S^z$ is the effective Zeeman interaction driven by a circular polarized wave. This is expected to lead to an IFE in magnetic insulators (see IFE of metal in Sec.~\ref{sec:IFE}). In fact, recent theoretical studies~\cite{Takayoshi2014a,Takayoshi2014b,Sato2016,Ikeda2020,Ikeda2021} 
have shown that quantum spin systems can be magnetized by applying a THz circularly polarized wave if the systems satisfy a certain symmetry condition.

\begin{figure}[t]
\begin{center}
\includegraphics[width=8cm]{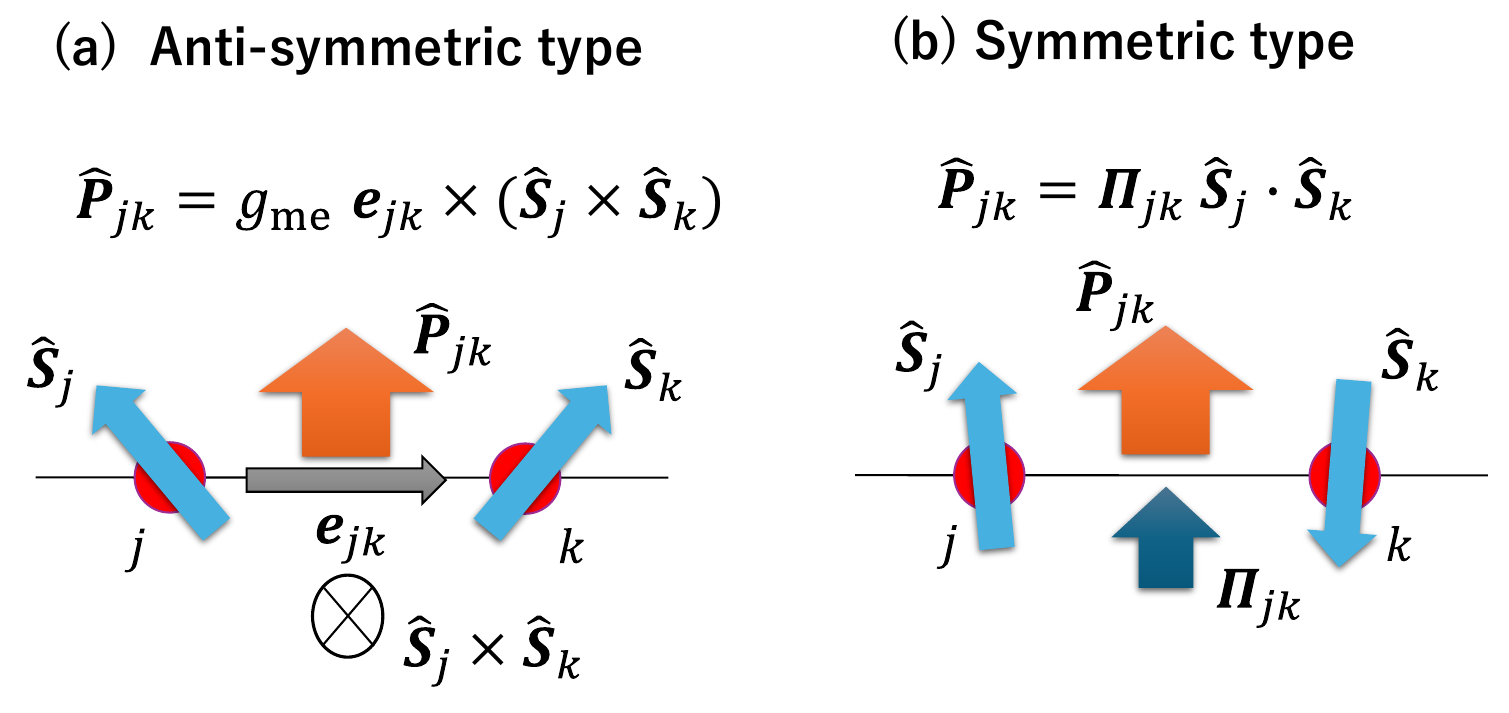}
\caption{Two representative ME couplings in multiferroic magnets~\cite{Tokura2014}. (a) The anti-symmetric ME coupling is also referred to spin-current type or inverse DM interaction type. Its origin is usually an SO coupling. As shown in panel (a), if neighboring spins form a non-collinear configuration, the vector spin chirality  appears (in the classical-spin view) and the ME coupling becomes larger. (b) The symmetric type is also called magneto-striction type. Its emergence does not requires an SO coupling and it is usually originated from a spin-phonon coupling. This ME coupling grows up when neighboring spins take a collinear ordering.}
\label{fig:ME}
\end{center}
\end{figure}
There are several types of ME interactions~\cite{Tokura2014} in magnetic materials. 
Figure~\ref{fig:ME} shows two representative ME couplings. The first in panel (a) is the asymmetric exchange type (also known as inverse Dzyaloshinskii-Moriya interaction, spin current type, etc.), in which the electric polarization localized on the $j$-$k$ bond is given by the outer product of two neighboring spins: $\hat{\bm P}=g_{\rm em}{\bm e}_{jk}\times(\hat{\bm S}_{j}\times\hat{\bm S}_{k})$ with $g_{\rm em}$ and ${\bm e}_{jk}$ being the ME coupling constant and the unit vector connecting the $j$ and $k$ sites, respectively. The second type in panel (b) refers to the symmetric exchange type (it is also called the magneto-striction type), in which the polarization is proportional to the exchange interaction: $\hat{\bm P}={\bm \Pi}_{jk}\hat{\bm S}_{j}\cdot\hat{\bm S}_{k}$ with ${\bm \Pi}_{jk}$ being the ME-coupling vector. 

Let us consider a simple two-spin multiferroic model with an inverse Dzyaloshinskii-Moriya (DM) interaction. If such a system is irradiated by THz or GHz circularly polarized wave, the Hamiltonian is given by $\hat H(t) = \hat H_{\rm mag}-{\bm B}(t)\cdot\hat{\bm S}-{\bm E}(t)\cdot \hat{\bm P}$, where $\hat H_{\rm mag}$ is the magnetic interaction between two spins $\hat{\bm S}_{1,2}$ and the polarization is given by $\hat{\bm P}=g_{\rm em}{\bm e}_{12}\times(\hat{\bm S}_{1}\times\hat{\bm S}_{2})$. Suppose that the two spins and the EM fields of applied wave are located in the $x$-$y$ plane, that is, ${\bm e}_{12}=(\cos\theta,\sin\theta,0)$, $\bm E(t)=E_0(\cos(\omega t), \sin(\omega t),0)$ and $\bm B(t)=B_0(\sin(\omega t),-\cos(\omega t),0)$, the effective Hamiltonian~\cite{Sato2016} is computed as 
\begin{align}
\label{eq:eff_2spin}
\hat H_{\rm eff} &= \hat H_{\rm mag}-\frac{B_0^2}{2\hbar\omega}\hat S^z
-\frac{g_{\rm me}E_0B_0}{2\hbar\omega}\bm e_{12}\cdot\hat {\bm V},
\end{align}
where $\hat {\bm V}=\hat{\bm S}_{1}\times\hat{\bm S}_{2}$ is the vector spin chirality of two spins. 
The last term $\propto E_0B_0$ may be called the photo-induced synthetic DM interaction, whose DM vector is proportional to ${\bm e}_{12}$. 
Spatially uniform DM interactions~\cite{Dzyaloshinsky1958,Moriya1960} 
generally induce a spiral spin structure. In fact, a theoretical study~\cite{Sato2016} shows that when we apply a circularly polarized THz laser to a quantum spin chain with antisymmetric ME coupling, a quantum state with a vector spin chirality emerges due to the photo-induced DM interaction. This is a typical FE in quantum spin systems.

In addition to the antisymmetric ME coupling, symmetric ME coupling can lead to non-trivial FE. 
For instance, as we briefly mentioned in Sec.~\ref{sec:FloquetTI},
it has been shown that a Floquet topological spin liquid is expected to appear if we apply a circularly polarized THz laser to a Kitaev honeycomb model with a symmetric ME coupling~\cite{Sato2014}.

\section{Floquet theory for open quantum systems}
\label{sec:Open}
In the last section, we have reviewed several fundamentals of Floquet theory in closed quantum systems, focusing on the high-frequency expansion method. 
This section is devoted to the Floquet theory in open (dissipative) quantum systems. We concentrate on the method based on the equation of motion (EOM) for the density matrices. 

\begin{figure}[t]
\begin{center}
\includegraphics[width=8cm]{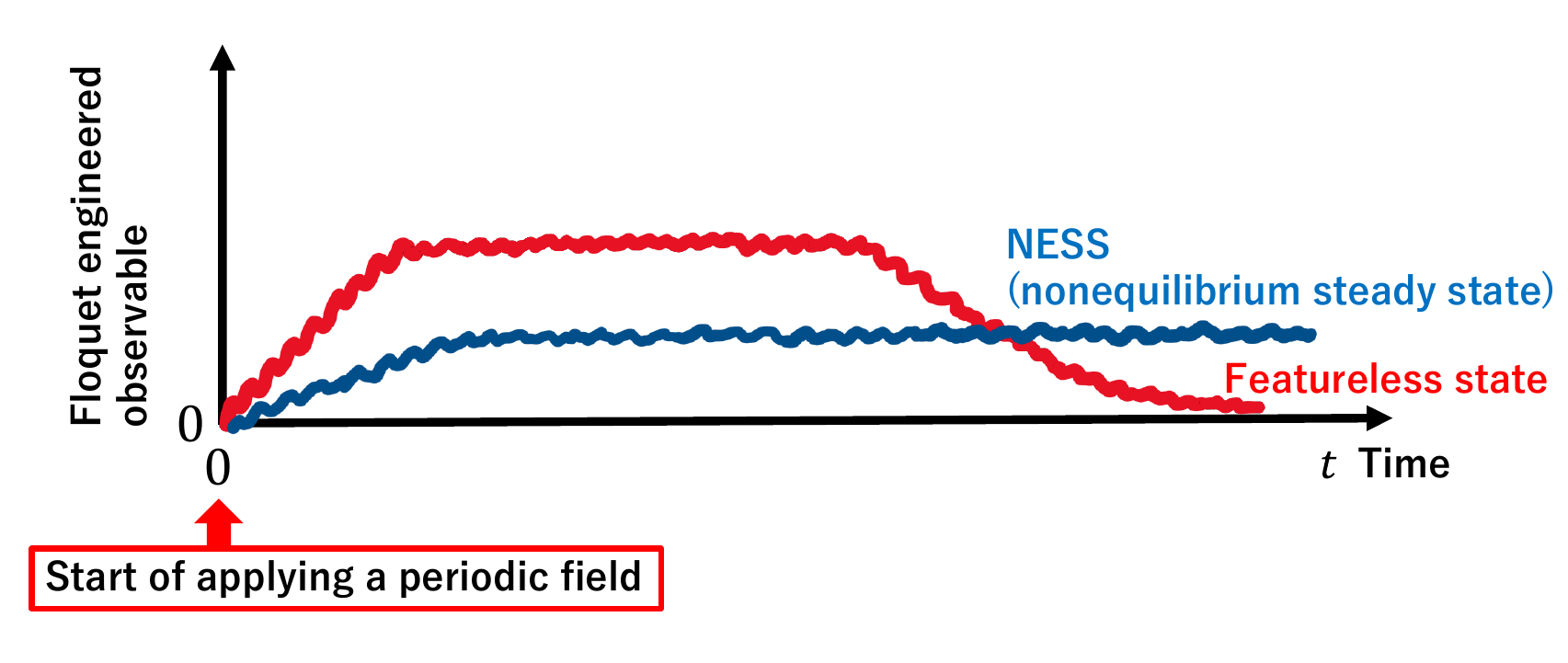}
\caption{Typical time evolution of Floquet engineered quantities in closed (red line) and open (blue line) quantum systems. 
At the beginning of applying a periodic field, 
the quantum state tends to approach an FG state in both the closed and open systems. 
After a long driving, the engineered quantity in an open system is expected to have a finite non-trivial value. It means that the FE survives in open systems even after the long driving.}
\label{fig:FE_Open}
\end{center}
\end{figure}

As we discussed in the Introduction and Sec.~\ref{sec:Closed}, the asymptotic states of periodically driven systems after long drive strongly depend on the system-bath interaction, as shown in Fig.~\ref{fig:LongLaserApplication}. 
In closed (isolated) systems, 
they reaches featureless state after long drive, 
while periodically driven open systems with dissipation are expected to approach a NESS due to the balance between the energy injection by the driving and the energy dissipation to the bath. Therefore, one can expect that in open systems, typical time evolution of the Floquet engineered observables obey the blue line shown in Fig.~\ref{fig:FE_Open}.  
In this section, we mainly consider how we can theoretically describe the properties of the NESS.

\subsection{GKSL (Lindblad) equation}
There are several types of EOMs for density matrices, depending on the accuracy of the approximations and the setups of open systems. 
Here, we consider the so-called Gorini–Kossakowski–Sudarshan–Lindblad (GKSL) equation~\cite{Lindblad1976,Gorini1976,BreuerBook,Alicki2007book} as the EOM for density matrices. 
This equation is also referred to as Lindblad equation or quantum master equation. 
The GKSL equation is expressed as 
\begin{align}
    \dv{\hat{\rho}(t)}{t} = -i\comm{\hat{H}(t)}{\hat{\rho}(t)} 
    + \hat{\mathcal{D}}(t)[\hat{\rho}(t)],
    \label{eq:GKSL}
\end{align}
where $\hat{\rho}(t)$ is the density matrix of the target system and the second term with a super operator $\hat{\cal D}$ is defined as 
\begin{align}
\hat{\mathcal{D}}[\hat{\rho}]=
\sum_j\gamma_j\left(\hat{L}_j\hat{\rho}\hat{L}_j^\dagger-\frac{1}{2}\{\hat{L}_j^\dagger\hat{L}_j,{\hat{\rho}}\}\right).
\label{eq:Dissipation}
\end{align}
The first term of Eq.~\eqref{eq:Dissipation}, 
$-i[\hat{H}(t), \hat{\rho}(t)]$, describes the standard unitary time evolution and $\hat H(t)$ is the Hamiltonian of the system. 
The second term 
describes the dissipative dynamics induced by the weak interaction between the system and the environment. The operator $\hat{L}_j$ is called the jump (Lindblad) operator, and each integer index $j$ corresponds to a sort of dissipation process. The parameter $\gamma_j$ represents the strength of $j$-th dissipation process and $1/\gamma_j$ can be interpreted as the relaxation time scale of the $j$-th dissipation process. 

This GKSL (quantum master) equation has some nice properties. First, this EOM is of Markovian type, namely, the time evolution of $\hat \rho(t)$ does not depend on the long history of $\hat \rho(t')$ with $t'<t$, and it is determined only by the information from the immediate past at $t-\Delta t$ ($\Delta t\to +0$). 
This nature makes numerical (or analytical) analysis of the GKSL equation easier. 
On the other hand, we should note that the GKSL equation cannot describe non-Markovian time evolutions, which would be important, for example, in the case with a strong interaction between the system and the environment.  
Secondly, if the density matrix is evolved within the GKSL equation, it always satisfies hermicity [$\hat\rho(t)^\dagger=\hat\rho(t)$], positivity, and ${\rm Tr}[\hat\rho(t)]=1$. 
From these natures, the density matrices can have probabilistic meaning at arbitrary time. 

The GKSL equation can be derived from the equations of motion for a composite system consisting of the target system and its environment~\cite{BreuerBook,Blumel1991,Hone1997,Ikeda2021}.
We begin by assuming that this composite system is isolated, so its total density matrix $\hat{\rho}_{\mathrm{tot}}(t)$ obeys the standard von Neumann equation, which is formally equivalent to the GKSL equation but without the dissipator $\hat{\mathcal{D}}\bigl[\hat{\rho}_{\mathrm{tot}}(t)\bigr]$.
By treating the system–environment coupling perturbatively, tracing out the environmental degrees of freedom, and applying the Markovian (secular) approximation, we obtain the GKSL equation for the reduced density matrix of the target system.
Section \ref{sec:microGKSL} presents this derivation in greater detail.

This derivation is justified under the condition that the relaxation time scale of the environment is sufficiently shorter than the time scale of typical relaxation dynamics of the target system. This is often satisfied if we consider a finite-size system with discrete energy levels, such as molecules, ultracold atoms, quantum-optics systems, etc. On the other hand, many-body systems such as solid electrons or spin systems usually include continuous energy levels; therefore, they sometimes violate the above condition. 
Recent theoretical studies have applied the GKSL equation to even many-body systems. In such a case, we should consider that the dissipation effects described by $\hat{\mathcal{D}}[\hat{\rho}(t)]$ are understood in a phenomenological sense.

Let us mention a few examples of the systems described by the GKSL equation. If we consider a simplified two-level atom or a nuclear spin with spin magnitude $S=\frac{1}2$, the ground and excited states before periodic driving are, respectively, given by $|g\rangle={}^t(0,1)$ and $|e\rangle={}^t(1,0)$. The density matrix is expressed by using $2\times 2$ matrix, and any physical operator is given by a linear combination of Pauli and unit matrices. If the jump operator is chosen as $\hat L\propto\sigma^-=\frac{1}{2}(\sigma_x-i\sigma_y)$ in the two-level atom, the dissipation process corresponds to spontaneous emission of the atom~\cite{BreuerBook}. If we consider two jump operators $\hat L_1\propto\sigma^-$ and $\hat L_2\propto\sigma_z$ in the single nuclear spin, these two dissipation processes are interpreted as longitudinal and transverse (dephasing) relaxations, respectively. In fact, the GKSL equation for the nuclear spin can be exactly mapped to the so-called Bloch equation for the magnetic resonance~\cite{Haug2008}.  

When we consider non-interacting many fermion or boson systems on a lattice, the Hilbert space is block diagonalized in the wave vector $\bm k$ space. In these systems, we can consider the GKSL equation for each $\bm k$ point, that is, we can define the density matrix $\hat\rho_{\bm k}(t)$ under the assumption that the dissipation process is also $\bm k$-independent. This $\bm k$-decomposed method has been applied in semiconductor and metallic systems~\cite{Haug2008,Tanaka2024,Kanega2021,Kanega2025}.

\subsection{Phenomenological GKSL equation for periodically driven systems}
In this section, we focus on periodically-driven open system described by the following GKSL equation,
\begin{align}
    \dv{\hat{\rho}(t)}{t} = -i\comm{\hat{H}(t)}{\hat{\rho}(t)} 
    + \hat{\mathcal{D}}[\hat{\rho}(t)],
    \label{eq:GKSL2}
\end{align}
where the Hamiltonian has time periodicity $\hat H(t)=\hat H(t+T)$ 
and the jump operators in the dissipation term are all time-independent. The Hamiltonian consists of two parts: $\hat H(t)=\hat H_{\rm st}+\hat H_{\rm ext}(t)$, where $\hat H_{\rm st}$ is the static Hamiltonian before periodic driving and $\hat H_{\rm ext}(t)=\hat H_{\rm ext}(t+T)$ is the periodic-driving induced part. 
Even under the condition of time-independent jump operators, one can adopt various types of $\hat L_j$. Among many types of $\hat L_j$, we choose a reasonable setup of jump operators: We determine the operators such that 
the system relaxes to an equilibrium state (canonical distribution) of the static part $\hat H_{\rm st}$. 
To discuss this setup, it is useful to use the energy bases of $\hat H_{\rm st}$,
\begin{align}
\hat H_{\rm st}|E_j\rangle=E_j|E_j\rangle
    \label{eq:Energy}
\end{align}
where $|E_j\rangle$ is the $j$-th eigenenergy state with the eigenvalue $E_j$ and we have set $E_{j+1}\geq E_j$. 
Using these bases, we assume that the dissipation term $\hat{\mathcal{D}}[\hat{\rho}(t)]$ is given by the following GKSL form
\begin{align}
\hat{\mathcal{D}}[\hat{\rho}]=
\sum_{ij}\Gamma_{ij}\left(\hat{L}_{ij}\hat{\rho}\hat{L}_{ij}^\dagger-\frac{1}{2}\{\hat{L}_{ij}^\dagger\hat{L}_{ij},{\hat{\rho}}\}\right),
    \label{eq:Dissipation2}
\end{align}
where $\hat{L}_{ij}=|E_i\rangle\langle E_j|$. Each pair of two integers $(i,j)$ corresponds to each integer $j$ in Eq.~\eqref{eq:Dissipation}. 
The operator ${\hat L}_{ij}$ represents a decay (excitation) process for $i<j$ ($i>j$). The real number $\Gamma_{ij}\geq0$
denotes the rate for the corresponding process and $\Gamma_{ii}=0$. 
If we assume that all $\hat{L}_{ij}$ satisfy the detailed balance condition, 
\begin{align}
\Gamma_{ij}e^{-\beta E_j}=\Gamma_{ji}e^{-\beta E_i} \,\,\,\,\,
({\rm for}\,\,\,\,i\neq j),
    \label{eq:DetailedBalanceCondition}
\end{align}
the static system of $\hat H_{\rm st}$ is known to relax to the equilibrium state with inverse temperature $\beta=1/(k_{\rm B}T)$.  
Here, the temperature $T$ should be considered as that of the environment. 
The GKSL equation satisfying Eqs.~\eqref{eq:Dissipation2} and \eqref{eq:DetailedBalanceCondition} is the setup of this section. This GKSL equation with a phenomenological dissipation term has been often used to analyze materials driven by a periodic field such as a laser.

\subsubsection{Generic formalism for high-frequency expansion}
Using the above GKSL equation with time-independent jump operators, 
we consider the NESS shown in Figs.~\ref{fig:LongLaserApplication} and \ref{fig:FE_Open}. 
We first symbolically rewrite the GKSL equation~\eqref{eq:GKSL} as 
\begin{align}
    \dv{\hat{\rho}(t)}{t} =\hat{\cal L}(t)[\hat\rho(t)],
    \label{eq:GKSL3}
\end{align}
where the super operator $\hat{\cal L}(t)$ is called the Liouvillian. We may call $\hat{\cal L}(t)$ the Lindbladian when we emphasize that it is of the GKSL form. 
The point is that (i) the Liouvillian is the linear operator with respect to the density matrix and (ii) it is time periodic, $\hat{\cal L}(t)=\hat{\cal L}(t+T)$, because of $\hat H(t)=\hat H(t+T)$ and the time independence of $\hat L_{ij}$. 
Therefore, we can apply the Floquet theorem to the GKSL equation of Eq.~\eqref{eq:GKSL3} like the Schr\"odinger equation in closed systems. At the beginning, let us define the Fourier series of $\hat{\cal L}(t)$ and $\hat H(t)$ as
\begin{align}
\label{eq:Fourier2}
 \hat{\cal L}(t)=\sum_{m\in \mathbb{Z}} e^{-im\omega t} \hat{\cal L}_m, \hspace{0.5cm}
 \hat H(t)=\sum_{m\in \mathbb{Z}} e^{-im\omega t} \hat H_m.
\end{align}
Secondly, like Eq.~\eqref{eq:3decomp}, the time evolution operator of the GKSL equation is decomposed as 
\begin{align}
    \hat V(t_2-t_1)&\equiv T_{t}\exp\left(\int^{t_2}_{t_1} d\tau\hat {\cal L}(\tau) \right)
    \nonumber\\
    &= e^{\hat {\cal G}(t_2)}e^{\hat {\cal L}_{\rm eff}(t_2-t_1)}e^{-\hat {\cal G}(t_1)},
    \label{eq:3decomp2}
\end{align}
where ${\cal G}(t)$ is periodic in time and $\hat {\cal L}_{\rm eff}$ is time-independent. Note that $\hat V(t)$ is generally not unitary. Also, it is known that the effective Liouvillian $\hat{\cal L}_{\rm eff}$ is not necessarily of the GKSL form~\cite{Haddadfarshi2015,Schnell2020,Schnell2021} although it has some favorable properties~\cite{Ikeda2021}. 
Similarly to the case of closed systems, we can solve the differential equation $\frac{d}{dt}\hat V(t,t')=\hat{\cal L}(t)V(t,t')$ in a recursive way under the condition of $\int_0^T dt\hat{\cal G}(t)=0$. As a result, we obtain the $1/\omega$ expansion formulas of $\hat {\cal L}_{\rm eff}$ and $\hat {\cal G}(t)$:
\begin{align}
\hat {\cal L}_{\rm eff}=\sum_{k=0} \hat {\cal L}_{\rm eff}^{(k)},
\hspace{0.5cm}
\hat {\cal G}(t)=\sum_{k=1} \hat {\cal G}^{(k)}(t),
    \label{eq:1/omega}
\end{align}
where $\hat {\cal L}_{\rm eff}^{(k)}$ and $\hat {\cal G}^{(k)}(t)$ are proportional to $\omega^{-k}$. 
The lower-order solutions are given by 
\begin{align}
\hat {\cal L}_{\rm eff}^{(0)}&=\hat {\cal L}_0,\nonumber\\
\hat {\cal G}^{(1)}(t)&=\frac{i}{\hbar\omega}\sum_{m\neq 0}\frac{e^{-im\omega t}}{m}\hat {\cal L}_m,\nonumber\\
\hat {\cal L}_{\rm eff}^{(1)}&=-\frac{i}{\hbar\omega}\sum_{m> 0}\frac{[\hat {\cal L}_m,\hat {\cal L}_{-m}]}{m}.
    \label{eq:lower}
\end{align}

\subsubsection{Nonequilibrium steady state (NESS)}
Using these results of the high-frequency expansion, we consider how the NESS is described in the sufficiently high-frequency regime. The slow dynamics of the density matrix is captured by the following expression, 
\begin{align}
 \hat\rho(t)
    &= e^{\hat {\cal G}(t)}e^{\hat {\cal L}_{\rm eff}t}e^{-\hat {\cal G}(0)}\hat\rho(0),
    \label{eq:Slow}
\end{align}
where $\hat\rho(0)$ denotes the initial state. To obtain the asymptotic behavior of $\hat\rho(t)$, we focus on the part $\hat q(t)=e^{\hat{\cal L}_{\rm eff}t} e^{-{\cal G}(0)}\hat\rho(0)$. The operator $\hat q(t)$ is the solution of the time-independent GKSL equation $d\hat q(t)/dt= \hat{\cal L}_{\rm eff}\hat q(t)$ from the initial state $e^{-{\cal G}(0)} \hat\rho(0)$.
If we assume that $\hat q(t)$ approaches the unique steady state regardless of the initial state, the asymptotic state 
$\hat q(t\to\infty)=\hat q_{\infty}$ can be determined by the following equation,
\begin{align}
\hat{\cal L}_{\rm eff}\hat q_{\infty}=0. 
    \label{eq:PreNESS}
\end{align}
Using the solution of $\hat q_{\infty}$, we can express the density matrix for the NESS as 
\begin{align}
\hat\rho_{\rm ness}(t)=e^{\hat{\cal G}(t)}\hat q_{\infty} \,\,\,\,\,({\rm for}\,\,\,\,t\to\infty).  
    \label{eq:NESS}
\end{align}
As $\hat{\cal G}(t)$ is periodic in time, the density matrix of the NESS is also periodic.

\subsubsection{Floquet engineering formula in the NESS}
From Eqs.~\eqref{eq:lower}, \eqref{eq:PreNESS}, and \eqref{eq:NESS}, we obtain the explicit formula of the density matrix in the NESS in the high-frequency regime. The result is 
\begin{align}
\hat\rho_{\rm ness}(t)=\hat\rho_{\rm can}+
\hat\rho_{\rm mm}(t)+\hat\rho_{\rm FE}+
{\cal O}(\omega^{-2}), 
    \label{eq:NESS_1/omega}
\end{align}
where both $\hat\rho_{\rm mm}(t)$ and $\hat\rho_{\rm FE}$ are ${\cal O}(\omega^{-1})$, and we call $\hat\rho_{\rm mm}(t)$ and $\hat\rho_{\rm FE}$ the micromotion and time-independent FE parts, respectively. The first term $\hat\rho_{\rm can}$ is the canonical distribution of the static Hamiltonian $\hat H_{\rm st}$: $\hat\rho_{\rm can}=e^{-\beta \hat H_{\rm st}}/Z_{\rm st}$ with $Z_{\rm st}$ being the partition function ${\rm Tr}[e^{-\beta \hat H_{\rm st}}]$. The micromotion part is estimated as 
\begin{align}
\hat\rho_{\rm mm}(t)=\frac{1}{\omega}\sum_{m\neq 0}\frac{e^{-im\omega t}}{m}[\hat H_m,\hat\rho_{\rm can}], 
    \label{eq:NESS_mm}
\end{align}
This result indicates that the micromition part indeed satisfies the periodicity $\hat\rho_{\rm mm}(t)=\hat\rho_{\rm mm}(t+T)$, and it does not contribute to the time averaged physical quantities because $\int_t^{t+T} d\tau {\rm Tr}[\hat\rho_{\rm mm}(t)\hat A]=0$ for arbitrary observable $\hat A$. 

The matrix elements of the FE part is explicitly given by 
\begin{align}
\langle E_k|\hat\rho_{\rm FE}|E_\ell\rangle =\frac{\langle E_k|\Delta {\hat H}_{\rm eff}|E_\ell\rangle}{(E_k-E_\ell)-i\gamma_{k\ell}}(\hat\rho_{\rm can}^{(k)}-\hat\rho_{\rm can}^{(\ell)})
    \label{eq:NESS_FE}
\end{align}
and $\langle E_k|\hat\rho_{\rm FE}|E_k\rangle=0$. Here, $\hat\rho_{\rm can}^{(k)}=e^{-\beta E_k}/Z_{\rm st}$ is the Boltzmann weight of the static Hamiltonian $\hat H_{\rm st}=\hat H_0$, $\Delta {\hat H}_{\rm eff}={\hat H}_{\rm eff}-\hat H_0={\cal O}(\omega^{-1})$, and 
$\gamma_{k\ell}=\sum_j (\Gamma_{jk}-\Gamma_{j\ell})/2$.
The reason why we call $\hat\rho_{\rm FE}$ the FE part is that it describes how the effective Hamiltonian changes physical observables from their values in thermal equilibrium. In contrast to the micromotion part, the FE part contributes to the time-averaged
quantities. 

Equations~\eqref{eq:NESS_1/omega}, \eqref{eq:NESS_mm}, and \eqref{eq:NESS_FE} are the main conclusion of the FE in the NESS generated by the phenomenological GKSL equation~\eqref{eq:GKSL2} with a time-independent jump operator. 
We note that ${\rm Tr}[\hat\rho_{\rm ness}(t)]=1$ holds, at least, up ${\cal O}(\omega^{-1})$ because both $\hat\rho_{\rm mm}(t)$ and $\hat\rho_{\rm FE}$ are traceless. 
In Table~\ref{tab:comparison1}, we summarize the comparison between FEs in closed systems and open systems described by Eq.~\eqref{eq:GKSL2}.
A significant advantage of the formula~\eqref{eq:NESS_1/omega} is that one can in principle estimate all observables in the NESS without directly computing the time evolution of the density matrix. Only the energy eigenvalues $\{E_j\}$ of the static Hamiltonian $\hat H_0$ and the Fourier components $\{\hat H_{m\neq 0}\}$ are enough to determine the density matrix of Eq.~\eqref{eq:NESS_1/omega}.

\begin{table*}
\caption{\label{tab:comparison1} 
Comparison between typical properties of FEs in closed systems described by the Schr\"odinger equation [Eq.~\eqref{eq:Sch}] and open systems described by the phenomenological GKSL equation [Eq.~\eqref{eq:GKSL2}]. }
\begin{ruledtabular}
\begin{tabular}{lll}
 Driving
    & Periodically driven closed systems    
    & Periodically driven open systems \\
    & Schr\"odinger Eq. $i\hbar \frac{\partial}{\partial t} |\psi(t)\rangle=\hat H(t) |\psi(t)\rangle$ 
    & Phenomenological GKSL Eq. $\frac{d}{dt}\hat{\rho}(t)= -i[\hat{H}(t),\hat{\rho}(t)] + \hat{\mathcal{D}}[\hat{\rho}(t)]$  \\
    & $\hat H(t)=\hat H(t+T)$
    & $\hat L_j$ satisfy the detailed balance condition \\
    \hline
 Short time & 
$\cdot$ Approach a Floquet-Gibbs (FG) state 
 &  $\cdot$ Basically the same as closed systems\\
    &  \hspace{0.5cm} $\hat\rho_{\rm FG}(t)=\frac{1}{Z_{\rm FG}}\sum_j e^{-\beta_{\rm FG}\hat H_{\rm eff}}|\phi_j(t)\rangle\langle\phi_j(t)|$
    & \hspace{0.5cm} (if driving time is shorter than relaxation time)\\
    &  $\cdot$ Difficult to estimate/control $\beta_{\rm FG}$   & \\
    &  $\cdot$ Difficult to estimate the pre-thermalization range   & \\
    &     & \\
 \hline
 Long time & $\cdot$ Heated up and burned 
 &  $\cdot$ Approach a NESS \\
   & $\cdot$ Approach a featureless state 
   & $\cdot$ FE formula in the NESS at high-frequency regime \\
   & \hspace{0.5cm} (high-temperature limit)  
   &  \hspace{0.5cm} $\hat\rho_{\rm ness}(t)=\hat\rho_{\rm can}+\hat\rho_{\rm mm}(t)+\hat\rho_{\rm FE}+
{\cal O}(\omega^{-2})$ \\
   &  $\cdot$ FE fails.
   &  \hspace{0.5cm}
  $\hat\rho_{\rm mm}(t)=\frac{1}{\omega}\sum_{m\neq 0}\frac{e^{-im\omega t}}{m}[\hat H_m,\hat\rho_{\rm can}]$ \\
   &  
   &  \hspace{0.5cm} $\langle E_k|\hat\rho_{\rm FE}|E_\ell\rangle =\frac{\langle E_k|\Delta {\hat H}_{\rm eff}|E_\ell\rangle}{(E_k-E_\ell)-i\gamma_{k\ell}}(\hat\rho_{\rm can}^{(k)}-\hat\rho_{\rm can}^{(\ell)})$ \\
   &  
   &  $\cdot$ $\beta$ is the inverse temperature of the environment \\
   &  
   &   $\cdot$ NESS slightly differs from FG state \\
   &  
   &   $\cdot$ $\hat\rho_{\rm cfss}(t)=\frac{1}{Z_{\rm FG}}\sum_j e^{-\beta E_j} |\phi_j(t)\rangle\langle\phi_j(t)|$ in weak dissipation limit. \\
\\
\end{tabular}
\end{ruledtabular}
\end{table*}

\subsubsection{NESS in the weak-dissipation limit}
\begin{figure}[t]
\begin{center}
\includegraphics[width=8cm]{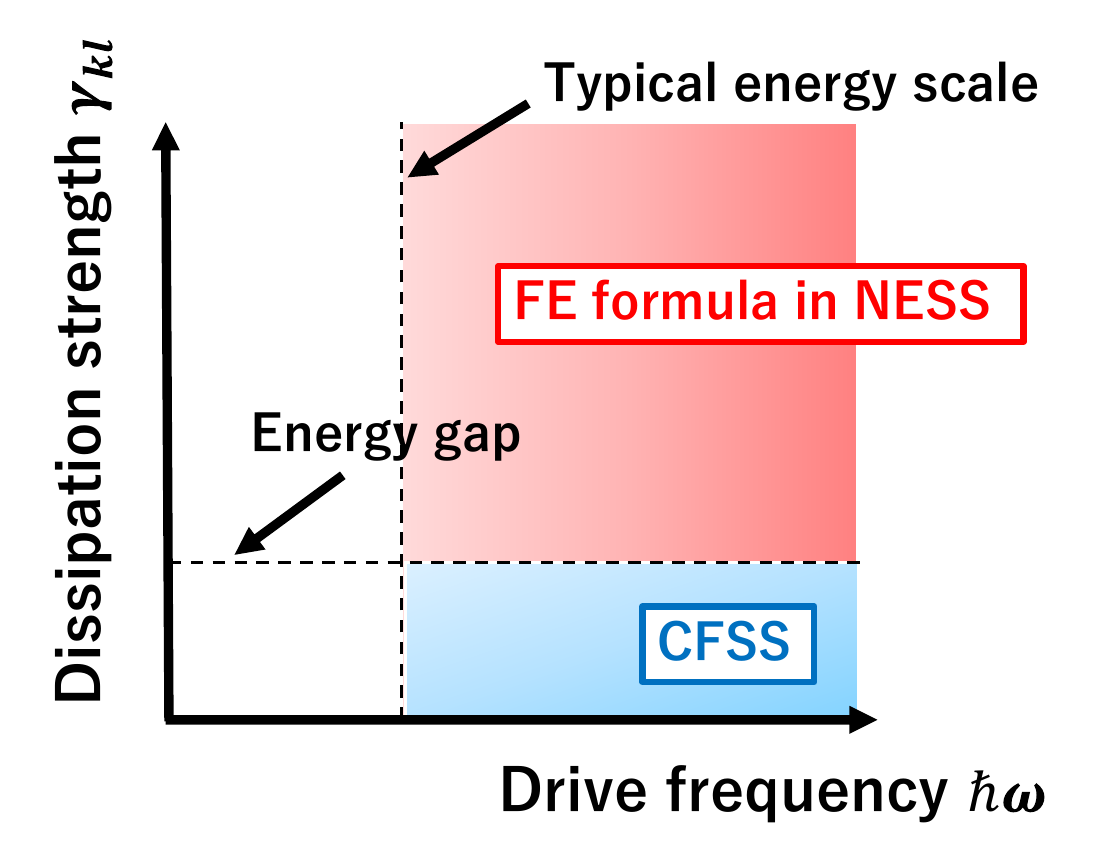}
\caption{FE formula of Eq.~\eqref{eq:NESS_1/omega} is valid when the driving frequency $\omega$ (more correctly $\hbar\omega$) is larger than the typical energy scale of the system. When the dissipation
strength is smaller than the (nonzero) minimum energy gap, the NESS reduces to the CFSS. Adapted from Ref.~\cite{Ikeda2020}.}
\label{fig:FEFormula_Open}
\end{center}
\end{figure}

One might expect that the NESS reaches a FG state when the dissipation becomes weak ($\Gamma_{ij}\to 0$ and $\gamma_{k\ell}\to 0$) because at least in a short time range, the state in driven closed systems with $\Gamma_{ij}=0$ generally approaches a FG state. However, this expectation is not true 
in the open system described by the phenomenological GKSL equation. Instead, the actual NESS coincides with another state, which we name the canonical Floquet steady state (CFSS). Its density matrix is given by 
\begin{align}
    \label{eq:CFSS}
    \hat\rho_{\rm cfss}(t)=\hat\rho_{\rm ness}(t)|_{\gamma_{k\ell}\to0}=\sum_j \frac{e^{-\beta E_j}}{Z} |\phi_j(t)\rangle\langle\phi_j(t)|.
\end{align}
Comparing this with Eq.~\eqref{eq:FG}, 
one sees that $\epsilon_j$ and $\beta_{\rm FG}$ in the FG state are respectively replaced with $E_j$ and $\beta$ in the CFSS. 
Therefore, the CFSS is similar to the FG state, but more accurately speaking, the former differs slightly from the FG state. In Fig.~\ref{fig:FEFormula_Open}, we depict (i) the regime that Eq.~\eqref{eq:NESS_1/omega} is valid and (ii) where the CFSS appears. 

\subsubsection{IFE of spin-1 in NV center}
\label{sec:example_NVcenter}
In Secs.~\ref{sec:example_NVcenter} and \ref{sec:example_IFE}, we briefly discuss the GKSL equation for two simple models as examples of the NESSs. In this subsection, we focus on the single spin-1 system driven by an EM wave as the model for an impurity spin embedded in a diamond NV center~\cite{Rondin2014,Suter2017}.
The NESS of this system may be viewed as a typical example of the NESS realized in a finite-level system.  
The Hamiltonian is defined as 
\begin{align}
    \label{eq:NVCenter}
    \hat H_{\rm nv}(t)=& -B_s{\hat S}^z+ N_z(\hat S^z)^2 
    -N_{xy}((\hat S^x)^2-(\hat S^y)^2) \nonumber\\
    &+ \hat H_{\rm cpl}(t),
\end{align}
where $\hat S^{x,y,z}$ is the spin-1 operator in the NV center, $B_s$ is an applied static magnetic field, 
$N_{z,xy}$ are the strength of the single-ion (spin quadrupole) anisotropy interactions, and $\hat H_{\rm cpl}(t)$ is the ac Zeeman interaction by a circularly polarized magnetic field: $\hat H_{\rm cpl}(t)=-B_d (\hat S^x\cos(\omega t)+\hat S^y\sin(\omega t))$, in which the ac magnetic field is in the $x$-$y$ plane. In the following, we set $N_z=1$ and $N_{xy}=0.05$ as the relation $N_z\gg N_{xy}$ holds in the spin of the NV center. 
Moreover, the magnitudes of the dc and ac magnetic fields are set to $B_s=0.3$ and $B_d=0.1$. The dissipation parameters $\Gamma_{ij}$ are set to $\Gamma_{ij}=\gamma e^{-\beta E_i}/(e^{-\beta E_i}+e^{-\beta E_j})$ for $i\neq j$ with $\gamma=0.2$ and $\Gamma_{ii}=0$ so that they satisfy the detailed balance condition. 

\begin{figure}[t]
\begin{center}
\includegraphics[width=8cm]{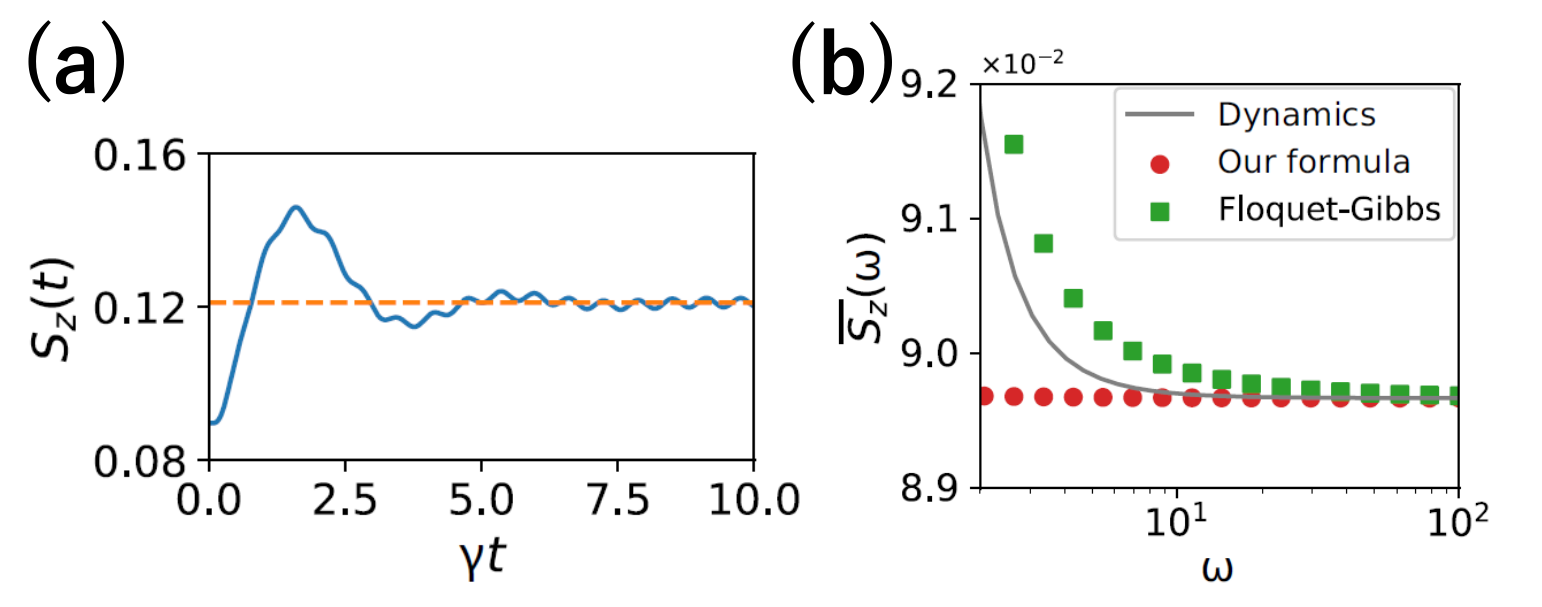}
\caption{(a) Time evolution of $\langle \hat S^z(t)\rangle$ described by the GKSL equation with the Hamiltonian $\hat H_{\rm nv}(t)$. We set the frequency of the circular polarized magnetic field to $\omega=1.0$. Other parameters are shown in the main text.  
The dashed line shows the one-cycle average of $\langle \hat S^z(t)\rangle$ at a sufficiently long time. (b) One-cycle average $\langle \hat S^z(t)\rangle$, $\bar S_z(\omega)$, as a function of the frequency of the applied ac magnetic field. The average is taken after a long driving. 
Solid line, red points, and green points are respectively calculated from the real-time evolution by the GKSL equation, the FE formula of Eq.~\eqref{eq:NESS_1/omega}, and the FG state of Eq.~\eqref{eq:FG}. Adapted from Ref.~\cite{Ikeda2021}.}
\label{fig:DissipativeNVCenter}
\end{center}
\end{figure}

From the discussion in Secs.~\ref{sec:IFE} and \ref{sec:spin_multiferro}, we can expect that an IFE occurs in this spin-1 model because of the circularly polarized field, and the impurity spin becomes magnetized along the $S^z$ direction. Figure~\ref{fig:DissipativeNVCenter}(a) shows the numerically computed time evolution of $\langle \hat S^z(t)\rangle$, in which the initial state at $t=0$ is set to the ground state of the static part of the Hamiltonian $\hat H_{\rm nv}(t)$ without the ac field $\hat H_{\rm cpl}(t)$. 
From panel (a), one finds that an IFE indeed takes place and the magnetization approaches a certain value 
$\langle \hat S^z(t)\rangle\simeq 0.12$ after long driving, i.e., a NESS is realized. Figure~\ref{fig:DissipativeNVCenter}(b) depicts the $\omega$ dependence of the time-averaged magnetization $\bar S_z(\omega)=\frac{1}{T}\int_{t_0}^{t_0+T}d\tau \langle \hat S^z(\tau)\rangle$ for a long enough $t_0$. 
It clearly shows that the FE formula of Eq.~\eqref{eq:NESS_1/omega} and the FG state of Eq.~\eqref{eq:FG} both can capture the photo-induced magentization in a sufficiently high-frequency range. In addition, one finds that the result of Eq.~\eqref{eq:NESS_1/omega} is slightly better than Eq.~\eqref{eq:FG} in a moderate frequency range. 

\subsubsection{IFE of SO-coupled metal}
\label{sec:example_IFE}
\begin{figure}[t]
\begin{center}
\includegraphics[width=8cm]{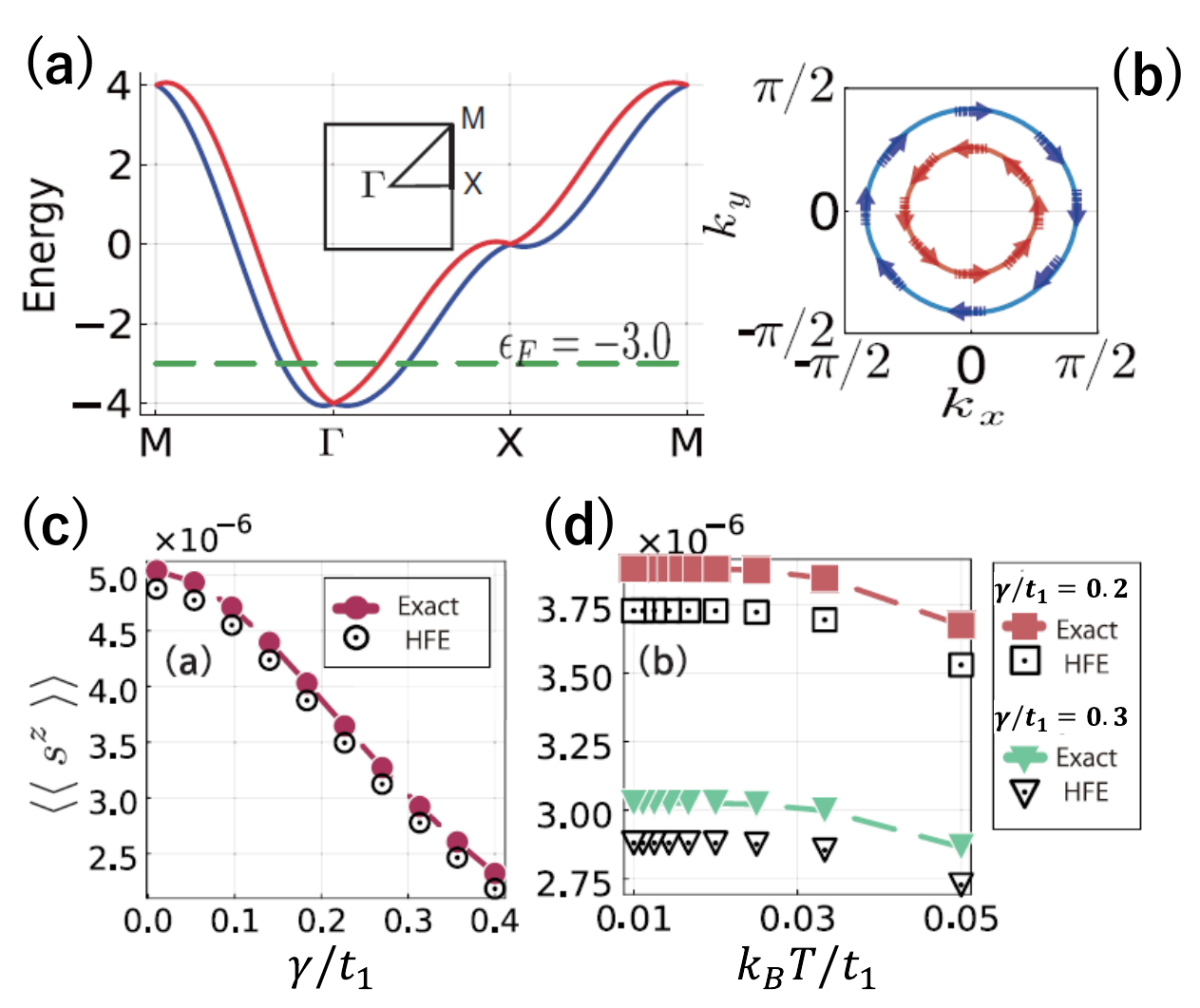}
\caption{(a) Energy band structure of the SO-coupled electron model $\hat H_{\rm R}=\hat H_{\rm K}+\hat H_{\rm so}$ of Eqs.~\eqref{eq:Rashba1} and \eqref{eq:Rashba2}. We have set the Fermi energy $\epsilon_F=-3 t_1$. 
(b) Spin polarization direction of electrons on the Fermi surface at $\epsilon_F=-3 t_1$ in the model $\hat H_{\rm R}$. The spin-momentum locking is realized because of the Rashiba SO coupling. 
(c,d) Laser-driven time-averaged magnetization $\langle\langle S^z\rangle\rangle$ per a site in the NESS as a function of (c) the dissipation magnitude $\gamma$ and (d) temperature $T$. Other parameters are chosen to be 
$eaE_0/t_{1} = 0.1$ ($a$ is the lattice constant and $E_0$ is the magnitude of the ac electric field), $\alpha R/t_{1} = 0.1$, and $\hbar\omega/t_{1} = 1.0$. 
In (c), $\gamma$ changes in the range of 
$0 < \gamma/t_{1} \leq 0.4$ at $T = 0$. Red
points and white circles, respectively, correspond to the result of the numerically solved GKSL equation and the FE formula of Eq.~\eqref{eq:NESS_1/omega}. 
In (d), we depict the magnetization in the cases of $\gamma/t_{1} = 0.2$ and 0.3 as afunction of $T$. 
Red and green points are numerical results of the GKSL equation, while white squares and triangles
are those computed by the FE formula of Eq.~\eqref{eq:NESS_1/omega}. 
Temperature is changed from $k_{\rm B}T/t_{1} = 0.0$ to 0.05, which corresponds to about 500K for
$t_{1}= 1$ eV. Reproduced from Ref.~\cite{Tanaka2024} with permission.}
\label{fig:DissipativeMetal}
\end{center}
\end{figure}

In this subsection, we consider the dissipative version~\cite{Tanaka2024} of the metallic IFE discussed in Sec.~\ref{sec:IFE}. Since the Rashiba-SO-coupled model of Eqs.~\eqref{eq:Rashba1} and \eqref{eq:Rashba2} is a free-electron type, 
its Hamiltonian is block diagonalized in the $\bm k$ space even after the laser application. 
Under the assumption that dissipation effects in two different $\bm k$ points are independent of each other, we may independently consider and solve the GKSL equation at each $\bm k$ point: 
\begin{align}
        \dv{\hat{\rho}_{\bm k}(t)}{t} = -i\comm{\hat{H}_{\bm k}(t)}{\hat{\rho}_{\bm k}(t)} 
    + {\hat{\mathcal{D}}}_{\bm k}[\hat{\rho}_{\bm k}(t)],
    \label{eq:GKSL_k}
\end{align}
where $\hat{\rho}_{\bm k}(t)$ is the density matrix in a fixed-$\bm k$ subspace, $\hat H_{\bm k}(t)$ is the Hamiltonian in the same subspace [see Eq.~\eqref{eq:Rashba_k}], and the final term ${\hat{\mathcal{D}}}_{\bm k}$ represents the standard dissipation term of the GKSL equation. 
In the present model with two energy bands [see Fig.~\ref{fig:DissipativeMetal} (a) and (b)], 
each $\bm k$ point possesses four independent states (bases): empty state, two kinds of one-electron states (upper- and lower-band occupied states), and two-electron (fully occupied) states. 
Within this tight-binding model, only one-electron states contribute to the laser-driven dynamics, whereas other two states have no time evolution. 
Therefore, it is enough to consider one-electron states, and the density matrix $\hat\rho_{\bm k}$ is reduced to the $2\times 2$ form.

The one-electron states at $T=0$ are located in the donut area surrounded by two Fermi surfaces in Fig.~\ref{fig:DissipativeMetal}. 
In a sufficiently low temperature range, we expect that it is enough to consider only one-electron states to estimate the laser-driven physical quantities. In the following, we discuss the IFE under this approximation. In each $\bm k$ point, 
two energy levels of $E_1^{(\bm k)}$ and $E_2^{(\bm k)}$, respectively, correspond to the energies of lower- and upper-band occupied states. 
Using $E_{1,2}^{(\bm k)}$, we set the dissipation strength $\Gamma_{ij}^{(\bm k)}$ at a $\bm k$ point to $\Gamma_{ij}^{(\bm k)}=\gamma e^{-\beta E_i^{(\bm k)}}/(e^{-\beta E_i^{(\bm k)}}+e^{-\beta E_j^{(\bm k)}})$ for $i\neq j$ and $\Gamma_{ii}^{(\bm k)}=0$.

Figures~\ref{fig:DissipativeMetal}(c) and (d) show the photo-induced magnetization in the NESS. Both panels (c) and (d) clearly indicate that the magnetization computed by the FE formula of Eq.~\eqref{eq:NESS_1/omega} well captures the numerically exact result. We emphasize that the FE formula of Eq.~\eqref{eq:NESS_1/omega} enables us to compute the temperature dependence of laser-driven physical quantities in the NESS [see panel (d)].  
Many theoretical studies for laser-driven dynamics in many-body systems have started from the zero-temperature ground state because it is generally difficult to analyze finite-temperature states in laser-driven nonequilibrium setups. On the other hand, the GKSL equation and Floquet theory provide a method of treating laser-driven systems at finite temperatures (although we have applied this method only to a simple free-electron system). In addition, (as we already mentioned) the temperature of the NESS can be controlled as that of the environment, and this nature is in contrast to the fact that it is difficult to control the inverse temperature $\beta_{\rm FG}$ of the FG states in periodically driven closed systems.

\subsection{Microscopically derived GKSL equation for periodically driven systems}\label{sec:microGKSL}
In this subsection, we review time-dependent dissipators that are microscopically derived for the system-bath coupling setup~\cite{Blumel1991,Hone1997,Ikeda2021}.
Here we focus on the quantum master equation obtained by the rotating wave approximation (RWA), although several different variants have been derived without the RWA~\cite{Nathan2020,Mozgunov2020,Becker2020}.

\subsubsection{Derivation of Floquet-GKSL equation}\label{sec:derivationRWA}
We here summarize the Floquet GKSL equation obtained microscopically with the RWA and leave its detailed derivation in Ref.~\cite{Ikeda2021}. 
Suppose that we have a system-bath coupled Hamiltonian, $H_\mathrm{tot}(t) = \hat{H}_s(t) + \hat{H}_b+ \hsb$,
where $\hat{H}_s(t+T)=\hat{H}_s(t)$ is for the periodically driven system of interest, $\hat{H}_b$ for the heat bath (reservoir), and $\hsb=\sum_\alpha \hat{A}_\alpha\otimes \hat{B}_\alpha$ for the system-bath coupling.
The bath defines the correlator,
\begin{align}
	\Gamma_{\alpha\beta}(\epsilon)
	=\int_0^\infty ds \,\ee^{\ii\epsilon s}\langle \hat{B}_\alpha(s)\hat{B}_\beta(0)\rangle
	= \frac{1}{2}\gamma_{\alpha\beta}(\epsilon)+\ii S_{\alpha\beta}(\epsilon),
\end{align}
where $\hat{B}_\alpha(t)=e^{i\hat{H}_b t}\hat{B}_\alpha e^{-i\hat{H}_b t}$,
and $\gamma_{\alpha\beta}(\epsilon)$ and $S_{\alpha\beta}(\epsilon)$ are Hermitian matrices.
We assume that the bath is thermal $\rho_B\propto e^{-\beta \hat{H}_b}$ at inverse temperature $\beta$,
which implies the Kubo-Martin-Schwinger (KMS) condition
\begin{align}\label{eq:KMS}
	\gamma_{\alpha\beta} (-\epsilon) = e^{-\beta\epsilon}\gamma_{\beta\alpha}(\epsilon).
\end{align}
The natural basis for the operator $\hat{A}_\alpha$ is the Floquet eigenstates:
\begin{align}
	\quad \ket{\psi_m(t)} = e^{-\ii \epsilon_m t}\ket{\phi_m(t)}
\end{align}
where $\ket{\phi_m(t+T)} = \ket{\phi_m(t)}$ and $\epsilon_m$ is the quasienergy.

Under appropriate Born and Markov approximations~\cite{Blumel1991,Hone1997}, we obtain the Floquet-GKSL equation with the Lindbladian
\begin{align}
	\mL_t(\rho) = -\ii [\hat{H}_s(t)+\lamb(t),\rho ] +\mD_t(\rho)
    \label{eq:MicroLindbladian}
\end{align}
with the Lamb shift $\lamb(t)$ and the dissipator $\mD_t$ given by
\begin{align}
	\lamb(t) &= \sum_{\alpha,\beta,\epsilon}S_{\alpha\beta}(\epsilon)\hat{A}^{\alpha\dag}_\epsilon(t) \hat{A}^\beta_\epsilon(t),\label{eq:Lambdef}\\
	\mD_t(\rho) &= \sum_{\alpha,\beta,\epsilon} \gamma_{\alpha\beta}(\epsilon)\left[
		\hat{A}^\beta_\epsilon(t) \rho \hat{A}^{\alpha\dag}_\epsilon(t)
		-\frac{1}{2}\left\{ \hat{A}^{\alpha\dag}_\epsilon(t) \hat{A}^\beta_\epsilon(t),\rho  \right\}\right].\label{eq:Dtdep}
\end{align}
The subscripts $t$ on $\mD_t$ and $\mL_t$ emphasize that the jump operator is time-dependent. 
In Eqs.~\eqref{eq:Lambdef} and \eqref{eq:Dtdep}, the jump operators are given by
\begin{align}\label{eq:jumpA}
	\hat{A}^\alpha_\epsilon(t) = e^{-i\epsilon t}\sum_{m,n} \Asf^\alpha_{mn}(\epsilon)\ket{\psi_m(t)}\bra{\psi_n(t)},
\end{align}
where the matrix elements $\Asf^\alpha_{mn}(\epsilon)$ are defined by the Fourier expansion
\begin{align}\label{eq:AFourier}
\braket{\psi_m(t)| \hat{A}_\alpha|\psi_n(t)}
=\sum_{\epsilon} \Asf^\alpha_{mn}(\epsilon)\ee^{-\ii \epsilon t}.
\end{align}
and periodic in time as $\hat{A}^\alpha_\epsilon(t)$ are periodic.
Since $\braket{\psi_m(t)| \hat{A}_\alpha|\psi_n(t)}=e^{i(\epsilon_m-\epsilon_n)t}\braket{\phi_m(t)|\hat{A}_\alpha|\phi_n(t)}$
and $\ket{\phi_n(t+T)}=\ket{\phi_n(t)}$, the sums over $\epsilon$ in the above equations are taken for
\begin{align}
	\epsilon=\epsilon_{nm;k}\equiv \epsilon_n-\epsilon_m+k\omega \quad (k\in\mathbb{Z}).
\end{align}
Thus, one can also rewrite Eq.~\eqref{eq:AFourier} as
\begin{align}\label{eq:AFourier_mn}
\braket{\psi_m(t)| \hat{A}_\alpha|\psi_n(t)}
&=\sum_{k} \Asf^\alpha_{mn}(\epsilon_n-\epsilon_m+k\omega)\ee^{-\ii (\epsilon_n-\epsilon_m+k\omega) t}\\
&\equiv\sum_{k} \Asf^\alpha_{mn;k}\ee^{-\ii (\epsilon_n-\epsilon_m+k\omega) t}.
\end{align}
As shown below, the jump operator $\hat{A}^\alpha_\epsilon(t)$ ($\hat{A}^{\alpha\dag}_\epsilon(t)$) lowers (raises) quasienergy by $\epsilon$.

One remarkable property of this Floquet GKSL equation is that the Lindbladian is time-independent in the interaction picture. Since the Lamb shift is diagonal in the Floquet eigenbasis and induces no transition~\cite{Ikeda2021}, we here ignore it for simplicity. To obtain the GKSL equation in the interaction picture, we express $\rho(t)$ as
\begin{align}
	\rho(t) = \sum_{m,n}\ket{\psi_m(t)}\sigma_{mn}(t)\bra{\psi_n(t)},
\end{align}
we have the following equation for $\sigma(t)$:
\begin{align}
	\frac{d\sigma(t)}{dt}=
	\sum_{\alpha,\beta,\epsilon}\gamma_{\alpha\beta}(\epsilon)\left[ \Asf^\beta(\epsilon)\sigma(t)\Asf^{\alpha\dag}(\epsilon)
	-\frac{1}{2}\left\{ \Asf^{\alpha\dag}(\epsilon) \Asf^{\beta}(\epsilon),\sigma(t)  \right\} \right].
	\label{eq:FLEsigmat}
\end{align}
Thus, the Lindbladian in this frame is time-independent.

\subsubsection{Nonequilibrium steady states}
In the rest of Sec.~\ref{sec:microGKSL}, we assume that the quasienergies are not degenerate
and analyze the NESS using the Floquet-GKSL in the time-independent frame~\eqref{eq:FLEsigmat}.
Then, the diagonal elements, $P_n(t)\equiv \sigma_{nn}(t)$, obey the following classical master equations
\begin{align}
\frac{dP_n(t)}{dt} = \sum_m \left[ W_{nm}P_m(t) - W_{mn} P_n(t) \right], \label{eq:cme}
\end{align}
where
\begin{align}
	W_{nm} 
	=\sum_{\alpha,\beta,k} \gamma_{\alpha\beta}(\epsilon_m-\epsilon_n-k\omega)\Asf_{nm;k}^{\alpha*} \Asf_{nm;k}^\beta \label{eq:defW2}
\end{align}
denotes the transition rates from the Floquet state $\ket{\psi_m(t)}$ to another $\ket{\psi_n(t)}$.
The off-diagonal elements, forming another closed set of equations, are guaranteed to vanish in the NESS~\cite{Ikeda2021}.

\begin{figure}
\begin{center}
	\includegraphics[angle=0,width=\columnwidth]{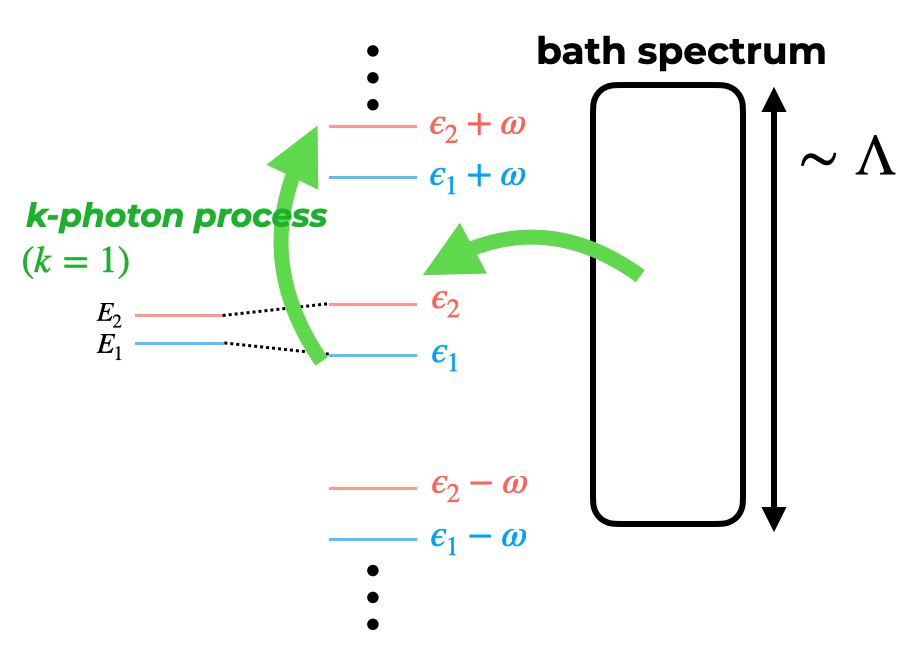}
	\caption{Schematic illustration of the $k$-photon process for $k=1$ in a periodically driven two-level system.
	The curved arrow indicates a dissipation-induced transition between the Floquet states $m=1$ and $n=2$, entailing the quasienergy change $\epsilon_2+\omega-\epsilon_1$. In this process, the bath's energy changes by $\epsilon_1-\epsilon_2-\omega$, which compensates the system's quasienergy change.
    Adapted from Ref.~\cite{Ikeda2021}.
	}
	\label{fig:diagram}
\end{center}
\end{figure}

Here we make a physical interpretation of the transition rate $W_{nm}$.
Equation~\eqref{eq:defW2} dictates that the transition rate $W_{nm}$ consists of contributions labeled by integer $k$.
For each $k$, in addition to the transition amplitude $\Asf^\beta_{nm;k}$, there appears the bath's spectral weight $\gamma_{\alpha\beta}(\epsilon_m-\epsilon_n-k\omega)$.
For $k=0$, the argument of the spectral weight is
$\epsilon_m-\epsilon_n$ corresponding to the quasienergy difference between the initial $\ket{\psi_m(t)}$ and final $\ket{\psi_n(t)}$ Floquet states in the transition.
For $k\neq0$, the argument involves an extra energy $k\omega$, which corresponds to the energy of $k$ photons.
From these observations, we interpret that $k$ in Eq.~\eqref{eq:defW2} stands for the number of photons involved in the Floquet-state transitions, and the quasienergy difference and the photon energy are exchanged between the system and bath (see Fig.~\ref{fig:diagram} for illustration).
Thus, in the following, we will refer to each $k$ contribution in the sum of Eq.~\eqref{eq:defW2} as the $k$-photon process.

To study the NESS, we assume that $W_{mn}$ is irreducible so that the classical master equation~\eqref{eq:cme} has the unique steady-state solution $P_n^\mathrm{ss}$,
which $P_n(t)$ approaches as $t\to\infty$.
This solution is characterized by Eq.~\eqref{eq:cme} as
\begin{align}
	\sum_m \left( W_{nm}P_m^\mathrm{ss} - W_{mn} P_n^\mathrm{ss} \right)=0,
\end{align}
which is equivalent to
\begin{align}\label{eq:ssdet}
	\sum_m S_{nm}P_m^\mathrm{ss} = 0,\qquad
	S_{nm}\equiv W_{nm} -\delta_{mn}\sum_m W_{mn}.
\end{align}
Equation~\eqref{eq:ssdet} means that $P_m^\mathrm{ss}$ is the zero-eigenvalue eigenvector for the matrix $S_{mn}$.
Let us also assume that the NESS density matrix is unique.
Once having $P_n^\mathrm{ss}$, we obtain the NESS density matrix as
\begin{align}\label{eq:nessRWA}
	\rhoness(t) = \sum_n P_n^\mathrm{ss} \ket{\psi_n(t)}\bra{\psi_n(t)}.
\end{align}

\subsubsection{Conditions for Floquet-Gibbs state}\label{sec:fgs}
Here we study when the NESS resembles the FG state
\begin{align}
     P^\mathrm{ss}_n = P^\mathrm{FG}_n\equiv \frac{e^{-\beta \epsilon_n}}{ \sum_m e^{-\beta \epsilon_m}} \label{eq:fgs}
\end{align}
for high-frequency drives.
Whereas $P_n^\mathrm{ss}$ are determined microscopically by $W_{mn}$ as in Eq.~\eqref{eq:ssdet}, $P_n^\mathrm{FG}$ are the simple canonical distribution for the quasienergies.
As we discussed above, the quasienergies $\epsilon_m$ are defined modulo $\omega$, and the FG state is, in general, ill-defined.
Nevertheless, when $\omega$ is much larger than the system's energy scale, there exists a one-to-one correspondence between the quasienergies in $-\omega/2 < \epsilon_n <\omega/2$ and the eigenenergies of the time-averaged Hamiltonian.
When we refer to the FG state in this paper, we implicitly assume the one-to-one correspondence.

The FG state well approximates the NESS if the driving frequency $\omega$ is larger than both the system's energy scale and the high-frequency cutoff $\Lambda$ of the bath spectral function.
When $\omega$ is much larger than the bath spectral cutoff $\Lambda$, we have
$\gamma(\epsilon_m-\epsilon_n+k\omega)\approx0$ for $k\neq0$.
Then, Eq.~\eqref{eq:defW2} can be approximated as
\begin{align}
	W_{nm}
	\approx\sum_{\alpha,\beta} \gamma_{\alpha\beta}(\epsilon_m-\epsilon_n)\Asf_{nm;0}^{\alpha*}\Asf_{nm;0}^\beta.\label{eq:defW_nok}
\end{align}
Within this approximation, we obtain the detailed balance condition $e^{-\beta\epsilon_m}W_{nm} = e^{-\beta\epsilon_n} W_{mn}$ with the help of the KMS condition~\eqref{eq:KMS}. 
The detailed balance condition means that the FG state population $P^\mathrm{FG}_n$ satisfies Eq.~\eqref{eq:ssdet}, meaning that the FG state~\eqref{eq:fgs} becomes the NESS $\rhoness(t) \approx \rhofg(t)$.

Physically, all $k$-photon processes except $k=0$ are negligible in this limiting case of $\omega\gg\Lambda$.
As we discussed earlier, for a $k$-photon process to occur, the accompanying photon energy $k\omega$ must be compensated by the bath.
However, if the bath spectral cutoff $\Lambda$ is much smaller than the photon energy $\omega$, such a compensation cannot happen.

In more realistic situations with $\omega\lesssim\Lambda$, however, the $k$-photon $(k\neq0)$ processes become relevant. In those situations, the transition rates $W_{mn}$~\eqref{eq:defW2} do not satisfy the detailed balance condition, and the FG state no longer approximates the NESS accurately except for some special cases~\cite{Shirai2016}.

\begin{table*}
\caption{\label{tab:comparison2} 
Comparison between typical properties of FEs in open systems described by the phenomenological GKSL equation [Eq.~\eqref{eq:GKSL2}] and microscopically derived GKSL equation [Eq.~\eqref{eq:MicroLindbladian}]. For ter short-time behavior, see Table~\ref{tab:comparison1}. }
\begin{ruledtabular}
\begin{tabular}{lll}
 Driving
    & Periodically driven open systems 
    & Periodically driven open systems \\
    & Phenomenological GKSL Eq.
    & Microscopically derived GKSL equation \\
    & $\frac{d}{dt}\hat{\rho}(t)= -i[\hat{H}(t),\hat{\rho}(t)] + \hat{\mathcal{D}}[\hat{\rho}(t)]$  
    & $\frac{d}{dt}\hat{\rho}(t)= -i[\hat{H}(t),\hat{\rho}(t)] + \hat{\mathcal{D}}_t[\hat{\rho}(t)]$  
    \\
    & $\hat L_j$ satisfy the detailed balance condition 
    & $\hat L_j$ is time-dependent
    \\
    \hline
 Long time &  $\cdot$ Approach a NESS 
    & $\cdot$ Approach a NESS 
    \\
   & $\cdot$ FE formula in the NESS at high-frequency regime 
   & $\cdot$ In the high-frequency regime ($\hbar|\omega|\gg$ typical energy scale)
   \\
   &  \hspace{0.5cm} $\hat\rho_{\rm ness}(t)=\hat\rho_{\rm can}+\hat\rho_{\rm mm}(t)+\hat\rho_{\rm FE}+
{\cal O}(\omega^{-2})$ 
   & \hspace{0.5cm} $\cdot$ NESS $\simeq$ FG state for $\hbar|\omega| > \Lambda$
   \\
   &  \hspace{0.5cm}
  $\hat\rho_{\rm mm}(t)=\frac{1}{\omega}\sum_{m\neq 0}\frac{e^{-im\omega t}}{m}[\hat H_m,\hat\rho_{\rm can}]$ 
   & \hspace{1cm} $\hat\rho_{\rm ness}(t)=\hat\rho_{\rm FG}(t)=\sum_j P^\mathrm{FG}_j \ket{\phi_j(t)}\bra{\phi_j(t)}$
   \\
   &  \hspace{0.5cm} $\langle E_k|\hat\rho_{\rm FE}|E_\ell\rangle =\frac{\langle E_k|\Delta {\hat H}_{\rm eff}|E_\ell\rangle}{(E_k-E_\ell)-i\gamma_{k\ell}}(\hat\rho_{\rm can}^{(k)}-\hat\rho_{\rm can}^{(\ell)})$ 
   &  \hspace{1cm} $\beta$ is the inverse temperature of the environment
   \\
   &  $\cdot$ $\beta$ is the inverse temperature of the environment 
   & \hspace{0.5cm} $\cdot$ NESS $\neq$ FG state for $\hbar|\omega| <\Lambda$
   \\
   &   $\cdot$ NESS slightly differs from FG state 
   & 
   \\
   &   $\cdot$ $\hat\rho_{\rm cfss}(t)=\frac{1}{Z_{\rm FG}}\sum_j e^{-\beta E_j} |\phi_j(t)\rangle\langle\phi_j(t)|$ 
   & 
   \\
   &   \hspace{0.5cm} in weak dissipation limit. 
   & 
   \\
\\
\end{tabular}
\end{ruledtabular}
\end{table*}

\subsection{Dissipation assisted Floquet engineering}
As an interesting phenomenon in periodically driven dissipative systems, 
we can consider the Floquet engineering (FE) that occurs with the help of a dissipation effect. We call such a FE the dissipation-assisted FE. In this subsection, we review two examples of dissipation-assisted FEs.

\subsubsection{Photo-induced spin nematic moment in NV center}
In this subsection, we discuss the dissipation-assisted FE realized in the phenomenological GKSL equation with a time-independent jump operator. To this end, we again pay attention to the periodically driven spin-1 model discussed in Sec.~\ref{sec:example_NVcenter}. 
In addition to the photo-induced magnetization of the IFE, there are several physical observables in this model. Among them, we focus on the following spin nematic operator 
\begin{align}
    \label{eq:nematic}
    \hat{\cal O}_{xy}=\hat S^x\hat S^y+\hat S^y\hat S^x. 
\end{align}
To discuss the symmetry related to this nematic operator, we introduce the anti-unitary time-reversal operator $\hat V$ that satisfies $\hat V\hat S^y\hat V^\dagger = -\hat S^y$ and $\hat V\hat S^{x,z}\hat V^\dagger = \hat S^{x,z}$. The operator $\hat{\cal O}_{xy}$ obeys 
\begin{align}
    \label{eq:nematic2}
    \hat V\hat{\cal O}_{xy}V^\dagger 
    = -\hat{\cal O}_{xy}
\end{align}
In addition, using $\hat V$, one can find the following dynamical symmetry relation, 
\begin{align}
    \label{eq:dynamicalsym}
    \hat V\hat H_{\rm nv}(T-t)\hat V^\dagger = \hat H_{\rm nv}(t).
\end{align}
One says that the time-dependent system possesses a dynamical symmetry when the time-periodic Hamiltonian is invariant under combination of a symmetry operation and a time shift. 
If a system has a dynamical symmetry and a related symmetry equation for an observable such as Eq.~\eqref{eq:nematic2}, one can often obtain a certain constraint for the observable. 
We can find such a constraint in the present case. Equation~\eqref{eq:dynamicalsym} indicates that $|\tilde\phi_j(t)\rangle=\hat V|\phi_j(T-t)\rangle$ is also a Floquet state with quasi energy $\epsilon_j$. If there is no degenerate quasi energy in the spin-$1$ model,  $|\tilde\phi_j(t)\rangle$ and $|\phi_j(T-t)\rangle$ are shown to be equivalent up to an overall phase shift. Then, using Eq.~\eqref{eq:nematic2}, we show that one-cycle averages of $\hat{\cal O}_{xy}$ in Floquet states $|\tilde\phi_j(t)\rangle$ and $|\phi_j(T-t)\rangle$ differ only by their signs. As a result, the expectation value of $\hat{\cal O}_{xy}$ in FG states vanishes. 
Similar arguments can be applied to other observables $\hat A$ satisfying $\hat V \hat A \hat V^\dagger=-\hat A$. 

\begin{figure}[t]
\begin{center}
\includegraphics[width=8cm]{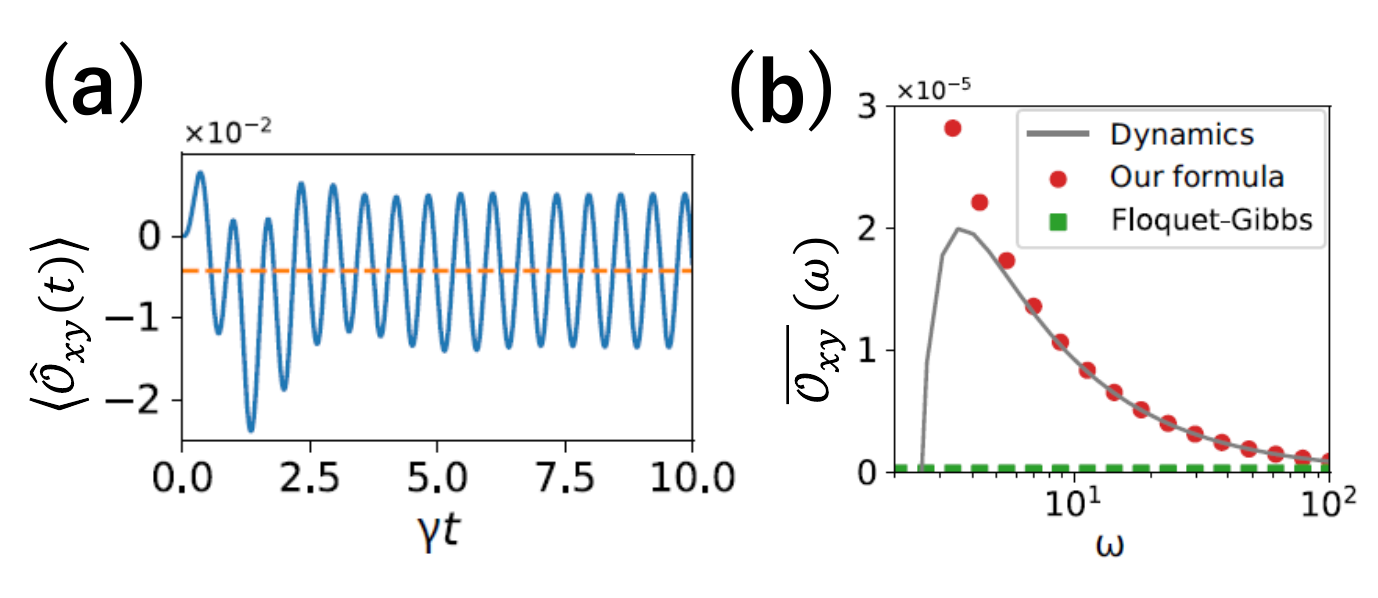}
\caption{(a) Time evolution of the expectation value of the nematic operator $\hat{\cal O}_{xy}$. The blie curve shows the numerically exact result and the dotted red line corresponds to the value of one-cycle average in the NESS after long driving. (b) Laser frequency $\omega$ dependence of one-cycle time average $\bar{\cal O}_{xy}(\omega)=\frac{1}{T}\int_{t_0}^{t_0+T}d\tau \langle\hat{\cal O}_{xy}(\tau)\rangle$. The black line is the numerically exact result, while red and green points respectively correspond to the values computed from the FE formula of Eq.~\eqref{eq:NESS_1/omega} and the FG state. The parameters in the spin-1 model are the same as those of Sec.~\ref{sec:example_NVcenter}. Reprinted from Ref.~\cite{Ikeda2020}.}
\label{fig:Nematic_NVCenter}
\end{center}
\end{figure}

On the other hand, it is worth noticing that dissipation can break the above anti-unitary dynamical symmetry, and this generally leads to a nonzero one-cycle average in the NESS of Eq.~\eqref{eq:NESS_1/omega}. The factor $-i\gamma_{k\ell}$ in the denominator of Eq.~\eqref{eq:NESS_FE} symbolically shows the breakdown of the anti-unitary dynamical symmetry. 
Figure~\ref{fig:Nematic_NVCenter} shows (a) the time evolution of $\langle \hat{\cal O}_{xy}(t)\rangle$ and (b) the $\omega$ dependence of the one-cycle average $\bar{\cal O}_{xy}(\omega)=\frac{1}{T}\int_{t_0}^{t_0+T}d\tau \langle\hat{\cal O}_{xy}(\tau)\rangle$. 
Panel (b) clearly demonstrates that the average $\bar{\cal O}_{xy}(\omega)$ in the FG state is always zero, whereas that computed by the FE formula takes a finite value and coincides well with the numerical result. 
Since $\bar{\cal O}_{xy}(\omega)$ appears only when the dissipation effect is introduced, one may say that the emergence of $\bar{\cal O}_{xy}(\omega)$ is a dissipation-assisted FE.

\subsubsection{Dissipation assisted IFE in Heisenberg antiferromagnet}
In this subsection, we consider another dissipation-assisted FE in a nonequilibrium system following the microscopically derived GKSL equation. The target is the IFE in a Heisenberg antiferromagnetic spin-$\frac{1}2$ chain irradiated by a circularly polarized THz laser. 
The Hamiltonian of the driven spin chain is given by 
\begin{align}
    \label{eq:SpinChain}
    \hat H_{\rm Hei}(t)=&\sum_{j=1}^{L}J \hat{\bm S}_j\cdot\hat {\bm S}_{j+1}
    -{\bm B}(t)\cdot\hat{\bm S}_{\rm tot},
\end{align}
where $\hat{\bm S}_j$ is the spin-$\frac{1}2$ operator residing on the $j$-th site, $L$ is the total number of sites, $\hat {\bm S}_{\rm tot}=\sum_{j=1}^L \hat {\bm S}_j$ is the total spin, and we have imposed the periodic boundary condition $\hat{\bm S}_{L+1}=\hat{\bm S}_{1}$. The first term is the antiferromagnetic exchange interaction between neighboring spins with strength $J>0$, and the second is the THz-laser driven ac Zeeman interaction with frequency $\omega$. We consider the setup that the ac magnetic field is in the $x$-$y$ plane: $\bm B(t)=B_d(\cos(\omega t),\sin(\omega t),0)$. Before laser application, the ground (or equilibrium) state of the Heisenberg chain has no magnetization $\langle\hat {\bm S}_{\rm tot}\rangle=0$.  

\begin{figure}[t]
\begin{center}
\includegraphics[width=8cm]{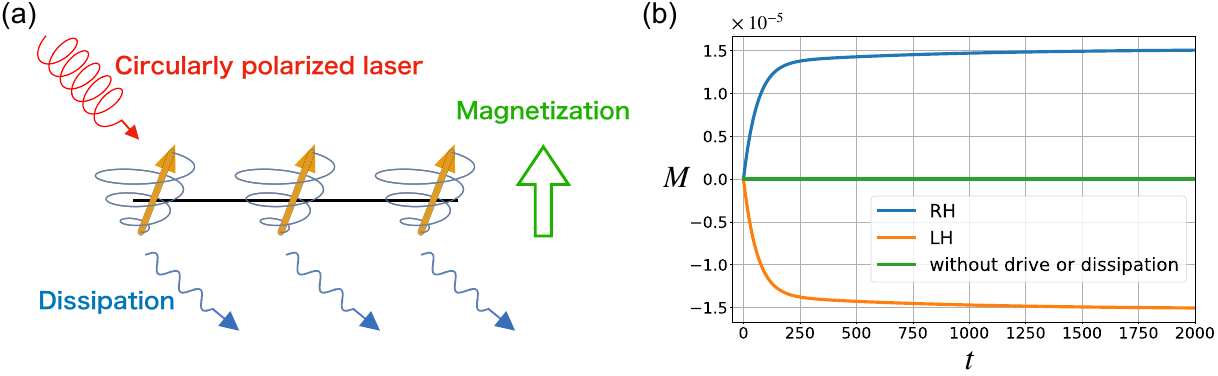}
\caption{(a) Image of circularly polarized THz-laser driven IFE in a Heisenberg spin-$\frac{1}{2}$ chain with dissipation effect. (b) Photo-induced magnetization of the Heisenberg spin-${1}{2}$ antiferromagnetic chain as a function of the laser application time $t$. Blue, orange and green lines are respectively the numerical results of the setups with right-handed polarized ($\omega=5$), left-handed polarized ($\omega=-5$) and linearly polarized lasers. 
Parameters are set to $J=1$, $B_d=0.1$, $\beta=10$, $\lambda=5$, and $L=8$. 
Reprinted from Ref.~\cite{Ikeda2021}.}
\label{fig:IFE_Heisenberg}
\end{center}
\end{figure}

From the discussion in Secs.~\ref{sec:spin_multiferro} and \ref{sec:example_NVcenter}, one expects that an IFE occurs and the laser generates a finite magnetization along the $S^z$ axis after its long application. 
However, if we consider an isolated spin chain without dissipation, 
the photo-induced magnetization does not grow up. 
This is because even under the laser application, the total spin satisfies 
\begin{align}
    \label{eq:spin_conservation}
    [(\hat {\bm S}_{\rm tot})^2,\hat H_{\rm Hei}(t)]=0. 
\end{align}
Thus, the magnitude of the total spin does not change through the time evolution. 
This symmetry argument also implies that if we introduce a dissipation dynamics breaking the above conservation law, the IFE, i.e., a THz-laser induced magnetization, is expected to occur like Fig.~\ref{fig:IFE_Heisenberg} (a). 
Therefore, we numerically solve the GKSL equation for the periodically driven spin chain of Eq.~\eqref{eq:SpinChain}. 
We start from the total system consisting of the spin chain and an environment of free boson models $\hat H_{\rm b}$. 
We assume that each spin $\hat{\bm S}_j$ is coupled to an independent boson model, that is, we prepare $L$ independent free-boson models for the spin chain with length $L$. 
The weak spin-boson interaction is set to  
\begin{align}
    \label{eq:spin_boson}
    \hat H_{\rm sb}=\sum_j g_{\rm sb} 
    \hat S_j^x \otimes (\hat a_j+\hat a_j^\dagger),
\end{align}
where $\hat a_j$ is a boson annihilation operator in the $j$-th bath and $g_{\rm sb}$ is the coupling constant. 
The Hamiltonian of the whole system is given by $\hat H_{\rm tot}=\hat H_{\rm Hei}(t)+\hat H_{\rm b}+\hat H_{\rm sb}$. 
We note that from the relation $S_j^x=S_j^++S_j^-$, the system-bath interaction $\hat H_{\rm sb}$ never tends to cause the magnetization to grow in a positive or negative direction. However, this interaction 
$\hat H_{\rm sb}$ breaks the conservation law of Eq.~\eqref{eq:spin_conservation}, i.e., $[(\hat {\bm S}_{\rm tot})^2,\hat H_{\rm tot}]\neq 0$, we can expect that the IFE occurs. 
Using $\hat H_{\rm sb}$, we derive the GKSL equation from the whole system of $\hat H_{\rm tot}$. 
The resultant GKSL equation is of the same form as Eq.~\eqref{eq:MicroLindbladian}. 
For simplicity, we ignore the Lamb shift $\lamb(t)$ because it essentially does not change the IFE. In the dissipation term, the spectral function $\gamma_\alpha(\omega)$ is estimated as  
\begin{align}
    \label{eq:gamma_Hei}
    \gamma_\alpha(\omega)=\gamma_0 
    \frac{\omega e^{-\omega^2/(2\Lambda^2)}}{1-e^{-\beta\omega}},
\end{align}
where $\gamma_0$ is a constant, $\beta$ is the temperature of the bath, and we have introduced 
the high-energy cutoff $\Lambda$ of the bath. 
The numerically computed photo-induced magnetization is given in Fig.~\ref{fig:IFE_Heisenberg} (b). As expected, one finds that a positive (negative) magnetization is generated by right-handed (left-handed) polarized laser. Since this phenomenon occurs through the coupling between the spin chain and the bath (free bosons), it can be called a typical dissipation-assisted FE.

Note that (as discussed in Ref.~\cite{Takayoshi2014b}) 
a magnetic anisotropy breaking the above conservation law of Eq.~\eqref{eq:spin_conservation} is necessary
to realize an IFE in isolated quantum spin systems.

\section{Floquet theory for open classical systems}
\label{sec:Open_classical}
In the last section, we have reviewed open (dissipative) quantum systems subject to a time-periodic driving, based on the EOM for the density matrix. 
In this section, we explain the Floquet theory for open (dissipative) classical systems subject to a time-periodic drive. Such driven classical systems have a long research history in which several intriguing phenomena have been explored~\cite{Chirikov1981,Gammaitoni1998,Jung1993,Kapitza1951,Paul1990,Courant1952,Feynman1964,Saito2003,Abdullaev2003}. 
As we show below, another type of the high-frequency expansion method~\cite{Higashikawa2018}, which differs from that in Secs.~\ref{sec:Closed} and \ref{sec:Open}, is developed through the distribution function of classical variables. 

\subsubsection{Langevin equation}
\label{sec:Langevin}
First, we define the EOM for periodically driven classical systems. We consider the setup that $N$ classical variables $\bm{\phi}(t) = \left(\phi_1(t), \phi_2(t),...,\phi_N(t)\right)$ obey the EOM
\begin{align}
  &\dot{\bm{\phi}} = \bm{f}(\bm{\phi},t) + G(\bm{\phi},t)\bm{h}(t),
  \label{eq:ClassicalEoM}
\end{align}
where the vector $\bm{f}=(f_1,\cdots,f_N)$ is a generalized force, $\bm{h}=(h_1,\cdots,h_N)$ is the white-Gaussian noise satisfying 
\begin{align}
  \ev{h_{i}(t)} &= 0,
  \nonumber\\
  \ev{h_{i}(t)h_{j}(t')} &= 2D\delta_{ij}\delta(t-t'),
\label{eq:noise}
\end{align}
with $D$ being a diffusion constant, and $G=(g_{ij})$ is an $N\times N$ matrix. 
The dot symbol $(\,\,\dot{}\,\,)$ stands for time derivative $\frac{d}{dt}$. 
We may call Eq.~\eqref{eq:ClassicalEoM} a generalized Langevin equation~\cite{Kubo1991}. For instance, let us briefly consider a standard Langevin equation in one dimension, $m\frac{d^2 x}{dt^2}=F(x)-\gamma\frac{dx}{dt}+{\xi}$, where $m$, $x$, $F(x)$, $-\gamma\frac{dx}{dt}$ and $\xi$ are respectively the mass of the particle, the coordinate, the external force, the friction, and the random force. 
If the Langevin equation is re-expressed as $\dot{x}(t)=v(t)$ and $\dot{v}(t)=(F(x)-\gamma v(t)-{\xi})/m$, one finds that the equation indeed follows the form of Eq.~\eqref{eq:ClassicalEoM}, by defining $\bm{\phi}(t)=(x(t),v(t))$. 

Since we focus on periodically driven systems, 
we suppose that the vector $\bm f$ and the matrix $G$ satisfy 
\begin{align}
  f_i(\bm{\phi},t+T) &= f_i(\bm{\phi},t),
  \nonumber\\
  g_{ij}(\bm{\phi},t+T) &= g_{ij}(\bm{\phi},t).
  \label{eq:period_cl}
\end{align}
However, we should note that Eq.~\eqref{eq:ClassicalEoM} is generally nonlinear with respect to the variable $\bm \phi$, and it is not temporally periodic due to the random force $\bm h$. Therefore, a simple application of the Floquet theorem is forbidden. 
In the next section, we will explain how to avoid this difficulty. Note that the deterministic EOM is included in Eq.~\eqref{eq:ClassicalEoM} by taking the limit of $g_{ij}\to 0$. 

\subsubsection{Fokker-Planck (master) equation}
\label{sec:Fokker-Planck}
As we mentioned above, the EOM of Eq.~\eqref{eq:ClassicalEoM} is outside the scope of Floquet theorem. However, it is well known that the Langevin equation is generally equivalent to the EOM for the distribution function of the classical variable. The latter EOM is often referred to as the Fokker-Planck (FP) or master equation. When the random force $\bm h$ is of a white-Gaussian noise, the distribution function $P({\bm \phi}, t)$ is known to obey the FP equation, 
\begin{align}
  \frac{\partial P(\bm{\phi},t)}{\partial t} =& \frac{\partial}{\partial\phi_i}\left[\mathcal{F}_i(\bm{\phi},t)P(\bm{\phi},t)\right]\nonumber\\
  &+
  \frac{\partial^2}{\partial\phi_i \partial\phi_j}\left[\mathcal{D}_{ij}(\bm{\phi},t)P(\bm{\phi},t)\right],
  \label{eq:FP-eq}
\end{align}
where $\mathcal{F}_i$ and $\mathcal{D}_{ij}$ are given by 
\begin{align}
  \mathcal{F}_i (\bm{\phi},t) &= -f_i (\bm{\phi},t) - Dg_{kl}(\bm{\phi},t)\frac{\partial g_{il}(\bm{\phi},t)}{\partial\phi_k},
  \nonumber\\
  \mathcal{D}_{ij}(\bm{\phi},t) &=  Dg_{ik}(\bm{\phi},t)g_{jk}(\bm{\phi},t).
  \label{eq:FandD}
\end{align}
In Eq.~\eqref{eq:FP-eq} we have used Einstein summation convention, 
and we will often use it hereafter. 
By definition, $\mathcal{F}_i$ and $\mathcal{D}_{ij}$ are time periodic: $\mathcal{F}_i({\bm \phi},t)=\mathcal{F}_i({\bm\phi},t+T)$ and $\mathcal{D}_{ij}({\bm \phi},t)=\mathcal{D}_{ij}({\bm\phi},t+T)$. Equation~\eqref{eq:FP-eq} can be re-written as 
\begin{align}
\label{eq:FP-eq_2}
   \partial_t P(\bm{\phi},t)=& 
  \hat{\mathcal{L}}_{\rm fp}(\bm{\phi},t)P(\bm{\phi},t)
  \\
  =&
  \mathrm{div}\left[\bm{\mathcal{F}}(\bm{\phi},t)P(\bm{\phi},t)\right] + \mathrm{div_2}\left[\mathcal{D}(\bm{\phi},t)P(\bm{\phi},t)\right],\nonumber
\end{align}
where the vector $\bm{\mathcal{F}} = (\mathcal{F}_1,\cdots,\mathcal{F}_N)$, the matrix $\mathcal{D}=(\mathcal{D}_{ij})$, and the symbols $\mathrm{div}$ and $\mathrm{div}_2$ are defined as 
\begin{align}
\label{eq:div}
  \mathrm{div}[\bm{A}] = \frac{\partial A_i}{\partial\phi_i},
  \hspace{1cm}
  \mathrm{div_2}[B] = \frac{\partial^2 B_{ij}}{\partial\phi_{i}\partial\phi_{j}},
\end{align}
for arbitrary vector $\bm A$ and matrix $B$. 
\begin{figure}[t]
\begin{center}
\includegraphics[width=8.5cm]{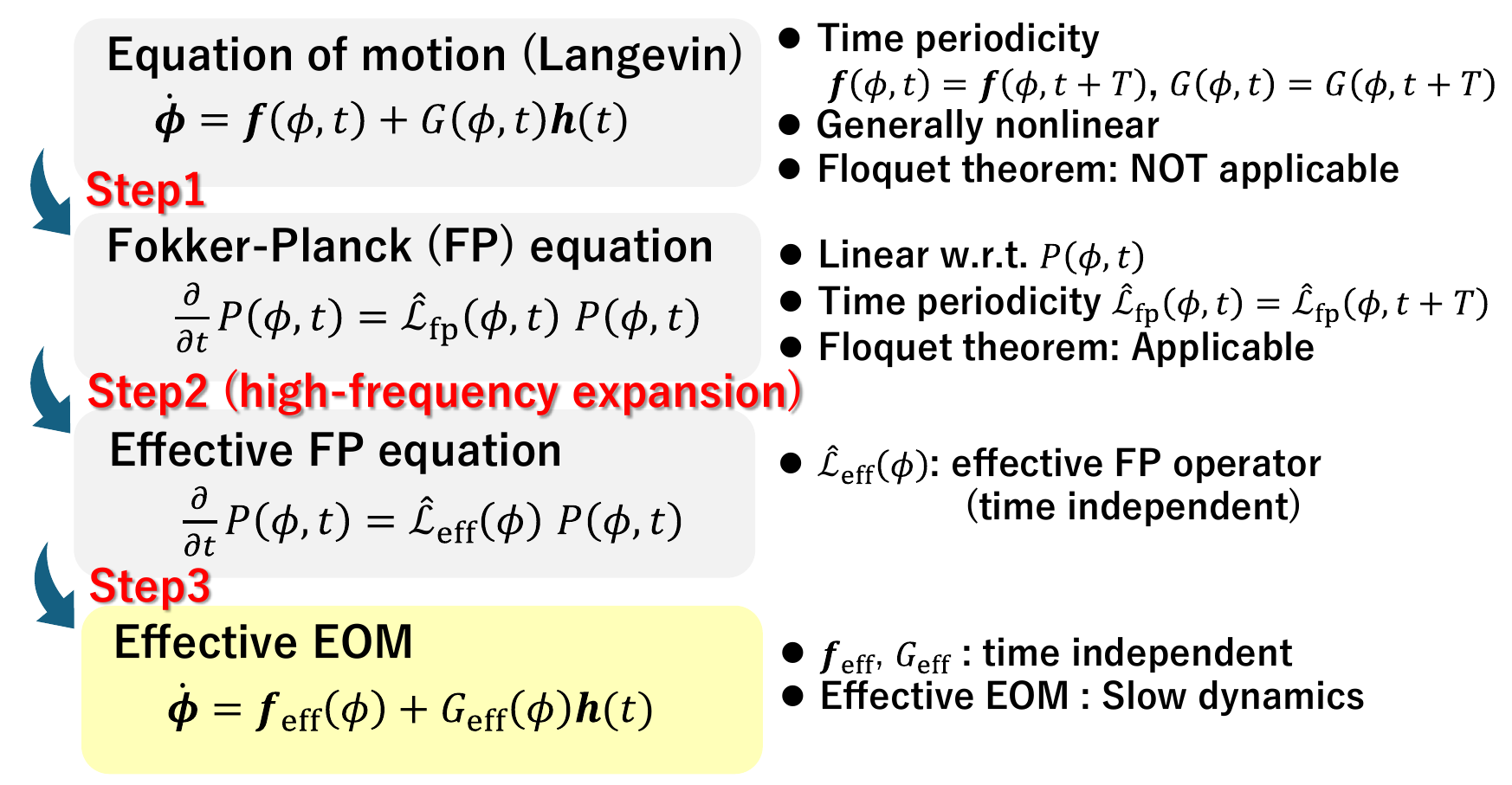}
\caption{Strategy of applying the Floquet theory to classical systems subject to a time periodic drive. Since the EOM is often nonlinear with respect to variables, the Floquet theorem cannot be applied to it. Instead, one can overcome this difficulty to consider the FP equation for the distribution function.}
\label{fig:Floquet_Classical}
\end{center}
\end{figure}
The FP operator (Liouvillian) $\hat{\mathcal{L}}_{\rm fp}$ is time periodic, $\hat{\mathcal{L}}_{\rm fp}({\bm\phi},t)=\hat{\mathcal{L}}_{\rm fp}({\bm\phi},t+T)$, and the FP equation~\eqref{eq:FP-eq_2} is linear with respect to the distribution $P({\bm \phi},t)$. Therefore, the Floquet theorem is applicable to the FP equation like the Schr\"odinger or GKSL equations. 
In fact, $P({\bm \phi},t)$ and $i\hat{\mathcal{L}}_{\rm fp}({\bm\phi},t)$ respectively correspond to the wave function $|\psi(t)\rangle$ and the Hamiltonian $\hat H(t)$ of the Schr\"odinger equation. Note that the "Hamiltonian" $i\hat{\mathcal{L}}_{\rm fp}({\bm\phi},t)$ is generally non-hermitian. Figure~\ref{fig:Floquet_Classical} shows the basic strategy of application of the Floquet theory to classical systems. 
Like the quantum cases, if the non-unitary time evolution operator is decopmosed into the product of three terms, $\hat U_{\rm fp}(t_2-t_1)=e^{\hat {\cal G}_{\rm fp}(t_2)}e^{\hat {\cal L}_{\rm eff}(t_2-t_1)}e^{-\hat {\cal G}_{\rm fp}(t_1)}$, 
one can obtain the high-frequency expansion formulas for the effective FP operator $\hat {\cal L}_{\rm eff}$ and kick operator $\hat {\cal G}_{\rm fp}(t)=\hat {\cal G}_{\rm fp}(t+T)$ as functions of Fourier components of $\hat{\mathcal{L}}_{\rm fp}({\bm\phi},t)$, which are defined as 
\begin{align}
  \hat{\mathcal{L}}_{\rm fp}(\bm{\phi},t) &= \sum_{m\in\mathbb{Z}}\hat{\mathcal{L}}_{m}(\bm{\phi})e^{-im\omega t}.
  \label{eq:Fourier_cl}
\end{align}
Then, we can obtain the effective FP equation describing the slow dynamics, 
\begin{align}
  \frac{\partial}{\partial t}P(\bm{\phi},t) 
  \simeq  \hat{\mathcal{L}}_{\mathrm{eff}}P(\bm{\phi},t)
  =\sum_{m=0}^{m_{0}}\hat{\mathcal{L}}_{\mathrm{eff}}^{(m)}P(\bm{\phi},t),
  \label{eq:eff-FP} 
\end{align} 
where $\hat{\mathcal{L}}_{\mathrm{eff}}(\bm{\phi})$ has no explicit time dependence, $\hat{\mathcal{L}}_{\mathrm{eff}}^{(m)}(\bm{\phi})$ is the operator of ${\cal O}(\omega^{-m})$, composed of $\{\hat{\mathcal{L}}_{n}\}$, and $m_0$ is the truncated order of the high-frequency expansion. 
From Eq.~\eqref{eq:eff-FP}, we can also lead to the effective EOM for the slow dynamics. 
In the following two subsections, we explain how to compute the effective Liouvillian $\hat{\mathcal{L}}_{\mathrm{eff}}$ and the effective EOM for the classical variables $\bm\phi(t)$.

\subsubsection{High-frequency expansion for $g_{ij}\to 0$}
\label{sec:Effective1}
In this subsection, we focus on the deterministic limit of zero random force $g_{ij}\to 0$ ($D\to 0$ and $\bm h\to \bm 0$), in which the EOM is given by 
\begin{align}
  \dot{\bm{\phi}} = \bm{f}(\bm{\phi},t). 
  \label{eq:EOM_ZeroRandom}
\end{align}
The Fourier transform of $\bm{f}(\bm{\phi},t)$ is defined as 
\begin{align}
  \bm{f}(\bm{\phi},t) = \sum_{m\in\mathbb{Z}}
  \bm{f}_{m}(\phi)e^{-im\omega t}.
  \label{eq:f-Fourier}
\end{align}
Then, we have the following relation between $\hat {\cal L}_{m}$ and $\bm{f}_{m}$:
\begin{align}
  \hat{\mathcal{L}}_{m}(\bm{\phi})P(\bm{\phi},t) 
  = -\mathrm{div}\left[\bm{f}_{m}(\bm{\phi})P(\bm{\phi},t)\right].
  \label{eq:Lm_fm}
\end{align}
Moreover, the commutator $[\hat{\mathcal{L}}_{m},\hat{\mathcal{L}}_{n}]$ is computed as 
\begin{align}
  [\hat{\mathcal{L}}_{m}(\bm{\phi}),\hat{\mathcal{L}}_{n}(\bm{\phi})]P 
  = -\mathrm{div}\left[-\left[\bm{f}_{m}(\bm{\phi}),\bm{f}_{n}(\bm{\phi})\right]_{\mathrm{cl}}P\right],
  \label{eq:commutator_cl}
\end{align}
where the "commutator" $\left[\bm{A},\bm{B}\right]_{\mathrm{cl}}$ is 
the Lie bracket defined as 
\begin{align}
  \left[\bm{A},\bm{B}\right]_{\mathrm{cl}} &= (\bm{A}\cdot\nabla_{\bm{\phi}})B_{j}\bm{e}_{j} - (\bm{B}\cdot\nabla_{\phi})A_{j}\bm{e}_{j}\notag\\
  &= A_{i}\frac{\partial B_{j}}{\partial\phi_{i}}\bm{e}_{j} - B_{i}\frac{\partial A_{j}}{\partial\phi_{i}}\bm{e}_{j},
  \label{eq:Lie-bracket}
\end{align}
where 
$\bm{e}_j$ is the orthonormal basis for vectors $\bm A$ and $\bm B$, and $\nabla_{\bm{\phi}} = \left(\frac{\partial}{\partial\phi_{1}},\frac{\partial}{\partial\phi_{2}},...,\frac{\partial}{\partial\phi_{N}}\right)$. 
Using these instruments and the high-frequency expansion method discussed in Sec.~\ref{sec:High-frequency}, one can estimate the lower-order terms $\hat{\mathcal{L}}_{\rm eff}^{(n)}$, which can be expressed in a compact fashion of $\hat{\mathcal{L}}_{\rm eff}^{(n)}(\bm{\phi})P(\bm{\phi},t)=
-\mathrm{div}[\bm{f}_{\rm eff}^{(n)}(\bm{\phi})P(\bm{\phi},t)]$. 
The effective forces $\bm{f}_{\rm eff}^{(n)}$ are given by
\begin{align}
  \bm{f}_{\mathrm{eff}}^{(0)}&= \bm{f}_{0},
  \label{eq:f(0)}\\
  \bm{f}_{\mathrm{eff}}^{(1)} &= -i\sum_{m\neq0}\frac{\left[\bm{f}_{-m},\bm{f}_{m}\right]_{\mathrm{cl}}}{2m\omega},
  \label{eq:f(1)}\\
  \bm{f}_{\mathrm{eff}}^{(2)} &= -\sum_{m\neq0}\Big\{\frac{\left[\bm{f}_{-m},\left[\bm{f}_{0},\bm{f}_{m}\right]_{\mathrm{cl}}\right]_{\mathrm{cl}}}{2m^2\omega^2} \nonumber\\
  &+ \sum_{n\neq0,m}\frac{\left[\bm{f}_{-n}\left[\bm{f}_{n-m},\bm{f}_{m}\right]_{\mathrm{cl}}\right]_{\mathrm{cl}}}{3mn\omega^2}\Big\},
  \label{eq:f(2)}
\end{align}
Therefore, the effective FP operator and the effective EOM are written as 
\begin{align}
\label{eq:effectiveFP}
      \hat{\mathcal{L}}_{\mathrm{eff}}(\bm{\phi})P &= -\mathrm{div}\Big[\Big(\sum_{n=0}^{m_0}\bm{f}_{\mathrm{eff}}^{(n)}(\bm{\phi}) \Big)P\Big],\\
      \dot{\bm{\phi}} &= \bm{f}_{\mathrm{eff}}(\bm{\phi})=\sum_{n=0}^{m_0}\bm{f}_{\mathrm{eff}}^{(n)}(\bm{\phi}).
      \label{eq:effectiveEOM}
\end{align}
By solving these effective equations, one can describe the slow dynamics
of the periodically driven classical systems in the deterministic limit. 
Similarly, one can estimate the lower-order terms of $\hat {\cal G}_{\rm fp}(t)$: For example, the lowest-order term is 
$\hat {\cal G}_{\rm fp}^{(1)}(t)=-i\sum_{m\neq 0} \frac{\hat {\cal L}_{-m}}{m\hbar\omega}e^{im\omega t}$ [see Eq.~\eqref{eq:1nd2rdkicked}].
These results of the high-frequency expansion are consistent with those of non-autonomous differential equations $\dot{\bm{\phi}} = \bm{f}(\bm{\phi},t)$~\cite{Blanes2009,Agrachev1979,Spirig1979,Agrachev1981}.

\subsubsection{High-frequency expansion for time-independent $g_{ij}$}
\label{sec:Effective2}
In this subsection, we consider classical systems subject to a random force $\bm h(t)$ with a time-independent coefficient matrix $G(\bm\phi)=(g_{ij}(\bm\phi))$. The Langevin equations within this setup have been often studied in statistical physics. 
Here, let us define the Fourier transform of the vector $\bm{\mathcal{F}}$ as 
\begin{align}
  \bm{\mathcal{F}}(\bm{\phi},t) 
  = \sum_{m}\bm{\mathcal{F}}_{m}(\phi)e^{-im\omega t}.
  \label{eq:F-Fourier}
\end{align}
In addition, we introduce two vectors $\bm{\mathcal{F}}_{m}=\left(\mathcal{F}_{m,1},\mathcal{F}_{m,2},
\cdots,\mathcal{F}_{m,N}\right)$ and 
$\bm{f}_{m}=\left(f_{m,1},f_{m,2},\cdots,f_{m,N}\right)$. 
From the time-independent nature of $G(\bm\phi)$, the components of $\bm{\mathcal{F}}_{m}$ are given by 
\begin{align}
  \mathcal{F}_{0,i}(\bm{\phi}) &= -f_{0,i}(\bm{\phi})-Dg_{kl}(\bm{\phi})\frac{\partial g_{il}(\bm{\phi})}{\partial\phi_{k}},
  \label{eq:FourierF_0}\\
  \mathcal{F}_{m,i}(\bm{\phi}) &= -f_{m,i}(\bm{\phi}).
  \qquad(\mathrm{for}\quad m\neq0)
  \label{eq:FourierF_m}
\end{align}
From these equations and the fact that the matrix ${\cal D}_{ij}$ is independent of time, 
the Fourier components of the FP operator are expressed as  
\begin{align}
    \hat{\mathcal{L}}_{0}(\bm{\phi})P(\bm{\phi},t) &= \mathrm{div}\left[\bm{\mathcal{F}}_{0}(\bm{\phi})P(\bm{\phi},t)\right] + \mathrm{div}_{2}\left[\mathcal{D}(\bm{\phi})P(\bm{\phi},t)\right],
    \nonumber\\
    \hat{\mathcal{L}}_{m}(\bm{\phi})P(\bm{\phi},t) &= 
    -\mathrm{div}\left[\bm{f}_{m}(\bm{\phi})P(\bm{\phi},t)\right]
    \qquad(\mathrm{for}\,\,\, m\neq0),
    \label{eq:Fourier-L_m}
\end{align}
Using these tools, we compute the high-frequency expansion form of the effective FP operator $\hat{\mathcal{L}}_{\rm eff}=\hat{\mathcal{L}}_{\rm eff}^{(0)}+\hat{\mathcal{L}}_{\rm eff}^{(1)}+\hat{\mathcal{L}}_{\rm eff}^{(2)}+\cdots$. We attempt to pack the lower-order terms $\hat{\mathcal{L}}_{\rm eff}^{(n)}$ into the following compact form:
\begin{align}
  \hat{\mathcal{L}}_{\mathrm{eff}}^{(n)}(\bm{\phi})P(\bm{\phi},t) =& \mathrm{div}\left[\bm{\mathcal{F}}_{\mathrm{eff}}^{(n)}(\bm{\phi})P(\bm{\phi},t)\right]
  \nonumber\\
  &+ \mathrm{div}_{2}\left[\mathcal{D}_{\mathrm{eff}}^{(n)}(\bm{\phi})P(\bm{\phi},t)\right].
  \label{eq:random-L}
\end{align}
The zero-th order term is easily computed due to the relation $\hat{\mathcal{L}}_{\mathrm{eff}}^{(0)}=\hat{\mathcal{L}}_{0}$ [see Eq.~\eqref{eq:0th_1/omega}]. The result is 
\begin{align}
  \bm{\mathcal{F}}_{\mathrm{eff}}^{(0)}(\bm{\phi}) = \bm{\mathcal{F}}_{0}(\bm{\phi}), \,\,\,\,\,
  \mathcal{D}_{\mathrm{eff}}^{(0)} (\bm{\phi}) = \mathcal{D}(\bm{\phi}).
  \label{eq:0th}
\end{align}
In order to estimate the first-order term [see Eq.~\eqref{eq:1st_1/omega}],  
we need the commutator,  
\begin{align}
\left[\hat{\mathcal{L}}_{m},\hat{\mathcal{L}}_{n}\right]P
  &= \mathrm{div}\left[\left[\bm{\mathcal{F}}_{m},\bm{\mathcal{F}}_{n}\right]_{\mathrm{cl}}P\right].
  \label{eq:L-commutator}
\end{align}
Using this result, we obtain 
\begin{align}
  \bm{\mathcal{F}}_{\mathrm{eff}}^{(1)} 
  = i\sum_{m\neq0}\frac{\left[\bm{f}_{-m},\bm{f}_{m}\right]_{\mathrm{cl}}}{2m\omega},\,\,\,\,\,\,
  \mathcal{D}_{\mathrm{eff}}^{(1)} = 0.
  \label{eq:1st}
\end{align}
More complicated commutators are needed for the second-order terms.
They are computed as 
\begin{widetext}
\begin{align}
\left[\hat{\mathcal{L}}_{0},\hat{\mathcal{L}}_{m}\right]P &= \mathrm{div}\left[\mathrm{drf}\left[\bm{\mathcal{F}}_{0},\bm{\mathcal{F}}_{m},\mathcal{D}\right]P\right] + \mathrm{div}_{2}\left[\mathrm{diff}\left[\bm{\mathcal{F}}_{m},\mathcal{D}\right]P\right],
    \nonumber\\
\left[\hat{\mathcal{L}}_{n},\left[\hat{\mathcal{L}}_{0},\hat{\mathcal{L}}_{m}\right]\right]P&= -\mathrm{div}\left[\mathrm{drf}\left[\mathrm{drf}\left[\bm{\mathcal{F}}_{0},\bm{\mathcal{F}}_{m},\mathcal{D}\right],\bm{\mathcal{F}}_{n},\mathrm{diff}\left[\bm{\mathcal{F}}_{m},\mathcal{D}\right]\right]P\right]
  -\big.\mathrm{div}_{2}\left[\mathrm{diff}\left[\bm{\mathcal{F}}_{n},\mathrm{diff}\left[\bm{\mathcal{F}}_{m},\mathcal{D}\right]\right]P\right].
  \label{eq:L-commutator3}
\end{align}
Here we have introduced new symbols of the vector "drf" and the matrix "diff" as follows:
\begin{align}
\mathrm{drf}\left[\bm{\mathcal{F}}_{0},\bm{\mathcal{F}}_{m},\mathcal{D}\right]
=&\left\{\mathrm{drf}_{i}\left[\bm{\mathcal{F}}_{0},\bm{\mathcal{F}}_{m},\mathcal{D}\right]\right\}_{i=1}^{N},\\
\mathrm{drf}_{i}\left[\bm{\mathcal{F}}_{0},\bm{\mathcal{F}}_{m},\mathcal{D}\right] 
=& (\bm{\mathcal{F}}_{0}\cdot\nabla_{\bm{\phi}})\mathcal{F}_{m,i} - (\bm{\mathcal{F}}_{m}\cdot\nabla_{\bm{\phi}})\mathcal{F}_{0,i} 
\nonumber\\
&- \frac{\partial^{2}\mathcal{F}_{m,i}}{\partial\phi_{j}\partial\phi_{k}}\mathcal{D}_{jk},
    \nonumber\\
\mathrm{diff}\left[\bm{\mathcal{F}}_{m},\mathcal{D}\right]=&
    \left\{\mathrm{diff}_{ij}\left[\bm{\mathcal{F}}_{m},\mathcal{D}\right]\right\}_{ij}^{N},
    \nonumber\\
\mathrm{diff}_{ij}\left[\bm{\mathcal{F}}_{m},\mathcal{D}\right] =& \frac{\partial\mathcal{F}_{m,i}}{\partial\phi_{k}}\mathcal{D}_{kj} + \frac{\partial\mathcal{F}_{m,j}}{\partial\phi_{k}}\mathcal{D}_{ki} - \mathcal{F}_{m,k}\frac{\partial\mathcal{D}_{ij}}{\partial\phi_{k}}.
    \label{eq:dif-diff}\nonumber
\end{align}
After some algebra with these commutation relations, we arrive at 
\begin{align}
  \bm{\mathcal{F}}_{\mathrm{eff}}^{(2)} &= \sum_{m\neq0}\Bigg\{\frac{\mathrm{drf}\left[\mathrm{drf}\left[\bm{\mathcal{F}}_{0},\bm{\mathcal{F}}_{m},\mathcal{D}\right],\bm{\mathcal{F}}_{-m},\mathrm{diff}\left[\bm{\mathcal{F}}_{m},\mathcal{D}\right]\right]}{2(m\omega)^2}\notag\\
  &\qquad\qquad + \sum_{m'\neq0,m}\frac{\left[\bm{f}_{-m'},\left[\bm{f}_{m'-m},\bm{f}_{m}\right]_{\mathrm{cl}}\right]_{\mathrm{cl}}}{3mm'\omega^2}\Bigg\},
  \\
  \mathcal{D}_{\mathrm{eff}}^{(2)} &= \sum_{m\neq0}\frac{\mathrm{diff}\left[\bm{\mathcal{F}}_{-m},\mathrm{diff}\left[\bm{\mathcal{F}}_{m},\mathcal{D}\right]\right]}{2(m\omega)^2}.
  \label{eq:dif.D(2)}
\end{align}
\end{widetext}
If we try to compute the third-order term $\hat{\mathcal{L}}_{\mathrm{eff}}^{(3)}$, we find that it cannot be expressed as in Eq.~\eqref{eq:random-L}. This implies that one has to consider a non-Gaussian correlated random force or non-Markovian dynamics~\cite{Kanazawa2012,Kanazawa2013,Zwanzig1961,Mori1973,Haake1973}. However, since such a complicated term first appears in ${\cal O}(\omega^{-3})$, we can expect that the effective FP equation and the corresponding EOM are valid in a sufficiently high-frequency regime. For the kick operator $\hat{\cal G}_{\rm fp}(t)$, see Ref.~\cite{Higashikawa2018}. 
In the following two subsections, we will apply the present high-frequency expansion method to two types of classical systems.

\subsubsection{Kapitza pendulum}
\begin{figure}[t]
\begin{center}
\includegraphics[width=8.5cm]{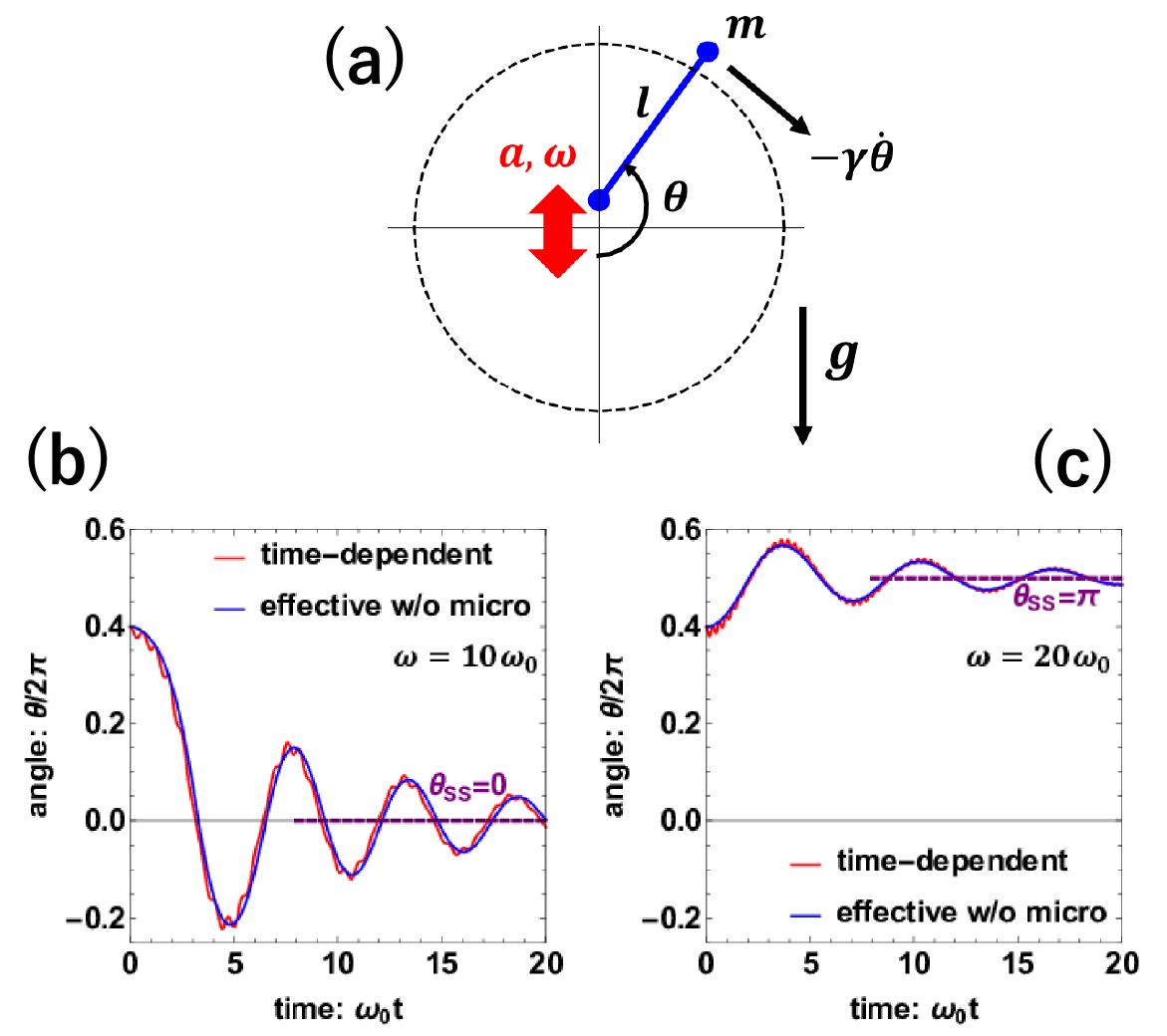}
\caption{(a) Image of the Kapitza pendulum. The angle $\theta(t)$ is the mechanical variable. By vibrating the fulcrum with high frequency $\omega$, we obtain additional stable point $\theta=\pi$. (b,c) Time evolutions of the angle $\theta$ for (b) slow ($\omega=10\omega_0$) and (c) fast ($\omega=20\omega_0$) drives, where the other parameters
are chosen to be $a/\ell=0.1$, $\gamma/\omega_0=0.2$ and $\theta(0)=0.4\pi$. Red curves with fast oscillations are the numerically exact results, while blue curves are computed by the effective EOM. Reprinted from Ref.~\cite{Higashikawa2018}.}
\label{fig:Kapitza}
\end{center}
\end{figure}

In this subsection, we apply the above Floquet theory to so-called Kapitza pendulum~\cite{Kapitza1951}, which is also referred to as the inverted pendulum. This is a classical rigid pendulum with a vertically oscillating point of the suspension as shown in Fig.~\ref{fig:Kapitza} (a), where the mechanical variable $\theta$ is the angle measured from the downward position, $\omega_0 = \sqrt{g/\ell}$ is the eigen frequency of the small oscillation around $\theta = 0$. Here, $g$ and $\ell$ are the gravitational constant and the length of the pendulum, respectively. 
The suspension point oscillates with amplitude $a$ and frequency $\omega$. 
The EOM of this model is given by~\cite{Kapitza1951,DAlessio2013, Bukov2015}
\begin{align}
\ddot{\theta} =-\gamma\dot{\theta}
 - \left[\omega_0^2 + 
 {a \over \ell} \omega^2 \cos(\omega t)\right]
\sin\theta. 
 \label{eq:Kapitza}
\end{align}
On the right-hand side, the first term with $\gamma$ is the friction force, the second corresponds to the gravitational force of the standard harmonic oscillator, and the third steams from the inertial force of the oscillation of the suspension point. 
Note that this model does not possess random force, that is, a deterministic model. 

To apply the formalism in this section, we rewrite Eq.~\eqref{eq:Kapitza} into a first-order ordinary differential equation of $\theta$ and the angular velocity $v$ as follows: 
\begin{align}
\dot{\theta} = v,\,\,\,\,\,\,
\dot{v} = 
- \gamma v
- \left[\omega_0^2 + {a\over\ell} \omega^2 \cos(\omega t)\right] \sin\theta. 
\label{eq:Kaptiza2}
\end{align}
Comparing this with Eq.~\eqref{eq:ClassicalEoM}, 
we find that the vector $\bm\phi$ is given by $\bm\phi=(\theta, v)$ and the matrix $G=0$. 
Namely, the model corresponds to a setup in Sec.~\ref{sec:Effective1}. 
The Fourier components of the drift force $\bm{f}(\bm\phi, t)= \bm{f}_0 + \bm{f}_1 e^{-i\omega t} + \bm{f}_{-1} e^{i\omega t}$ are estimated as
\begin{align}
    \bm{f}_0(v,\theta) & = 
    \begin{pmatrix}
    \bm{f}_{0,\theta}\\
    \bm{f}_{0,v}
    \end{pmatrix}
	 = 
    \begin{pmatrix}
    v  \\
    - \gamma v -\omega_0^2 \sin\theta
    \end{pmatrix}, 
    \nonumber \\
    \bm{f}_{\pm 1}(v,\theta) &= 
     \begin{pmatrix}
     \bm{f}_{\pm 1,\theta} \\
     \bm{f}_{\pm 1,v}
     \end{pmatrix}
	 = \begin{pmatrix}
     0\\
     - {a \omega^2 \over 2\ell} \sin \theta
     \end{pmatrix}.
\end{align}
From the high-frequency expansion in Eqs.\eqref{eq:f(0)}-\eqref{eq:f(2)}, we obtain $\bm{f}_{\rm eff}^{(0)}=\bm{f}_0$, $\bm{f}_{\rm eff}^{(1)}=\bm 0$ and 
\begin{align}
   \bm{f}_{\rm eff}^{(2)}(v,\theta) &= 
   \begin{pmatrix} 
   0\\
   - \left(a \omega \over 2\ell\right)^2 \sin(2\theta)
    \end{pmatrix}.
\end{align}
The resultant effective EOM is thus expressed as 
\begin{align}
 \dot{\theta} = v,\,\,\,\,\,\,\,\,
  \dot{v} = 
	- \gamma v - \omega_0^2 \sin\theta 
	- \left(a \omega \over 2 \ell \right)^2 \sin(2\theta). 
	\label{eq:Kapitza_eff}
\end{align}
We find that in this effective EOM, instead of the original static potential $-\omega_0^2\cos\theta$, the effective potential 
\begin{align}
	V_{\rm eff}(\theta) = 
	- \omega_0^2 \cos\theta
	- \left(a \omega \over 2 \ell\right)^2 \sin^2\theta,
	\label{eq:Kapituza_effpotential}
\end{align}
emerges due to the periodic drive, and the time-dependent force $\propto\cos(\omega t)$ disappears. 
We note that $V_{\rm eff}(\theta)$ is independent of the friction strength $\gamma$ and the same as the one obtained from the analysis without friction~\cite{Kapitza1951,DAlessio2013}. 
Due to the second term, a new local potential minimum at $\theta = \pi$ appears if the external frequency $\omega$ is larger than the critical value $\omega_c = (\sqrt{2}\ell \omega_0) / a$. 
Since $(\theta, v) = (0,0)$ and $(\pi,0)$ are stationary solutions of Eq.~\eqref{eq:Kapitza_eff}, it is expected that the system converges to either of these points after a sufficiently long drive with the help of friction $-\gamma v$. 
The steady-state angle $\theta_{SS}=\theta(t\to \infty)$, in general, depends on $\theta(t=0)$, $\omega$, $v(t=0)$, and $\gamma$.  
Panels (b) and (c) in Fig.~\ref{fig:Kapitza} show typical time evolutions of $\theta$ for (b) $\omega<\omega_c$ and (c) $\omega>\omega_c$, respectively. As expected, a sufficient fast drive leads to $\theta_{SS}=\pi$. 
The emergence of $\theta_{SS}=\pi$ should be referred to as a typical FE and it is also the origin of the name "inverted pendulum". 
Comparing the red and blue curves in the panels (b) and (c), 
we find that the effective EOM well captures the exact solution except for the fast oscillation motion of the red curves, which would correspond to the kick operator. 

Finally, we comment on a recent experiment of a THz-laser driven magnet~\cite{Zhang2025}. The above Kapitza pendulum is a prototypical example of periodically driven one-body classical systems. 
On the other hand, if we focus on a magnetically ordered state in many-spin systems, the effective EOM for the order parameter is sometimes approximated by a nonlinear differential equation like Eq.~\eqref{eq:Kapitza} within the mean-field level. 
One can find such an EOM for the order parameter and its FE by a THz laser pulse in Ref.~\cite{Zhang2025}.

\subsubsection{Landau-Lifshitz-Gilbert equation}
In this subsection, we discuss the FE in classical spin systems irradiated by THz or GHz wave, whose photon energy is comparable to the magnetic excitations. This is a typical application of the Floquet theory to classical many-body systems. It is well known that the 
Landau-Lifshitz-Gilbert (LLG) equation~\cite{Lakshmanan2011,Seki2015,Li2004,Aron2014} well describes time evolution of spins in ordered magnets like ferromagnetic or antiferromagnetic states within the semiclassical level. 

We start from the definition of the Hamiltonian (energy) of a classical spin model ${\cal H}(\{\bm {m}_{\bm r}\},t)$, where $\bm m_{\bm r}=(m_{\bm r}^x,m_{\bm r}^y,m_{\bm r}^z)$ is the local magnetization vector on the site $\bm r$. We here suppose that the time dependence of the Hamiltonian comes from a light-spin coupling such as an ac Zeeman interaction. The dominant interactions of ${\cal H}$ are usually an exchange interaction and/or magnetic anisotropies such as DM interaction, single-ion anisotropy, etc. From ${\cal H}$ and $\bm m_{\bm r}$, the LLG equation reads 
\begin{align}
\dot{\bm{m}}_{\bm r} =
	 - \gamma \bm{m}_{\bm r} \times
	\left(
	\bm{H}_{\bm r}(t) + \bm{h}_{\bm r}(t)\right)
	 + {\alpha \over m_s}
	 \bm{m}_{\bm r} \times \dot{\bm m}_{\bm r}, 
    \label{eq:LLG}
\end{align}
where $\bm{H}_{\bm r}(t) = - (\delta \mathcal{H}(t))/ (\delta \bm{m}_{\bm r})$ is an effective magnetic field (torque) generated by the spins surrounding $\bm m_{\bm r}$ and external fields, and $\bm{h}_{\bm r}(t)$ is the random magnetic field on the site $\bm r$ modeling the thermal fluctuation. The parameters $\alpha$, $\gamma=-g_{\rm e}\mu_{\rm B}/\hbar$, and $m_s=|\bm{m}_{\bm r}|$ are the (dimensionless) Gilbert damping constant, the electron's gyromagnetic ratio ($\mu_{\rm B}$ is the Bohr magneton and $g_{\rm e}\sim 2$ is the g factor), 
and the magnitude of the local magnetization, respectively. It is natural to determine the random field $\bm{h}_{\bm r}(t)$ so that the field relaxes the system to a thermal equilibrium state at a temperature $T$ in the absence of a THz or GHz wave. 
To this end, one usually makes the field satisfy \begin{align}
\langle h_{\bm r}^a(t)\rangle &=0,
    \nonumber\\
\langle h_{\bm r}^a(t)h_{\bm r'}^b(t') \rangle
    &= 2D \delta_{ab}
	 \delta_{\bm r, \bm r'}
	 \delta(t - t'), 
     \label{eq:randomLLG}
\end{align} 
where $D = 2 \hbar k_{\rm B} T \alpha/(\gamma\hbar)^2$ is the diffusion constant and we have assumed $g_{\rm e}=2$. 
For simplicity, we, respectively, normalize time, the local magnetization, the Hamiltonian, and the random force as $tJ/\hbar$, $\bm m_{\bm r}/(\gamma\hbar)$, ${\cal H}/J$, and ${\bm h}_{\bm r}/(J\gamma\hbar)$ in which $J$ is a typical energy scale of the Hamiltonian such as an exchange coupling. 
We again represent these dimensionless terms with the same symbols $t$, $\bm m_{\bm r}$, ${\cal H}$ and $\bm h_{\bm r}$. 
By substituting the left-hand side to the right-hand side in Eq.~\eqref{eq:LLG}, one arrives at 
\begin{align}
	\dot{\bm m}_{\bm r} &= 
	 - { \bm m_{\bm r} \over 1 + \alpha^2}
	 \times 
	\Big\{
	\bm{H}_{\bm r}(t) + \bm{h}_{\bm r}(t)
	\nonumber \\
	& 
	 \quad \quad + {\alpha \over m_s}
	\bm{m}_{\bm r} \times 
	\left(\bm{H}_{\bm r}(t) + \bm{h}_{\bm r}(t)\right)
	\Big\}.
    \label{eq:linearizedLLG}
\end{align}
The procedure from Eq.~\eqref{eq:LLG} to Eq.~\eqref{eq:linearizedLLG} is called "linearization"~\cite{Seki2015}. 
One finds that Eq.~\eqref{eq:linearizedLLG} belongs to the class of the generalized Langevin equation~\eqref{eq:ClassicalEoM} by identifying the set of local moments $\{\bm{m}_{\bm r}\}$ to $\bm\phi(t)$~\cite{Li2004,Aron2014}. In this sense, the LLG equation may be viewed as a spin analog of the Langevin equation. By comparing Eqs.~\eqref{eq:linearizedLLG} and ~\eqref{eq:ClassicalEoM}, we find
\begin{align}
{\bm f}_{\bm r}(t) & = - \frac{{\bm m}_{\bm r}}{1 + \alpha^2} 
	\times 
	\Big[ \bm{H}_{\bm r}(t) + {\alpha \over m_s} \bm m_{\bm r} \times \bm{H}_{\bm r}(t)\Big]
	\nonumber \\
	&\quad + {2D \over 1 + \alpha^2} {\bm m}_{\bm r}, \nonumber \\
	g_{\bm ra, \bm r'b} & = {\delta_{\bm r,\bm r'} \over 1 + \alpha^2} \epsilon_{abc} m_{\bm rc} 
    +{\alpha m_s \delta_{\bm r,\bm r'}\over 1 + \alpha^2} \Big[
	\delta_{a,b} - {m_{\bm ra} m_{\bm rb} \over (m_s)^2}
	\Big]. 
    \label{eq:Langevin_LLG}
\end{align}
Here, the vector $\bm f_{\bm r}$ and the matrix $G=(g_{\bm ra,\bm rb})$ describe the spin precession generated by $\bm{H}_{\bm r}$ and the spin diffusion induced by $\bm h_{\bm r}$ and perpendicular to $\bm m_{\bm r}$, respectively~\cite{Li2004,Aron2014}. 
The last term of $\bm f_{\bm r}$ comes from the term $-D g_{k\ell} \partial g_{i\ell}/\partial\phi_k$ in Eq.~\eqref{eq:FandD}.

The content of Eqs.~\eqref{eq:LLG} to \eqref{eq:Langevin_LLG} is the generic framework describing the LLG equation from the perspective of the Langevin equation~\eqref{eq:ClassicalEoM}. 
Based on it, we apply the Floquet theory (high-frequency expansion) to a classical spin system subject to THz or GHz wave. 
As such a simple model, we here focus on a laser-driven multiferroic spin chain whose Hamiltonian is given by 
\begin{align}
\mathcal{H}_{\rm chain}(t) 
  =& - \sum_{j}^L  J {\bm m}_{j} \cdot {\bm m}_{j+1}
  \nonumber\\
  & -{\bm E}(t) \cdot {\bm P}
    -[{\bm B}_s+{\bm B}(t)] \cdot {\bm M}, 
     \label{eq:classical_chain}
\end{align}
where ${\bm m}_{j}$ is the magnetization on the $j$-th site, $J$ is the exchange coupling constant, $L$ is the system size, $\bm M=\sum_j {\bm m}_j$ is the total magnetization and $\bm P$ is the total electric polarization. The vector $\bm B_s$ is a static magnetic field, while $\bm E(t)$ and $\bm B(t)$ are, respectively, the ac electric and magnetic fields of the applied wave. 
We assume that the system possesses an inverse-DM type ME coupling (see Fig.~\ref{fig:ME}) and the polarization is given by 
\begin{align}
\bm{P} 
 & = \sum_{j} \bm{P}_{j,j+1} 
	 = g_{\rm me}
	 \sum_{j} 
	 \bm{e}_{j,j+1} \times 
	(\bm m_{j} \times \bm m_{j+1}), \label{eq:Classical_invDM}
\end{align}
where $g_{\rm me}$ is the ME coupling constant, and ${\bm e}_{j,j+1}$ is the unit vector connecting the $j$- and $(j+1)$-th sites. The torque term ${\bm H}_{\bm r}(t)$ is estimated as 
\begin{align}
  \bm{H}_{\bm r}(t) &= 
  \sum_{k=j\pm 1} \Big[J \bm m_{k}
	 + \bm{D}_{j,k}(t) \times \bm m_{k}\Big] 
     + \bm{B}_s + \bm{B}(t), 
    \label{eq:Classical_Torque}
\end{align}
where $\bm{D}_{j,k}(t)=g_{\rm em}{\bm E}(t)\times {\bm e}_{j,k}$. This model is a classical analog of the quantum spin chain model studied in Ref.~\cite{Sato2016}. 
We set the spin chain along the $x$ direction ($\bm e_{j,j+1}\parallel \hat x$) and the ac fields are circularly polarized in the $x$-$y$ plane: $\bm E(t)=E_{\rm ac}(\sin(\omega t),\cos(\omega t),0)$ and $\bm B(t)=B_{\rm ac}(\cos(\omega t),-\sin(\omega t),0)$ with $\omega$ being the laser frequency. 
From the first-order high-frequency expansion, 
one obtains the following effective torque,
\begin{align}
 \bm{H}_{{\rm eff},j} = &
	\sum_{k=j\pm 1} 
	\Big[J \bm m_{k}
	 + \bm{D}_{{\rm eff},j,k} \times \bm m_{k}\Big] + \bm{B}_{\rm eff} + \bm{B}_s\nonumber \\
	 & - \alpha \bm{B}_{\rm eff} \times \bm m_{j}
	 \nonumber \\
	& - \sum_{k=j\pm 1} 
	{\alpha \epsilon_B\epsilon_E 
         \over 2 m_s(1+\alpha^2)\omega  }
\begin{pmatrix}
      m_s^2 + m_{k}^y\delta m_{j,k}^y\\
      - m_{k}^y\delta m_{j,k}^x\\
      0
\end{pmatrix},
\label{eq:Classical_EffTorque}
\end{align}
where $\epsilon_E=g_{\rm me}E_{\rm ac}$, $\epsilon_B=B_{\rm ac}$ are the light-matter coupling constants, and 
$\delta m_{j,k}^{\alpha}=m_j^\alpha-m_k^\alpha$. 
Vectors $\bm{D}_{{\rm eff},j,k}$ and $\bm B_{\rm eff}$ we have introduced are defined as
\begin{align}
    \bm{D}_{{\rm eff},j,k} &=
    {\epsilon_E \epsilon_B 
    \over  2 ( 1 + \alpha^2) \omega } 
    \bm{e}_{j,k}
    = D_{\rm eff} \bm{e}_{j,k}, \nonumber \\
    \bm{B}_{\rm eff} 
    &= {\epsilon_{B}^2 
    \over 2 ( 1 + \alpha^2) \omega} \hat{z}.
    \label{eq:ClassicalSpin_factors}
\end{align}
For mode detail of computing these high-frequency expansion terms, see Ref.~\cite{Higashikawa2018}. 
One finds that the effective magnetic field (torque) $\bm{H}_{{\rm eff},j}$ has no time dependence. If the Gilbert damping is small enough ($\alpha\ll 1$), the second and third lines in Eq.~\eqref{eq:Classical_EffTorque} can be ignored. 
In such a weak-damping regime, the original LLG equation for the periodically driven spin chain is approximated by the LLG equation for the static classical Hamiltonian,
\begin{align}
     \mathcal{H}_{\rm chain}^{\rm eff} = &
	- \sum_{j=1}^L 
    \Big[ J \bm m_j \cdot \bm m_{j+1}
	 + D_{\rm eff} (\bm m_j \times \bm m_{j+1})^x \nonumber\\
     &  + (\bm{B}_{\rm eff} + \bm{B}_s) \cdot {\bm m}_j
	\Big].
\label{eq:Classical_effChain}
\end{align}
The second term is the photo-induced synthetic DM interaction with the DM vector ${\bm D}_{\rm eff}\parallel \hat x$. From Eq.~\eqref{eq:ClassicalSpin_factors}, one finds that teh syntehtic DM term occurs only when both $\epsilon_E$ and $\epsilon_B$ are finite. This is reminiscent of the synthetic DM term in the quantum model, Eq.~\eqref{eq:eff_2spin}~\cite{Sato2016}. The effective static field $\bm B_{\rm eff}$ in the third term is also analogous to the IFE in quantum spin systems~\cite{Takayoshi2014a,Takayoshi2014b,Sato2016,Ikeda2020,Ikeda2021} (see Eq.~\eqref{eq:eff_2spin}).
For the ferromagnetic case ($J>0$), the effective model of Eq.~\eqref{eq:Classical_effChain} is equivalent to the famous spin chain model hosting a chiral-soliton-lattice ground state~\cite{Togawa2012,Togawa2016}. 
Therefore, if we apply an intense circularly polarized THz or GHz wave to the multiferroic spin chain~\eqref{eq:classical_chain}, the vector spin chirality,
\begin{align}
    V^x_{\rm tot}=\sum_j(\bm m_j \times \bm m_{j+1})^x,
\end{align}
is expected to emerge. This phenomenon may be referred to as a typical FE in classical many-spin systems.

\begin{figure}[t]
\begin{center}
\includegraphics[width=8.5cm]{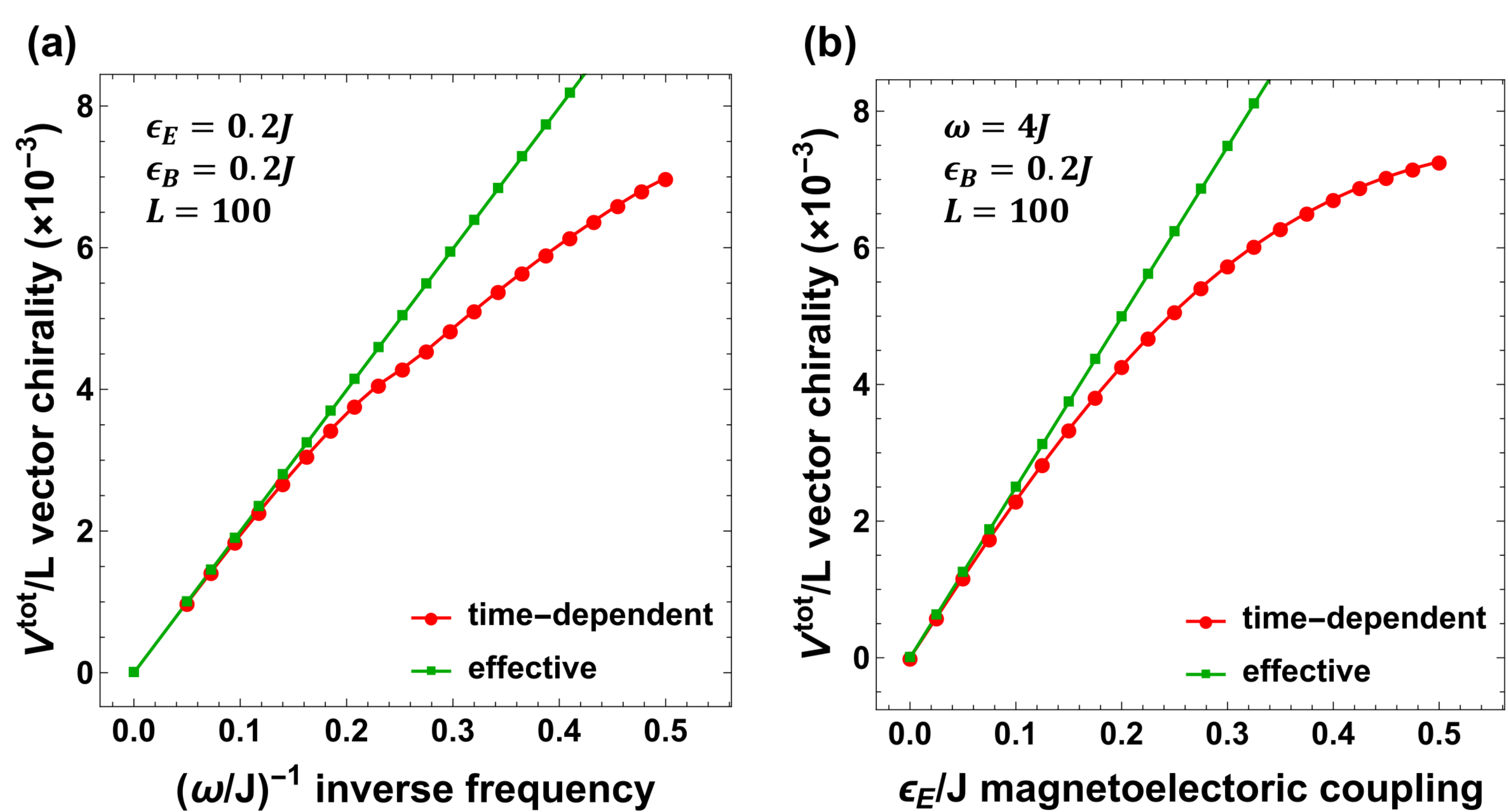}
\caption{(a) Laser frequency $\omega$ and (b) ME coupling $\epsilon_E$ dependences of the photo-induced vector spin chirality $V_{\rm tot}^x/L$ in the NESS of the periodically driven classical spin chain model with open boundary condition. Red and green curves are, respectively, the numerical results of the original LLG equation for the time-dependent Hamiltonian~\eqref{eq:Classical_effChain} and the effective LLG equation for the effective static Hamiltonian~\eqref{eq:Classical_effChain}. 
When the frequency $\omega$ (the coupling $\epsilon_E$) is small enough, the high-frequency expanded result (red curve) well coincides with the numerical exact result. For simplicity, we set zero temperature $T=0$. The system size is $L=100$. 
Reprinted from Ref.~\cite{Higashikawa2018}.}
\label{fig:FE_Chiral_Classical}
\end{center}
\end{figure}
To detect clear evidence for this expected FE, we numerically solve the LLG equations for both the original model~\eqref{eq:classical_chain} and the effective model~\eqref{eq:Classical_effChain}.  
Figure~\ref{fig:FE_Chiral_Classical} shows the numerically computed spin chirality $V_{\rm tot}^x$ in these two models after a long drive (namely, the system approaches a NESS).  
One finds that when the laser frequency $\omega$ (or the ME coupling $\epsilon_E$) is sufficiently small, the effective model well describes the spin chirality in the original model. 
This is an expected result because the high-frequency expansion is more reliable when $g/(\hbar\omega)$ becomes smaller, where $g$ is the typical strength of the light-matter interaction (see the discussions in Sec.~\ref{sec:FloquetETH}). 

Finally, we note that in recent years, the above high-frequency expansion technique for the LLG equation~\cite{Higashikawa2018} has been applied to several issues associated to opto-spintronics~\cite{Hirosawa2022,Yamabe2023}.

\section{Summary}
\label{sec:Summary}
In this final section, we summarize the contents of this review paper. 
We have reviewed the fundamentals of the Floquet theory in closed systems in Sec.~\ref{sec:Closed}, focusing on the high-frequency expansion method. 
As concrete examples of many-body FEs, we discuss the Floquet topological insulators, IFEs in metal, and FEs in magnets. 
The theoretical tools in Sec.~\ref{sec:Closed} also becomes the basis for open systems. Section~\ref{sec:Open} is devoted to the Floquet theory in open quantum systems. We have used the GKSL equation, the EOM for density matrices, to analyze the open systems. 
In particular, we argue how the NESS is described in the GKSL equation formalism. We compare the NESS and the FG state by changing the conditions of the systems (see Tables~\ref{tab:comparison1} and \ref{tab:comparison2}). 
We also discuss dissipation-assisted FEs in the spin-1 model of the diamond NV center and the Heisenberg spin chain model. 
In Sec.~\ref{sec:Open_classical}, we consider the Floquet theory in open classical systems. Generally, the Floquet theorem cannot be applied directly to the EOM in classical systems. However, we can overcome this issue by considering the distribution function of the variables, the Fokker-Planck equation. We have explained this method and applied it to the Kapitza pendulum and the LLG equation. 

Finally, we remark that theoretical studies for driven open quantum systems have continuously progressed and their research topics are diverse~\cite{Haddadfarshi2015,Schnell2020,Schnell2021,Hartmann2017,Dai2017,Mizuta2021,Michishita2021,Terada2024,Mori2023,Dinc2025}: High-frequency expansion methods, derivation of the GKSL type equations, non-Markovian effects, the condition for the existence of the Lindbladian, the limitation of the relaxation-time approximation, etc.

\vspace{1cm}
\section*{Acknowledgments}
We thank Minoru Kanega, Amane Takano, Reiya Suzuki, Miho Tanaka, Takeshi Hasebe, Sho Higashikawa, Koki Chinzei, Takashi Oka, Shintaro Takayoshi, Takeshi Mori, Martin Holthaus, Ryusuke Matsunaga, Yuta Murotani, Tomohiro Fujimoto, Zhenya Zhang and Hideki Hirori for fruitful discussions about Floquet theory and photo-induced phenomena. 
M. S. was supported by JSPS KAKENHI (Grants No.~JP25K07198, No.~JP25H02112, No.~JP22H05131, No.~JP25H01609 and No.~JP25H01251) and JST, CREST Grant No.~JPMJCR24R5, Japan.
T.~N.~I. was supported by JSPS KAKENHI Grant No.~25K07178.


%

\end{document}